\documentclass[11pt]{article}

\usepackage{amsthm,amsmath,amssymb,bbm}
\usepackage{natbib}
\usepackage{multirow}
\usepackage[pdftex]{graphicx}
\usepackage{subfigure}
\usepackage{makecell}
\usepackage{booktabs}
\usepackage{array}
\usepackage{tabularx}
\usepackage{tabulary}
\usepackage{caption}
\usepackage{booktabs}
\usepackage{url}
\usepackage{algorithm}
\usepackage{algorithmic}
\usepackage{bm}
\usepackage{wrapfig}
\usepackage{lipsum}	
\usepackage{mathrsfs}
\usepackage{dsfont}
\usepackage{titling}
\usepackage{paralist}
\usepackage{smile}

\usepackage{xr}
\usepackage[usenames,dvipsnames,svgnames,table]{xcolor}
\usepackage[colorlinks,
linkcolor=blue,
urlcolor = blue,
anchorcolor=blue,
citecolor=blue
]{hyperref}

\newcommand*{\var}{\textnormal{var}}

\newcommand{\nn}{\nonumber}

\def\##1\#{\begin{align}#1\end{align}}
\def\$#1\${\begin{align*}#1\end{align*}}

\usepackage{relsize}
\newcommand{\T}{{\mathsmaller {\rm T}}}

\def\sn{\sum_{i=1}^n}

\newcommand{\BB}{\mathbb{B}}

\newcommand{\wt}{\widetilde}

\newcommand{\Rom}[1]{\text{\uppercase\expandafter{\romannumeral #1\relax}}}

\usepackage{geometry}
\usepackage{setspace}

\usepackage{enumitem}

\numberwithin{equation}{section}
\newcommand{\ES}{{\rm ES}}

\begin{document}

\title{Robust Estimation and Inference for Expected Shortfall Regression with Many Regressors}

\author{Xuming He,~~Kean Ming Tan ~and~~Wen-Xin Zhou}

\date{}
\maketitle

\vspace{-.5in}

\begin{abstract}
Expected Shortfall (ES), also known as superquantile or Conditional Value-at-Risk, has been recognized as an important measure in risk analysis and stochastic optimization, and is also finding applications beyond these areas. In finance, it refers to the conditional expected return of an asset given that the return is below some quantile of its distribution. In this paper, we consider a recently proposed joint regression framework that simultaneously models the quantile and the ES of a response variable given a set of covariates, for which the state-of-the-art approach is based on minimizing a joint loss function that is non-differentiable and non-convex. This inevitably raises numerical challenges and limits its applicability for analyzing large-scale data. Motivated by the idea of using Neyman-orthogonal scores to reduce sensitivity with respect to nuisance parameters, we propose a statistically robust (to highly skewed and heavy-tailed data) and computationally efficient two-step procedure for fitting joint quantile and ES regression models. With increasing covariate dimensions, we establish explicit non-asymptotic bounds on estimation and Gaussian approximation errors, which lay the foundation for statistical inference. Finally, we demonstrate through numerical experiments and two data applications that our approach well balances robustness, statistical, and numerical efficiencies for expected shortfall regression.
\end{abstract}


 \let\thefootnote\relax\footnotetext{Xuming He: Department of Statistics, University of Michigan, Ann Arbor, MI, 48109, USA.  Email: \href{xmhe@umich.edu}{\textsf{xmhe@umich.edu}}. 
 Kean Ming Tan: Department of Statistics, University of Michigan, Ann Arbor, MI, 48109, USA.  Email: \href{mailto:keanming@umich.edu}{\textsf{keanming@umich.edu}}.
 Wen-Xin Zhou: Department of Mathematical Sciences, University of California, San Diego, La Jolla, CA, 92093, USA.  Email:\href{mailto:wez243@ucsd.edu}{\textsf{wez243@ucsd.edu}}.
}
 
\section{Introduction}
\label{sec:1}

Expected shortfall (ES), also known as superquantile or conditional Value-at-Risk, has been recognized as an important risk measure with versatile applications in finance \citep{AT2002, RU2002}, management science \citep{BT1986, DE2017}, operations research \citep{RU2000,RRM2014}, and clinical studies \citep{HHH2010}.
For example, in finance, expected shortfall refers to the expected return of an asset or investment portfolio conditional on the return being below a lower quantile of its distribution, namely its Value-at-Risk (VaR). In their Fundamental Review of the Trading Book \citep{BC2016, BC2019}, the Basel Committee on Banking Supervision confirmed the replacement of VaR with ES as the standard risk measure in banking and insurance.

Let $Y$ be a real-valued random variable with finite first-order absolute moment, $\EE|Y | <\infty$, and let $F_Y$ be its cumulative distribution function.
For any $\alpha \in (0, 1)$, the quantile and ES at level $\alpha$ are defined as
\#
	Q_{\alpha}(Y) = F_{Y}^{-1}(\alpha) =  \inf\{ y \in \RR : F_{Y}(y) \geq \alpha \}  
	~\mbox{ and }~ 
		\ES_{\alpha}(Y) = \EE \{  Y | Y \leq Q_{\alpha}(Y) \}, 
		\label{ES.def1}
\#
respectively. 
If $F_Y$ is continuous,  the $\alpha$-level ES is equivalently given by (see, e.g., Lemma~2.16 of \cite{MFE2015})
\#
	\ES_\alpha(Y) = \frac{1}{ \alpha} \int_0^\alpha  Q_u(Y) \,{\rm d} u.  
\#
For instance, in socioeconomics applications, $Y$ is the income and  $\ES_\alpha(Y)$ can be interpreted as the average income for the subpopulation whose income falls below the $\alpha$-quantile of the entire population.   
We refer the reader to Chapter 6 of \cite{SDR2014} and \cite{RR2013} for a thorough discussion of ES and its mathematical properties.

With the increasing focus on ES and its desired properties as a risk measure, it is natural to examine the impact of a $p$-dimensional explanatory vector $X$, on the tail behavior of $Y$ through ES.  
One motivating example is the Job Training Partnership Act (JTPA), a large publicly-funded training program that provides training for adults with barriers to employment and out-of-school youths.  The goal is to examine whether the training program improves future income for adults with low-income earnings \citep{BOBCDLB1997}, for which quantile regression-based approaches have been proposed \citep{AAI2002,CH2008}. 
For example, the 0.05-quantile of the post program income refers to the highest income earning of those who have the 5\% lowest income among the entire population, while the 0.05-ES concerns the average income earning within this subpopulation and therefore is more scientifically relevant in the JTPA study.

Compared to the substantial body of literature on quantile regression,  extant works on ES estimation and inference in the presence of covariates are scarce.   We refer to \cite{S2005}, \cite{CW2008}, \cite{K2012}, \cite{LX2013} and  \cite{MYT2018} for nonparametric conditional ES estimation,  and more recently to \cite{DB2019}, \cite{PZC2019} and \cite{B2020} in the context of (semi-)parametric models.  As suggested in \cite{PZC2019},  this is partly because regulatory interest in ES as a risk measure is only recent, and also due to the fact that this measure is not \emph{elicitable} \citep{G2011}. Let $\cP$ be a class of distributions on $\RR^d$. We say that a statistical functional $\theta: \cP \to D$ with $D\subseteq \RR^p$ ($p\geq 1$) is elicitable relative to the class $\cP$ if there exists a loss function $\rho: \RR^d \times \RR^p\to \RR$ such that $\theta(F) = \argmin_{\theta \in D}   \EE_{Z \sim F} \rho(Z , \theta )$ for any $F\in \cP$.  Here $ \EE_{Z \sim F}$ means that the expectation is taken with respect to the random variable $Z$ that follows the distribution $F$.
For example, the mean is elicitable using the $L_2$-loss, and the median is elicitable using the $L_1$-loss.
Although the ES is not elicitable on its own, it is jointly elicitable with the quantile using a class of strictly consistent joint loss functions \citep{FZ2016}. Based on this result, \cite{DB2019} and \cite{PZC2019} proposed a joint regression model for the conditional $\alpha$-level quantile and ES of $Y$, given the covariates $X \in \RR^p$. In this work, we focus on (conditional) linear joint quantile-ES models:
\#
	Q_\alpha( Y |X )  = X^\T \beta^*    ~~\mbox{ and }~~ {\rm ES}_\alpha(Y|X) = X^\T \theta^*   . \label{joint.model}
\#
Equivalently, we have $ \varepsilon=Y-X^\T \beta^* $ and $\xi=Y-X^\T \theta^* $, where  the conditional $\alpha$-quantile of $\varepsilon$ and the conditional $\alpha$-level expected shortfall of $\varepsilon$,  given $X \in \RR^p$, are zero. More generally,  one may allow the quantile and the ES models to depend on different covariate vectors $X_q$ and $X_e$, respectively. In this case,  the conditional $\alpha$-quantile and $\alpha$-ES of $\varepsilon $ and $\xi$, respectively given $X=(X_q^\T, X_e^\T)^\T$, are assumed to be zero.

To jointly estimate $\beta^*$ and $\theta^*$, \cite{DB2019} and \cite{PZC2019} considered an $M$-estimator, defined as the global minimum of any member of a class of strictly consistent joint loss functions over some compact set  \citep{FZ2016}. 
The joint loss function, which will be specified in~\eqref{def:rho}, 
is non-differentiable and non-convex.  \cite{DB2019} employed the derivative-free Nelder-Mead algorithm to minimize the resulting non-convex loss, which is a heuristic search method that may converge to non-stationary points. 
From a statistical perspective, they further established consistency and asymptotic normality for the global minima. 
However, from a computational perspective, finding the global minimum of a non-convex function is generally intractable: approximating the global minimum of a $k$-times continuously differentiable function $f:\RR^p \to \RR$ to $\epsilon$-accuracy requires at least as many as $(1/\epsilon)^{p/k}$ evaluations (ignoring problem-dependent constants) of the function and its first $k$-derivatives \citep{NY1983}. The lack of differentiability makes this problem even more challenging numerically.
To mitigate the computational effort, \cite{B2020} proposed a two-step procedure by first estimating the quantile parameters via standard quantile regression, followed by least squares regression with generated response variables. 
Although being computationally efficient, the ensuing estimator is sensitive to heavy-tailed error distributions due to the use of $L_2$-loss for fitting possibly highly skewed data in the second step; see Section~\ref{sec:3.1} for a rigorous statement.

In this paper, we propose a novel two-stage method for joint quantile and expected shortfall regression under model \eqref{joint.model},  with a particular focus on the latter. 
Compared to existing approaches, our proposed method is robust against heavy-tailed errors without compromising statistical efficiency under light-tailed distributions.  Computationally, our method can be implemented via fast and scalable gradient-based algorithms. 
The main contributions of this manuscript are summarized as follows.

\begin{itemize}
\item Our method is built upon the recent approach to quantile-ES regression via a two-step procedure \citep{B2020}, for which a general non-asymptotic theory has yet to be established.
We first establish such a finite-sample theoretical framework when the dimension of the model, $p$, is allowed to increase with the number of observations, $n$. Specifically, we provide explicit upper bounds, as a function of $(n, p)$, on the estimation error (under $L_2$-risk) and (uniform) Gaussian approximation errors for the two-step ES estimator. We further construct asymptotically valid (entrywise) confidence intervals for the ES parameters. The dominant computation effort of this two-step procedure is the QR regression fit in stage one. We thus recommend using the convolution-smoothed QR method \citep{FGH2021}, solvable by fast first-order algorithms  that are scalable to very large-scale problems \citep{HPTZ2021}. Our non-asymptotic theory allows the dimension $p$ to grow with the sample size, and hence paves ways for analyzing series/projection estimators under joint nonparametric quantile-ES models \citep{BCCF2019} and penalized estimators under high-dimensional sparse models.

\item The standard two-step estimator is a least squares estimator with generated response variables, and therefore is sensitive to the tails of the distribution of $Y$. To achieve sub-Gaussian deviation bounds when the (conditional) distribution of $Y|X$ only has Pareto-like tails, we propose a robust ES regression method that applies adaptive Huber regression \citep{ZBFL2018} in the second step. The idea is to use a diverging robustification parameter $\tau = \tau(n, p) >0$ for bias-robustness tradeoff. To choose this hyper-parameter in practice, we employ a recently developed data-driven mechanism \citep{WZZZ2021}, inspired by the censored equation approach introduced in \cite{HKW1990}. 
We also develop   efficient algorithms to compute both standard and robust two-step ES estimators under additional constraints that  the fitted ES does not exceed the fitted quantile at each observation; see Section~A of the Appendix.

\item Numerically, we compare both the two-step estimator and the proposed robust variant with the joint $M$-estimator of \cite{DB2019} on large synthetic datasets generated from a location-scale model with both light- and heavy-tailed error distributions.  The joint $M$-estimator is computed by the R package \texttt{esreg} \citep{BD2019}. The proposed robust two-step procedure can be implemented by a combination of R packages \texttt{quantreg} \citep{K2022} or \texttt{conquer} \citep{conquer2022} and \texttt{adaHuber} \citep{PZ2022}. We demonstrate through numerical experiments and a real data example that the proposed robust ES regression approach achieves satisfying statistical performance, a higher degree of robustness (against heavy-tailed data), and superior computational efficiency and stability.
\end{itemize}

\medskip
\noindent
{\sc Notation}:  For any two vectors $u=(u_1, \ldots, u_k)^\T$ and $v=(v_1, \ldots ,v_k)^\T \in \RR^k$, we define their inner product as $u^\T v = \langle u,  v \rangle= \sum_{j=1}^k u_j v_j$.
We use $\| \cdot \|_p$ $(1\leq p \leq \infty)$ to denote the $\ell_p$-norm in $\RR^k$: $\| u \|_p = ( \sum_{i=1}^k | u_i |^p )^{1/p}$ for $p\geq 1$ and $\| u \|_\infty = \max_{1\leq i\leq k} |u_i|$. 
 For $r>0$, define the Euclidean ball and unit sphere in $\RR^k$ as $\BB(r) = \BB^k(r) = \{ u \in \RR^k: \| u \|_2 \leq r \}$ and $\mathbb{S}^{k-1} = \{ u \in \RR^k: \| u \|_2 = 1 \}$, respectively. Given a positive semi-definite matrix $A \in \RR^{k\times k}$ and $u \in \RR^k$,  let $\| u \|_A := \| A^{1/2} u \|_2$. We write  $\BB_A(r) =  \big\{ u \in \RR^k: \| u \|_A \leq r  \big\}$ and  	$\BB_A(u, r) = u + \BB_A(r)$. Given an event/subset $\cA$, $\mathbbm{1}(\cA)$ or $\mathbbm{1}_{\cA}$ denotes the zero-one indicator function for $\cA$. For two real numbers $a$ and $b$, we write $a \wedge b = \min\{a, b\}$ and $a \vee b = \max\{a, b\}$. For two sequences $\{a_n\}_{n \geq 1}$ and $\{ b_n\}_{n \geq 1}$ of non-negative numbers,  we write $a_n  \lesssim b_n$ if $a_n \le C b_n$ for some constant $C > 0$ independent of $n$,  $a_n \gtrsim b_n$ if  $b_n \lesssim a_n$, and $a_n  \asymp b_n$ if $a_n  \lesssim b_n$ and $a_n \gtrsim b_n$.

\section{Preliminaries and Background}
\label{sec:2}

\subsection{The Joint Regression Framework}
\label{sec:2.1}
Assume we observe a sequence of data vectors $\{ (Y_i,  X_i) \}_{i=1}^n$, where $Y_i \in \RR$ is the response variable,  and $X_i \in \RR^p$ is a $p$-dimensional vector of explanatory variables (covariates). 
For some fixed probability level $\alpha \in (0 , 1)$, denote the conditional $\alpha$-level quantile and ES of $Y_i$ given the covariates $X_i$ as $Q_{\alpha}(Y_i| X_i )$ and $\ES_\alpha( Y_i|  X_i )$, respectively.   For the latter, we adhere to the definition  $ \ES_\alpha( Y_i|  X_i ) = \EE\{ Y_i | Y_i\leq  Q_\alpha(Y_i  | X_i ),  X_i   \} $.

We consider the joint regression framework introduced in \cite{DB2019} for modeling the conditional quantile and expected shortfall.  For some probability level $\alpha \in (0 , 1)$, assume  that
\#
	Q_\alpha( Y_i|  X_i ) = X_i^\T \beta^*  , \qquad  \ES_\alpha( Y_i |    X_i ) = X_i^\T \theta^* , \label{joint.reg.model}
\# 
where $\beta^*, \theta^* \in \RR^p$ are the unknown true underlying parameters for quantile and ES, respectively.  \cite{FZ2016} explained that quantile and ES are jointly {\it elicitable}, and proposed a class of {\it strictly consistent} joint loss functions for quantile and ES estimation. 
Let $G_1$ be an increasing and integrable function, and let $\cG_2$ be a three times
continuously differentiable function such that both $\cG_2$ and its derivative $G_2 = \cG_2'$ are strictly positive.   
The proposed joint loss function in \cite{FZ2016} takes the form 
\#
	S(\beta, \theta; Y, X)  & = \{  \alpha -  \mathbbm{1}( Y \leq X^\T \beta )   \} \{  G_1(Y) - G_1(X^\T \beta) \} \label{def:rho} \\
	&~~~~ + \frac{ G_2( X^\T \theta) }{\alpha}  \big\{  \underbrace{  \alpha   X^\T ( \theta - \beta )  -   (   Y - X^\T \beta ) \mathbbm{1} (Y\leq X^\T \beta)  }_{=: \, S_0(\beta, \theta; Y, X) } \big\}  - \cG_2(X^\T \theta)   . \nn
\#
This general form also includes the joint loss function proposed by  \cite{AS2014} by taking $G_1(x) = - (W/2) x^2$ for some $W\in \RR$ and $\cG_2(x) = \alpha x^2/2$.

In the regression framework with a fixed number of covariates, \cite{DB2019} established the consistency and asymptotic normality of the $M$-estimator $(\wt \beta^\T , \wt \theta^\T)^\T$,  defined as
\#
\begin{pmatrix}
  \wt \beta  \\ 
  \wt \theta
\end{pmatrix} \in \argmin_{\beta,\theta \in \,  \Theta } \frac{1}{n} \sn S(\beta, \theta; Y_i, X_i) ,  \label{def:M-est}
\#
 where $\Theta \subseteq \RR^p$ is the parameter space,  assumed to be compact, convex, and has nonempty interior. The main challenge of the aforementioned approach is that the objective function in \eqref{def:M-est} is non-differentiable and non-convex for any feasible choice of the functions $G_1$ and $G_2$ \citep{FZ2016}. 
Note from definition \eqref{ES.def1} that the expected shortfall depends on the quantile, not vice versa.  The estimation and inference of $\theta^*$ is thus   the main challenge.   It is, however, infeasible to estimate a single regression model for ES through $M$-estimation, that is, by minimizing some strictly consistent loss function \citep{DB2019}. 

In the joint regression framework, if the main goal is to estimate and forecast ES,  then $\beta^*$ can  be naturally viewed as a nuisance parameter.  Motivated by the idea of  using Neyman-orthogonal scores to reduce sensitivity with respect to nuisance parameters \citep{N1979, CCDDHNR2018}, \cite{B2020} proposed a two-stage procedure that bypasses non-convex optimization problems.  In the first stage, an estimate $\hat \beta$ of $\beta^*$ is obtained via standard quantile regression. The second step employs an orthogonal score with fitted thresholding quantiles to estimate $\theta^*$. The key observation is as follows.  Define the function
\#
\psi_0(\beta, \theta; X) & = \EE\{ S_0( \beta, \theta; Y, X ) | X \} \label{neyman.score}  \\
& =  \alpha X^\T \theta  - \PP( Y\leq X^\T \beta  | X )  \EE (   Y   |  Y\leq X^\T \beta , X )  + \big\{ \PP (Y\leq X^\T \beta | X )   - \alpha  \big\} X^\T \beta , \nn
\#
where $S_0$ is given in \eqref{def:rho}. 
Under model \eqref{joint.reg.model}, we have $\psi_0(\beta^*, \theta^*; X)=0$ almost surely over $X$. 
Let $F_{Y | X}$ be the conditional distribution function of $Y$ given $X$. Provided $F_{Y|X}$ is continuously differentiable, taking the gradient with respect to $\beta$ on both sides of the above equality yields
\$
	 \partial_\beta   \psi_0(\beta, \theta; X)   =     \{ F_{Y|X}(X^\T \beta ) -\alpha \}  X , ~\mbox{ for any }  \beta, \theta  \in \RR^p. 
\$	
We hence refer to the following property
\#
	 \partial_\beta  \psi_0(\beta, \theta; X)  \big|_{\beta = \beta^* } =   \{ F_{Y|X}(X^\T \beta^* ) -\alpha \} X =  0  \label{neyman.cond}
\#
as {\it Neyman orthogonality}.

\subsection{Two-Step ES Estimation via Neyman Orthogonal Score}
\label{sec:2.2}
We start with a detailed overview of the two-step approach proposed by \citep{B2020} using the Neyman orthogonal score \eqref{neyman.score} under the joint model~\eqref{joint.reg.model}. 
In Section~\ref{sec:3.1}, we will develop a non-asymptotic (finite-sample) theory for the two-step ES estimator, $\hat{\theta}$, under the regime in which $p$ is allowed to increase with the sample size $n$.
We further develop asymptotic normality results for individual coordinates, or more generally linear projections,  of $\hat \theta$, in the increasing-dimension regime ``$p^2/ n = o(1)$". 
Our non-asymptotic results and techniques pave the way for analyzing high-dimensional sparse quantile-ES models.

The first step involves computing the standard QR estimator of $\beta^*$:
\#
	\hat \beta \in \argmin_{\beta \in \RR^p} \frac{1}{n} \sn \rho_\alpha( Y_i - X_i^\T \beta),\label{qr.est}
\#
where $\rho_\alpha(u) = \{ \alpha - \mathbbm{1} (u<0) \} u$ is the check function \citep{KB1978}.  
The second step is motivated by the orthogonal score $S_0$ in \eqref{def:rho}.  
Specifically, let $\hat \cL(  \beta , \theta  )  =  (1/n) \sn S^2_i(  \beta, \theta )$ be the joint empirical loss with
\#
	S_i (\beta, \theta) := S_0(\beta, \theta; Y_i, X_i) =  \alpha X_i^\T \theta  -   \mathbbm{1}(Y_i\leq X_i^\T \beta) Y_i  +  \{   \mathbbm{1}(Y_i\leq X_i^\T \beta) - \alpha    \}  X_i^\T \beta    . \label{def:Si}
\# 
Given $\hat{\beta}$ obtained from the first step, the ES estimator $\hat \theta$ of $\theta^*$ is computed as
\#
	\hat \theta \in \argmin_{\theta \in \RR^p} \hat \cL\,( \hat \beta , \theta ). \label{ES.est}
\#

For any $\beta$ fixed, the function $\theta \mapsto \hat \cL(  \beta , \theta  )$ is convex with gradient and Hessian given by
\$
 \partial_\theta \hat \cL(  \beta , \theta  )  = \frac{2 \alpha }{n}\sn S_i (\beta, \theta) X_i  ~~\mbox{ and }~~ \partial^2_\theta\hat \cL(  \beta , \theta  ) = \frac{2 \alpha^2 }{n} \sn X_i X_i^\T ,
\$
respectively. 
By the first-order condition, the ES regression estimator $\hat \theta$ satisfies the moment condition $ \partial_\theta \hat \cL(  \hat \beta , \hat \theta  ) = 0$, and has a closed-form expression
\#
 \hat \theta =  \hat \beta + \bigg( \sn X_i X_i^\T \bigg)^{-1}  \frac{1}{\alpha }  \sn  ( Y_i - X_i^\T \hat \beta ) X_i \mathbbm{1}(Y_i\leq X_i^\T \hat \beta)    , \label{two-step.ES.est1}
\#
provided that $\mathbb{X} = (X_1, \ldots, X_n)^\T \in \RR^{n\times p}$ is full-rank.

\begin{remark} \label{rmk:conquer}
When $p$ is large,  we suggest using the convolution-smoothed quantile regression (conquer) estimator \citep{FGH2021, HPTZ2021} in the first step, which can be computed by fast and scalable gradient-based algorithms.  Given a smoothing parameter/bandwidth $h>0$, the conquer estimator $\hat \beta_h$ minimizes the convolution smoothed loss function $ \hat \cQ_h(\beta) = (1/n)\sn \rho_{\alpha, h}(Y_i - X_i^\T \beta)$ with 
\#
	\rho_{\alpha, h}(u) = (\rho_\alpha * K_h) (u) = \int_{-\infty}^\infty \rho_\alpha(u) K_h(v-u) \, {\rm d} v,  \label{conquer.est}
\#
where $K_h(u) := (1/h) K(u/h)$ for some non-negative kernel function $K$, and $*$ is the convolution operator. We refer to \cite{FGH2021} and \cite{HPTZ2021} for more details, including both asymptotic and finite-sample properties of $\hat \beta_h$ when $p$ is fixed and growing (with $n$) as well as the bandwidth selection.
\end{remark}

Define $p \times p$ matrices 
\$ 
	\Sigma=\EE(X X^\T)  ~~\mbox{ and }~~ \Omega = \EE ( \omega^2 X X^\T )  
\$
with $\omega := (Y - X^\T \beta^*) \mathbbm{1}(Y \leq X^\T \beta^*) + \alpha X^\T(\beta^* - \theta^*)$ satisfying $\EE(\omega | X) = 0$ under model \eqref{joint.reg.model}.  Provided that $p=p_n$ satisfies $p^2/n \to 0$, we will show in Theorem~\ref{thm:CLT} that $\hat{\theta}_j$ is asymptotically normal:
\$
	 \frac{ \alpha  \sqrt{n} (\hat \theta_j - \theta^*_j )}{\sqrt{( \Sigma^{-1} \Omega \Sigma^{-1} )_{jj}}} \xrightarrow {\rm d} \cN(0, 1) ~\mbox{ as }~ n,p \to \infty. 
\$
As a direct implication, an asymptotically valid entrywise confidence interval for  $\theta^*$ can be constructed as follows.
Recall that $(\hat \beta, \hat \theta)$ is the joint quantile-ES regression estimators given in \eqref{qr.est} and \eqref{ES.est}, respectively.  
Define the estimated ``residuals" as
\#
	\hat \varepsilon_i = Y_i - X_i^\T \hat \beta ~~\mbox{ and }~~ \hat \omega_i  = \hat \varepsilon_i  \mathbbm{1}(\hat \varepsilon_i  \leq 0) + \alpha X_i^\T(\hat \beta - \hat \theta )  . \label{def:est.residual}
\# 
We then use the sample analog of $\Sigma$ and a plug-in estimator of $\Omega$:
\#
	\hat \Sigma = \frac{1}{n} \sn X_i X_i^\T , \quad \hat \Omega  = \frac{1}{n} \sn \hat \omega_i^2  X_i X_i^\T .  \label{def:est.covariance}
\#
Consequently,  we construct (approximate)  95\% confidence interval for each coefficient as 
\begin{equation}
\label{eq:twostep:ci}
	\bigg[ \hat \theta_j  -   \frac{ 1.96}{\alpha \sqrt{n}}  ( \hat \Sigma^{-1} \hat \Omega  \hat \Sigma^{-1} )^{1/2}_{jj} ,    \, \hat \theta_j +    \frac{1.96}{\alpha \sqrt{n}}  ( \hat \Sigma^{-1} \hat \Omega  \hat \Sigma^{-1} )^{1/2}_{jj} \bigg] , \ \ j= 1,\ldots, p.
\end{equation}

\section{Robust Expected Shortfall Regression} 
\label{sec:resr}
\subsection{Motivation} 
\label{sec:3.1}

The two-step estimator $\hat \theta$ given in \eqref{two-step.ES.est1} is essentially a least squares estimator (LSE) with generated response variables.
While the two-step procedure is computationally efficient and enjoys nice asymptotic properties, due to the use of the least squares type loss, it is sensitive to outliers or heavy-tailed data  that is ubiquitous in various areas such as climate, insurance claims, and genomics data. In particular, heavy-tailedness has become a well-known stylized fact of financial returns and stock-level predictor variables \citep{C2001}.
Since the expected shortfall is a quantity that describes the tail behavior of a distribution, it is important to construct an estimator that is robust to the power-law or Pareto-like tails.

To motivate the need for a robust ES estimator, we start with the non-regression setting in which $X_i \equiv 1$. The two-step ES estimator \eqref{two-step.ES.est1} can then be simplified as
\#
	\hat{\ES}_\alpha =     \frac{1}{\alpha n} \sn  Y_i \mathbbm{1}\{ Y_i\leq \hat Q_\alpha \} +  \hat Q_\alpha \{ 1 - \hat F(\hat Q_\alpha ) / \alpha   \}  ,  \label{def:univariate.ES}
\#
where $\hat F$ is the empirical CDF of $Y$ and $\hat Q_\alpha = \hat F^{-1}(\alpha)$ is the sample quantile. The estimator $\hat{\ES}_\alpha$ \eqref{def:univariate.ES} coincides with the ES estimate (4) in \cite{BKK2004}, although the latter is motivated differently by the following property:
$$
	\ES_\alpha(Y) =  \EE(Y) - \frac{1}{\alpha} \min_{\beta \in \RR} \EE \rho_\alpha(Y - \beta)   .
$$
Since $| \hat  F(\hat Q_\alpha ) - \alpha| \leq 1/n$,  up to higher-order terms,  $\hat{\ES }_\alpha$ equals $(\alpha n )^{-1} \sn Y_i \mathbbm{1}\{ Y_i\leq \hat Q_\alpha \}$ which, by the consistency of sample quantiles,  is first-order equivalent to the ``oracle" ES estimator $\hat{\ES_\alpha^{{\rm ora}} }  := (\alpha n )^{-1} \sn Y_i \mathbbm{1}\{ Y_i\leq  Q_\alpha(Y)\}$.

Since the truncated variable $Y_i\mathbbm{1}\{Y_i\leq  Q_\alpha(Y)\}$ can be highly left-skewed with heavy tails, the corresponding empirical mean is sensitive to the (left) tails of the distribution of $Y$, and hence lacks robustness against heavy-tailed data. 
Specifically, let $X_1, \ldots, X_n$ be i.i.d.~random variables with mean $\mu$ and variance $\sigma^2>0$. When $X_i$ is sub-Gaussian (i.e., $\EE(e^{\lambda X_i}) \leq \lambda^2 \sigma^2/2$ for any $\lambda \in \RR$), it follows from the Chernoff bound \citep{C1952} that 
\#
	 \PP \big\{ |  \bar X_n - \mu | \geq   \sigma \sqrt{ 2 \log (2/\delta) /n} \big\} \leq \delta , \ \ {\rm valid~for~any~} \delta \in (0,1 ). \label{chernoff}
\#
In other words, the sample mean $\bar{X}_n =(1/n) \sn X_i$ satisfies the sub-Gaussian deviation bound.
On the other hand, the following proposition provides a lower bound for the deviations of the empirical mean $(1/n) \sn Y_i \mathbbm{1}\{ Y_i\leq  Q_\alpha(Y)\}$ when the  distribution of $Y$ is the least favorable among all heavy-tailed distributions with mean zero and variance $\sigma^2$.

\begin{proposition} \label{prop:lbd}
For any value of the standard deviation  $\sigma>0$ and any probability level $\delta \in (0, e^{-1}]$, there exists some distribution with  mean zero and variance $\sigma^2$ such that for any $\alpha \in (0,1)$, the i.i.d.~sample $\{ Y_i\}_{i=1}^n$ of size $n$ drawn from it satisfies 
\#
	\PP \bigg[  \frac{1}{n} \sn  Y_i \mathbbm{1}( Y_i \leq Q_\alpha ) - \EE  \{Y  \mathbbm{1}( Y  \leq Q_\alpha )\} \leq   -  \sigma \sqrt{\frac{1}{\delta n }}  \cdot \frac{  1- e\delta }{ \sqrt{2 e} }   \bigg] \geq \delta, \label{es.lbd}
\#
as long as $n\geq e\delta/\alpha$, where $Q_\alpha=Q_\alpha(Y)$ is the $\alpha$-th quantile of $Y$.
\end{proposition}
Together, the  upper and lower bounds \eqref{chernoff} and \eqref{es.lbd} show that the worst case deviations of the empirical mean are suboptimal when the underlying distribution is heavy-tailed (as opposed to having Gaussian-like thin tails). 
If $Y$ follows a heavy-tailed distributed, such as the $t$- or Pareto distributions, then the left-truncated variables $Z_i :=  Y_i  \mathbbm{1}\{ Y_i \leq Q_\alpha(Y) \}$ have not only heavy but also asymmetric tails. In this case, the empirical mean $(\alpha n)^{-1} \sn Z_i$ can be a sub-optimal estimator of $\ES_\alpha(Y)$.

\subsection{Robust Estimation and Inference via the Adaptive Huber Regression}
\label{sec:2.3}
To robustify the ES regression estimator~\eqref{two-step.ES.est1} in the presence of skewed heavy-tailed observations, we utilize the idea of adaptive Huber regression in \citet{ZBFL2018}.
For some $\tau>0$,  the Huber loss \citep{H1973} takes the form
\#
	\ell_\tau (u ) = \begin{cases}
   u^2 /2    & ~\mbox{ if }~ |u | \leq \tau,    \\
  \tau | u | - \tau^2 /2  &~\mbox{ if }~ |u| > \tau.
\end{cases}
\#
We propose a robust/Huberized ES regression estimator defined as
\#
 \hat \theta_\tau   \in   \argmin_{\theta \in \RR^p}  \frac{1}{n} \sn  \ell_\tau ( S_i(\hat \beta , \theta) )   , \label{robust.ES.est}
\# 
where $S_i(\hat \beta, \theta)$ is as defined in~\eqref{def:Si}, and  $\tau >0$ is a robustification parameter that should be calibrated adaptively from data.   

To see this, we consider the oracle Huber ES estimator defined as: 
\#
 \hat \theta^{{\rm ora}}_\tau   \in   \argmin_{\theta \in \RR^p}  \frac{1}{n} \sn  \ell_\tau ( S_i( \beta^* , \theta) )  = \argmin_{\theta \in \RR^p}  \frac{1}{n} \sn \ell_\tau(  Z_i - \alpha X_i^\T \theta )   , \label{robust.oracle.ES}
\#
where $Z_i = (Y_i - X_i^\T \beta^*) \mathbbm{1} (Y_i \leq X_i^\T \beta^* ) + \alpha X_i^\T \beta^*$.  For any $\tau>0$,  $\hat \theta^{{\rm ora}}_\tau$ is an $M$-estimator of its population counterpart
\$
	\theta^*_\tau = \argmin_{\theta \in \RR^p} \EE  \{ \ell_\tau (  Z_i - \alpha X_i^\T \theta   ) \}. 
\$
Let  $\psi_\tau(t)  =\ell_\tau'(t) = \sign(t) \min( |t|, \tau)$ be the Huber's score function. By the convexity of the Huber loss, $\theta^*_\tau$ must satisfy the first-order condition $\EE\{ \psi_\tau( Z_i - \alpha X_i^\T \theta^*_\tau) X_i \} = 0$.  On the other hand, define the ES deviations $\omega_i = Z_i -  \alpha X_i^\T \theta^*$, satisfying $\EE(\omega_i | X_i) = 0$ and $\EE(\omega_i) =0$.  Since the conditional distribution of $\omega_i$ given $X_i$ is asymmetric, in general   we have $\EE\{ \psi_\tau( Z_i - \alpha X_i^\T \theta^*) X_i \} = \EE \{\psi_\tau(\omega_i) X_i \} \neq 0$, which in turn implies that $\theta^*_\tau \neq \theta^*$. We thus refer to their difference under the $\ell_2$-norm,  $\| \theta^*_\tau - \theta^* \|_2$, as the robustification bias.  Proposition~\ref{prop:bias} provides an upper bound for the robustification bias,  which depends on $\tau$ and some moment parameter. In particular, $\tau$ needs to diverge for the robustification bias to diminish.  

\begin{proposition} \label{prop:bias}
Assume that $\varepsilon:= Y - X^\T \beta^*$ satisfies $\var_X\{\varepsilon \mathbbm{1} (\varepsilon\leq 0) \} \leq \overbar \sigma^2$ almost surely for some constant $\overbar \sigma^2$, and that $\kappa_4 =  \sup_{u\in \mathbb{S}^{p-1}} \EE \langle u, \Sigma^{-1/2} X \rangle^4 < \infty$, where $\Sigma = \EE(X X^\T ) $ is positive definite.  Then, for any $\tau \geq 2 \kappa_4^{1/4} \overbar \sigma$,  we have $\| \theta^*_\tau - \theta^* \|_\Sigma \leq 2 \overbar \sigma^2 / (\alpha \tau )$.
\end{proposition}

In Section~\ref{sec:3.2},  we investigate the finite-sample properties of the robust ES estimator $\hat \theta_\tau$ obtained via \eqref{qr.est} and \eqref{robust.ES.est}: our results include a deviation inequality for $\|\hat \theta_\tau - \theta^*\|_\Sigma$ (Theorem~\ref{thm:huber.ES}),  the Bahadur representation (Theorem~\ref{thm:ES.bahadur}), and a Berry-Esseen bound for linear projections of $\hat \theta_\tau$ and $\hat \theta^{{\rm ora}}_\tau$ (Theorem~\ref{thm:ES.CLT}).  With a properly chosen $\tau$ that is of order $\tau \asymp \overbar \sigma \sqrt{ n/p}$,  we will show that  $\alpha \|\hat \theta_\tau - \theta^*\|_\Sigma \lesssim \overbar \sigma \sqrt{  p / n}$ with high probability. {Moreover,  for any deterministic vector $a\in \RR^p$, the standardized statistic $\alpha  \sqrt{n}  \langle a, \hat \theta_\tau - \theta^* \rangle / \varrho_{a}$  converges in distribution to $\cN(0, 1)$, where $\varrho^2_{a } = a^\T \Sigma^{-1} \Omega \Sigma^{-1} a$ and $\omega = (Y - X^\T \beta^*)\mathbbm{1}(Y \leq X^\T \beta^*) + \alpha X^\T (\beta^* - \theta^*)$. Our theoretical analysis reveals two attractive properties of the adaptive Huberized ES estimator $\hat \theta_\tau$: (i) the non-asymptotic deviation upper bounds for $\hat \theta_\tau$ are much smaller in order than those for $\hat \theta$ at any given confidence level, and (ii) the asymptotic relative efficiency of $\hat \theta_\tau$ to $\hat \theta$ is one. 
Moreover,  Theorem~\ref{thm:ES.CLT} shows that  the two-step robust estimator (with estimated conditional quantiles) is asymptotically equivalent to the oracle Huberized estimator \eqref{robust.oracle.ES} (assuming $\beta^*$ were known).   This further justifies the usefulness of the Neyman orthogonal score, which makes the QR estimation error first-order negligible.}

{
Consistent estimators of $\Sigma$ and $\Omega = \EE(  \omega^2 X X^\T )$  are useful for statistical inference.  Given the pair of quantile-ES regression estimators $(\hat  \beta, \hat \theta_\tau)$,  with slight abuse of notation we use $\hat \varepsilon_i$ and $\hat \omega_i$ to denote the fitted QR and ES residuals as in  \eqref{def:est.residual} except with $\hat \theta$ replaced by $\hat \theta_\tau$. 
As discussed in Section~\ref{sec:2.2}, a naive estimate of $\Omega$ is $\hat \Omega = (1/n) \sn \hat w_i^2 X_i X_i^\T$. In the presence of heavy-tailed errors $\varepsilon_i$, even the ``oracle" estimate $\wt \Omega = (1/n) \sn   w_i^2 X_i X_i^\T$ performs poorly and tends to overestimate. Motivated by Huber regression, we further propose a simple truncated estimator of $\Omega$ given by
\#
	\hat \Omega_\gamma =  \frac{1}{n} \sn \psi_\gamma^2(\hat \omega_i) X_i X_i^\T, \label{robust.omega} 
\#
where $\gamma = \gamma(n,p) >0$ is a second robustification parameter. Consequently, we construct approximate 95\% robust confidence intervals for $\theta^*_j$'s as
\begin{equation}
\label{eq:ci}
	\bigg[ \hat \theta_{\tau , j } -   \frac{ 1.96}{\alpha \sqrt{n}}  ( \hat \Sigma^{-1} \hat \Omega_\gamma \hat \Sigma^{-1} )^{1/2}_{jj} ,    \, \hat \theta_{\tau , j } +    \frac{1.96}{\alpha \sqrt{n}}  ( \hat \Sigma^{-1} \hat \Omega_\gamma \hat \Sigma^{-1} )^{1/2}_{jj} \bigg] , \ \ j= 1,\ldots, p . 
\end{equation}
The convergence rate of $\hat \Omega_\gamma$ with a suitably chosen $\gamma$ will be discussed in Section~\ref{sec:3.2}.
}

As discussed previously,  the robustification parameter $\tau$ plays an important role in balancing the bias and robustness (against heavy-tailed error distributions). The former is due to the asymmetric nature of the ES residual $\omega = \varepsilon  \mathbbm{1}(\varepsilon  \leq 0) + \alpha X^\T(\beta^* - \theta^*)$ with $\varepsilon = Y - X^\T \beta^*$. 
Under a second moment assumption, that is, $\var_X\{ \varepsilon  \mathbbm{1}(\varepsilon  \leq 0) \} \leq \overbar \sigma^2$ (almost surely) for some $\overbar \sigma>0$, Theorem~\ref{thm:huber.ES} shows that $\tau$ should be of order $\overbar \sigma \sqrt{n/ p}$ so that the resulting ES estimator $\hat \theta_\tau$ satisfies sub-Gaussian deviation bounds. 
To choose the tuning parameter $\tau$ in practice, we employ a recently developed data-driven mechanism \citep{WZZZ2021}. This method is inspired by the censored equation approach proposed by \cite{HKW1990}, which was originally introduced as a proof technique for deriving robust weak convergence theory for self-normalized sums.

Given an initial QR estimator $\hat \beta$, define the generated response variables 
$$
	\hat Z_i =(Y_i - X_i^\T  \hat \beta) \mathbbm{1} (Y_i \leq X_i^\T \hat \beta ) + \alpha X_i^\T \hat \beta, \ \ i =1 ,\ldots , n .
$$
The proposed $\tau$-calibration procedure is iterative, starting at iteration 0 with an initial estimate $\theta^{ 0 } = \hat \theta$, which is the two-step ES estimator given in \eqref{ES.est} or equivalently \eqref{two-step.ES.est1}. At iteration $t = 0, 1, 2, \ldots$, it solves a censored equation to update its estimate $\theta^t \in \RR^p$, producing $\theta^{ t +1  } \in \RR^p$. The procedure involves two steps.  

\begin{enumerate}
\item[(i)] Using the current estimate $\theta^t$, compute the ES ``residuals" $\omega^t_i = \hat Z_i - \alpha X_i^\T \theta^t$. Let $\tau^t >0$ be the solution to the equation
$$
	\frac{1}{n} \sn \frac{ ( | \omega^t_i |  \wedge  \tau )^2}{\tau^2} = \frac{p + \log n }{n}   .
$$ 
By Proposition~3 of \cite{WZZZ2021}, this equation has a unique solution provided that $\sn \mathbbm{1}(| \omega^t_i | >0) > p + \log n$.

\item[(ii)] Compute the updated estimate $\theta^{t+1} \in \argmin_{\theta \in \RR^p}   \sn \ell_{\tau^t} (\hat Z_i - \alpha X_i^\T \theta)$.  This convex optimization problem can be solved via either the iteratively reweighted least squares (IRLS) algorithm or the Barzilai-Borwein gradient descent method \citep{BB1988}.
\end{enumerate}
Repeat the above two steps until convergence or until the maximum number of iterations is reached.

\section{Statistical Theory}
\label{sec:theory}

Throughout this section, we write $X=(x_1, \ldots, x_p)^\T \in \RR^p$ with $x_1 \equiv 1$ so that $\beta^*_1$ and $\theta^*_1$ denote the intercepts.
Without loss of generality, we assume that the random predictors $x_2, \ldots, x_p$ have zero means, that is, $\mu_j = \EE(x_j)=0$ for $j=2,\ldots, p$. This makes the later {\em sub-Gaussian} assumption more reasonable; see Condition~\eqref{cond:covariate}.
Otherwise, we set $\wt X = (1, \tilde{x}_2, \ldots, \tilde{x}_p)^\T = (1, x_2 - \mu_2, \ldots, x_p-\mu_p)^\T$. With this notation, the joint model \eqref{joint.reg.model} becomes
\$
 Q_\alpha( Y |  X  )    = \tilde{\beta}^*_0 +  \sum_{j=2}^p \tilde{x}_j \beta^*_j , \qquad 
  \ES_\alpha( Y  |    X  )   = \tilde{\theta}^*_0 +  \sum_{j=2}^p \tilde{x}_j \theta^*_j   
  ,
\$
where $\tilde{\beta}^*_1 = \beta^*_1 +  \sum_{j=2}^p  \mu_j \beta^*_j$ and $\tilde{\theta}^*_1 = \theta^*_1 +  \sum_{j=2}^p  \mu_j \theta^*_j$. The sub-Gaussian assumption can then be imposed on $\wt X$, and our analysis naturally applies to $\{(Y_i , \wt X_i ) \}_{i=1}^n$.

\subsection{Two-step Joint Quantile and ES Regression}
\label{sec:3.1}

Recall the two-step $\alpha$-ES estimator $\hat \theta$ in~\eqref{ES.est}. In this section, we establish the theoretical properties of $\hat{\theta}$ under the regime in which $p$ is allowed to grow with $n$. We start with some conditions on the covariates and the conditional distribution of $Y$ given $X$.

\begin{cond} \label{cond:density}
The conditional CDF $F_{\varepsilon | X}$ of $\varepsilon := Y - X^\T \beta^*$ given $X$ is continuously differentiable and satisfies $| F_{\varepsilon | X}(t) -F_{\varepsilon | X}(0) |  \leq \overbar  f |t|$ for all $t\in \RR$.  Moreover,  the negative part of $\varepsilon$, denoted by $\varepsilon_-  =   \varepsilon \wedge 0$, satisfies 
$$
	  \var_X(\varepsilon_-) \leq \overbar \sigma^2  ~\mbox{ almost surely (over } X),
$$ 
where $\var_X$ denotes the conditional variance given $X$.
\end{cond}

\begin{cond} \label{cond:covariate}
The random covariate vector $X\in \RR^p$ is sub-Gaussian, that is, there exists some (dimension-free) constant $\upsilon_1 \geq 1$ such that $\PP(|  u^\T W | \geq \upsilon_1 t ) \leq 2 e^{-t^2/2 }$ for all $t\geq 0$ and $u\in \mathbb{S}^{p-1}$, where $W  = \Sigma^{-1/2} X$ and $\Sigma = \EE(X X^\T)$ is positive definite. Let $\kappa_l = \sup_{u\in \mathbb{S}^{p-1}} \EE | u^\T W|^l$ for $l\geq 1$.  
\end{cond}

Several remarks are in order. Condition~\ref{cond:density} states that the negative part of the QR residual $\varepsilon=Y - X^\T \beta^*$ has bounded (conditional) variance.  For convenience,  we assume $\overbar \sigma$ is a constant in the technical analysis.  More generally, one may assume $\varepsilon=\sigma(X) \eta$, where $\sigma:\RR^p \to (0, \infty)$ is a positive function on $\RR^p$ (not necessarily bounded), and $\eta$ is independent of $X$ satisfying $\var(\eta \mathbbm{1} (\eta\leq 0) ) \leq \overbar \sigma^2$.  In this case,  we only need an additional moment assumption on $\sigma(X)$, say $\EE\{ \sigma(X)^4\}$ is bounded.
Condition~\ref{cond:covariate} is mainly used to guarantee that population and empirical quantities (e.g., the objective function or the gradient function) are uniformly close to each other in a compact region.  It can be replaced by a boundedness assumption, which will lead to similar results.  For example, $X= (x_1, \ldots, x_p)^\T$ is compactly supported with either $\| X\|_\infty   \leq C_X$ or $\| \Sigma^{-1/2} X \|_2 \leq B_X$, where $C_X$ is an absolute constant and $B_X$ is usually proportional to $\sqrt{p}$.

\begin{theorem} \label{thm:ES}
Assume Conditions~\ref{cond:density} and \ref{cond:covariate} hold. Conditioned on the event $\{\hat \beta \in \BB_\Sigma(\beta^*, r_0) \}$ for some $r_0>0$, and for any $\delta \in (0, 1/3]$, the two-step $\alpha$-ES estimator $\hat \theta$ satisfies that, with probability at least $1-3 \delta$,
\#
	 \alpha  \| \hat \theta  - \theta^* \|_\Sigma    \leq   2 \overbar \sigma   \sqrt{\frac{ p}{n\delta }} +  \overbar f   \kappa_3  r_0^2 + C_1 \upsilon_1^2   \sqrt{\frac{p + \log(1/\delta) }{n}}    \cdot r_0 \label{ES.est.bound}
\#
as long as $n \geq  C_2 \kappa_4\{ p + 2 \log(2/\delta) \}$, where $C_1, C_2>0$ are absolute constants.
\end{theorem}

From \eqref{ES.est.bound} we see that the first-stage QR estimation error enters the final convergence rate through higher-order terms, that is, $r_0^2 + r_0 \sqrt{p/n}$. This is a direct consequence of the Neyman orthogonality condition that $\partial_\beta \EE \{ S_i(\beta, \theta^* ) \} = 0$. In a joint (linear) quantile and ES regression framework, next we provide the explicit convergence rate, as a function of $n$ and $p$, for the QR estimator under standard regularity conditions.

\begin{cond} \label{cond:density2}
The conditional density function of $\varepsilon$ given $X$, denoted by $f_{\varepsilon|X}$, exists and is continuous on its support. Moreover, there exist constants $\underbar{$f$}, l_0 > 0$ such that $ f_{\varepsilon |X}(0) \geq \underbar{$f$}$ and $|f_{\varepsilon|X}(t) - f_{\varepsilon|X}(0) | \leq l_0 |t|$ for all $t\in \RR$ almost surely (over $X$).
\end{cond}

\begin{proposition} \label{prop:qr}
Assume Conditions~\ref{cond:covariate} and \ref{cond:density2} hold. For any $t\geq 0$, the QR estimator $\hat \beta$ given in \eqref{qr.est} satisfies that, with probability at least $1-e^{-t}$,
\$
	\| \hat \beta - \beta^* \|_\Sigma \leq  C_1 \underbar{$f$}^{-1} \sqrt{\frac{p+t}{n}} 
\$
as long as $n \geq C_2 l_0^2 \underbar{$f$}^{-4} (p+t)$, where $C_1, C_2>0$ are constants depending only on $\upsilon_1$.
\end{proposition}

Together, Theorem~\ref{thm:ES} and Proposition~\ref{prop:qr} show that with probability at least $1- \delta$, the two-step ES estimator $\hat \theta$ satisfies the bound
\$
	 \alpha \| \hat \theta - \theta^* \|_\Sigma \lesssim   \overbar \sigma \sqrt{\frac{ p}{n \delta}} + \frac{\overbar f}{  \underbar{$f$}^2} \frac{p+\log(1/\delta)}{n}  
\$
for all sufficiently large $n\gtrsim \max( 1, l_0^2 \underbar{$f$}^{-4} ) \{ p+ \log(1/\delta)\}$. Using the $\cO_{\PP}(1)$ notation, the previous non-asymptotic bound immediately implies $\alpha \| \hat \theta - \theta^* \|_\Sigma = \cO_{\PP} ( \sqrt{p/n})$.
Furthermore, from the proof of Theorem~\ref{thm:ES} we see that
\#
	 \alpha ( \hat \theta - \theta^* )  = \bigg( \frac{1}{n} \sn X_i X_i^\T \bigg)^{-1} \Bigg\{ \frac{1}{  n} \sn  \omega_i X_i  + \cO_{\PP}  \bigg(\frac{p}{n} \bigg) \Bigg\}, \label{ES.linear.approximation}
\# 
where $\omega_i = \varepsilon_i\mathbbm{1}(\varepsilon_i\leq 0) + \alpha X_i^\T (\beta^* - \theta^*)$.  Because of the Neyman orthogonality, the QR estimation error is first-order negligible, and therefore does not affect the asymptotic distribution of $\hat \theta$.
When $p$ is fixed, applying the multivariate CLT to the linear term $(1/n)\sn \omega_i X_i$ in \eqref{ES.linear.approximation}, we have
\$
	\sqrt{n} ( \hat \theta - \theta^*) \xrightarrow {\rm d}  \cN \big( 0 , \,    \alpha ^{-2} \Sigma^{-1} \Omega  \Sigma^{-1} \big) ~\mbox{ as } n \to \infty, 
\$
where  $\Omega  =   \EE \{ X X^\T \var_X  ( \varepsilon_- )  \}$ with $\varepsilon_- =  \varepsilon \wedge 0 $.

In comparison, consider the ``oracle" ES estimator $\hat \theta^{{\rm ora}}$, defined as 
\#
	\hat \theta^{{\rm ora}} \in \argmin_{\theta \in \RR^p} \frac{1}{n} \sn ( Y_i - X_i^\T \theta)^2 \mathbbm{1}(Y_i\leq X_i^\T \beta^* ) . \label{oracle.es}
\#
As shown in \cite{DB2019}, $\sqrt{n} ( \hat \theta^{{\rm ora}} - \theta^* )  \xrightarrow {\rm d}  \cN   ( 0 ,   \Sigma^{-1} \Omega^*_\alpha   \Sigma^{-1}  )$ as $n\to \infty$, where $\Omega^*_\alpha  = \alpha^{-1}  \EE  \{ X X^\T \var_X( \varepsilon  | \varepsilon \leq 0    ) \}$. By a straightforward calculation, we find that the two asymptotic covariance matrices $\alpha^{-2} \Omega $ and $\Omega^*_\alpha$ are closely connected through the identity
\$
\alpha^{-2} \Omega   =\Omega^*_\alpha    + \frac{1 - \alpha }{\alpha }   \EE \big\{  X X^\T   \langle X, \beta^* - \theta^* \rangle^2  \big\} ,
\$
which also quantifies the efficiency gap between the two-step estimator and the oracle.

In an increasing dimensional regime that $p=p_n\to \infty$ and $p =o(\sqrt{n})$ as $n\to \infty$,  we further establish two Berry-Esseen bounds for linear projections of $\hat \theta$.  Define the ES residual $\omega  = \varepsilon_-  - \EE_X (\varepsilon_-)  =   \varepsilon_-  + \alpha X^\T (\beta^* - \theta^*)$,  such that $\Omega = \EE ( \omega^2 X X^\T)$.

\begin{theorem} \label{thm:uniform.clt}
In addition to Conditions~\ref{cond:density}--\ref{cond:density2},  assume there exist  constants $\underbar{$\sigma$}, \alpha_3>0$ such that
\# \label{var.lbd}
	\var_X ( \varepsilon_-   )  \geq \underbar{$\sigma$}^2  ~\mbox{ and }~ 
\EE_X   (  | \varepsilon_- |^3    ) \leq \alpha_3  ~\mbox{ almost surely over $X$} .
\#
Let $G=(G_1, \ldots, G_p)^\T \in \RR^p$ be a centered Gaussian random vector with covariance matrix
$\Cov(G) = {\rm Corr}( \omega\Sigma^{-1} X )$.  Then we have
\#
	 	& \sup_{t\geq 0}    \Bigg| \PP \bigg(   \max_{1\leq j\leq p} \bigg|  \frac{  \alpha \sqrt{n} (   \hat \theta_j - \theta_j^* )   }{  \sqrt{(\Sigma^{-1}  \Omega \Sigma^{-1} )_{jj}}  } \bigg|  \leq t \bigg)   -  \PP\bigg(  \max_{1\leq j\leq p} |G_j| \leq t  \bigg)  \Bigg|    \nn \\
	&  \leq  C_1    \big\{ \rho_0^{-3/2} (\log n)^{3/2} + \rho_0^{-1} (\log n)^{7/2}  \big\} \frac{\alpha_3 \, p^{3/4} }{\underbar{$\sigma$}^3 \sqrt{n}}   + C_2 ( \overbar f  /\underbar{$f$}^2 \vee  \alpha_3^{1/3})(\log p)^{1/2}    \frac{p+\log n}{\underbar{$\sigma$} \sqrt{n}}  ,  \label{uniform.clt}
\#
where  $\rho_0 := \lambda_{\min}(\Cov(G)) \in (0, 1)$, and $C_1,  C_2>0$ are constants depending only on $\upsilon_1$.
\end{theorem}

\begin{theorem} \label{thm:CLT}
Under the same settings as Theorem~\ref{thm:uniform.clt},  we have
\#
	\sup_{a\in \RR^p, \, t\in \RR} \big| \PP  \big\{  \alpha \sqrt{n} \, a^\T ( \hat \theta  - \theta^* ) / \varrho_a  \leq t   \big\} - \Phi(t)   \big| \lesssim   ( \overbar f  /\underbar{$f$}^2 \vee  \alpha_3^{1/3})  \frac{p+\log n}{ \underbar{$\sigma$} \sqrt{n}},   \label{univariate.clt}
\#
where $\varrho_a^2 = a^\T \Sigma^{-1} \Omega \Sigma^{-1} a$.
\end{theorem}

From an asymptotic view, Theorem~\ref{thm:CLT} shows that any linear combination of the coordinates of $\alpha \sqrt{n}(\hat \theta - \theta^*)$  converges in distribution to the correspondent linear combination of $\cN(0, \Sigma^{-1} \Omega \Sigma^{-1})$ under array asymptotics $n, p \to \infty$ and the growth condition $p^2 = o(n)$. 
This constraint is as expected because known multivariate central limit theorems do no apply when $p^2/n \to \infty$. \cite{P1986} constructed a counterexample showing that a general central limit theorem cannot hold if $p^2 / n \to \infty$.
On the other hand, the best known growth condition on $p$ that ensures the asymptotic normality of linear combinations of the standard QR estimator $\hat \beta$ is $p^3 (\log n)^2 = o(n)$ \citep{W1989, HS2000}.
That is, for any given (deterministic) vector $a \in \RR^p$, 
$$
	\sqrt{n} \langle a, \hat \beta - \beta^* \rangle \xrightarrow{{\rm d}} \cN\big( 0,  \tau(1-\tau) a^\T \Xi  ^{-1} \Sigma \Xi^{-1} a  \big)
$$
as $n \to \infty$ subject to $p^3 (\log n)^2 = o(n)$,  where $\Xi = \EE\{ f_{\varepsilon | X}(0) X X^\T\}$ and $f_{\varepsilon | X}$ denotes the conditional density function of $\varepsilon$ given $X$.
As discussed in \cite{HS2000},  the order of  $p$, as  an integrated part of the design conditions,  crucially depends on the smoothness of the loss function, or equivalently, the score function. 
When $p$ is fixed, this together with the  Cram\'er–Wold theorem implies $\alpha \sqrt{n}(\hat \theta - \theta^*) \to \cN(0, \Sigma^{-1} \Omega \Sigma^{-1})$ in distribution as $n\to \infty$.

\subsection{Robust ES Regression}
\label{sec:3.2}

In this section, we provide non-asymptotic upper bounds on $\| \hat \theta_\tau - \theta^* \|_2$ for the Huberized two-step ES estimator $\hat \theta_\tau$ defined in \eqref{robust.ES.est}. Moreover, we establish a non-asymptotic Bahadur representation for $\hat \theta_\tau$, which is the key step towards a Berry-Esseen-type bound for Gaussian approximation.

\begin{theorem} \label{thm:huber.ES}
Assume Conditions~\ref{cond:density} and \ref{cond:covariate} hold.  For any $t>0$, let $r_0>0$ be such that $r_0 \lesssim \overbar \sigma$ and $\overbar f r_0^2 \lesssim \overbar \sigma \sqrt{(p+t) / n}$. Then, the two-step robust $\alpha$-ES ($0<\alpha \leq 1/2$) estimator $\hat \theta_\tau$ with $\tau \asymp  \overbar \sigma \sqrt{n/(p+t)}$ satisfies that, with probability at least $1-3 e^{-t}$ conditioned on the event $\{\hat \beta \in \BB_\Sigma(\beta^*, r_0) \}$,
\#
	\alpha  \| \hat \theta_\tau - \theta^* \|_\Sigma \leq C_1  \overbar \sigma  \sqrt{\frac{p+t}{ n}} +  C_2  \bigg( \sqrt{\frac{p+t}{n}}    r_0 +  \overbar f r_0^2     \bigg)
\#
provided the sample size obeys $n \geq C_3( p + t)$, where $C_1$--$C_3$ are positive constants depending only on $\upsilon_1$.
\end{theorem}

\begin{remark}[Bias-robustness tradeoff] \label{rmk4.1}
The choice of $\tau$ stated in Theorem~\ref{thm:huber.ES} is a reflection of the bias-robustness tradeoff.  As discussed in Section~\ref{sec:2.3}, the robust estimator $\hat \theta_\tau$ can be viewed as an $M$-estimator of $\theta^*_\tau = \argmin_{\theta } \EE \{ \ell_\tau(Z_i - \alpha X_i^\T \theta) \}$, which differs from the true ES regression coefficient $\theta^*$ due to the asymmetry of ES ``residuals" $\omega_i = Z_i - \alpha X_i^\T \theta^*$. Consider the decomposition
\$
	 \| \hat \theta_\tau - \theta^* \|_\Sigma \leq   \underbrace{  \| \hat \theta_\tau - \theta_\tau^* \|_\Sigma  }_{{\rm robustification~bias}} +  \underbrace{ \|  \theta^*_\tau - \theta^* \|_\Sigma }_{{\rm robust~estimation~error}} .
\$
As long as $\tau \gtrsim \overbar \sigma$ under Condition~\ref{cond:density}, by Proposition~\ref{prop:bias},  we have $\alpha \| \hat \theta_\tau - \theta_\tau^* \|_\Sigma \leq 2 \overbar \sigma^2 / \tau$. Moreover, conditioned on the event $\{\hat \beta \in \BB_\Sigma(\beta^*, r_0) \}$, we have
$$
\alpha \| \hat \theta_\tau - \theta^* \|_\Sigma \lesssim \overbar \sigma \sqrt{\frac{p+t}{n}} + \tau \frac{p+t}{n} + \frac{\overbar \sigma^2}{\tau } + r_0 \bigg( \sqrt{\frac{p+t}{n}}  + \frac{\overbar \sigma }{\tau} \bigg) + r_0^2
$$
with high probability. 
We thus  choose $\tau \asymp \overbar \sigma \sqrt{n/(p+t)}$ to minimize the upper bound as a function of $\tau$.
\end{remark}

\begin{remark} \label{rmk4.1b}
Recall from Proposition~\ref{prop:qr} that with probability at least $1-n^{-1}$,  $\| \hat \beta - \beta^* \|_\Sigma \lesssim  \sqrt{(p + \log n) / n}$ as long as $n\gtrsim p + \log n$.  Combining the proof of Theorem~\ref{thm:huber.ES} with a discretization argument,  we can obtain a more general result that holds for a range of $\tau$ values. 
In addition to Condition~\ref{cond:density}, assume $\EE_X ( | \varepsilon_- |^k ) \leq \alpha_k$ almost surely (over $X$) for some $k>2$.
Then, for all $\tau$ satisfying $\overbar \sigma \lesssim \tau \lesssim \overbar \sigma \sqrt{n/(p+\log n)}$, the corresponding ES regression estimator $\hat \theta_\tau$ satisfies  with probability $1-Cn^{-1}$ that
\$
	\alpha \| \hat \theta_\tau - \theta^* \|_\Sigma \lesssim  \overbar \sigma \sqrt{\frac{p+\log n}{n}} +  \frac{\alpha_k}{\tau^{k-1}} + \text{higher order terms} 
\$
holds uniformly over $\tau$,  where the ``higher order terms" stem from the first-step quantile regression estimation error. In other words, a data-adaptive choice of $\tau$ within the aforementioned range can be used. 
To achieve tight (finite-sample) concentration bounds,  the order of the robustification parameter $\tau=\tau(n,p)$ should be no larger than $\sqrt{n/(p+\log n)}$. On the other hand, $\tau$ should exhibit a sufficiently fast growth so that the bias term $\overbar \sigma \tau^{-1}$ or $\alpha_k \tau^{1-k}$ (if higher-order moments are bounded)  decays as fast as the stochastic error.

\end{remark}

For any $\delta \in (0,1)$, the robust estimator $\hat \theta_\tau$ with $\tau \asymp  \overbar \sigma \sqrt{n/(p+\log(1/\delta))}$ satisfies with probability at least $1-\delta$ conditioned on $\{\hat \beta \in \BB_\Sigma(\beta^*, r_0) \}$ that
\$
	\alpha \| \hat \theta_\tau - \theta^* \|_\Sigma \lesssim \overbar \sigma \sqrt{\frac{p+\log(1/\delta) }{n}} + \sqrt{\frac{p+\log(1/\delta)}{n}}    r_0 +  \overbar f r_0^2   .
\$
The above bound is proportional to $\log(1/\delta)$ as opposed to the bound in \eqref{ES.est.bound} which is proportional to $1/\delta$. This indicates that the Huberized estimator is much more robust to heavy tails from a non-asymptotic perspective: when the regression error only has finite variance, the worst case deviations of $\hat \theta$ are much larger than that of $\hat \theta_\tau$.

\begin{remark} \label{rmk4.2}
To achieve a tight deviation bound at $1-\delta$ confidence level for any given $\delta \in (0, 1)$, Theorems~\ref{thm:huber.ES} suggests that the robustification parameter $\tau=\tau(n,p)$ should be of order $\overbar \sigma \sqrt{n / (p + \log  \delta^{-1}  )}$, where $\overbar \sigma^2>0$ is an upper bound on the (conditional) variance of $\varepsilon_-= \min\{ Y - X^\T \beta^*, 0\}$.
Since $\overbar \sigma$ is typically unknown in practice,  a rule of thumb is to replace it by the sample standard deviation of the negative QR residuals $\varepsilon_{i,-} = \min\{ Y_i - X_i^\T \hat \beta, 0 \}$, denoted by $\hat \sigma$, where $\hat \beta$ is the first-stage QR estimator. 
By taking $\hat \tau = \hat \sigma \sqrt{n/(p+ \log \delta^{-1} )}$, the resulting estimator is also location and scale equivariant.
Recall that in Section~\ref{sec:2.3},  we describe a slightly more sophisticated  data-driven method for choosing $\tau$,  which is adapted from that in  \cite{WZZZ2021}.  All the numerical studies in Section~\ref{sec:6} are based on this tuning scheme because it consistently outperforms the aforementioned rule of thumb. 
\end{remark}

Unlike the standard two-step estimator $\hat \theta$, its robust counterpart $\hat \theta_\tau$ does not have a closed-form expression. As the main building block toward deriving Gaussian approximation results, the next theorem provides a non-asymptotic Bahadur representation for $\hat \theta_\tau$ with explicit error bounds that depend on $(n, p)$ and the first-stage QR estimation error.

\begin{theorem} \label{thm:ES.bahadur}
Assume the same conditions as in Theorem~\ref{thm:huber.ES}.  For any $t>0$, the $\alpha$-ES estimator $\hat \theta_\tau$ with $\tau \asymp  \overbar \sigma \sqrt{ n/(p+t)}$ satisfies that, with probability at least $1- 6 e^{-t}$ conditioned on $\{ \hat \beta \in \BB_\Sigma(\beta^*, r_0) \}$,
\#
 \bigg\| \alpha \Sigma^{1/2} ( \hat \theta_\tau - \theta^*  ) - \frac{1}{n} \sn \psi_\tau(\omega_i) W_i \bigg\|_2  \lesssim   \overbar \sigma \frac{p+t}{ n}  + \overbar f r_0^2 + r_0  \sqrt{\frac{p\log n + t}{n}}    \label{ES.bahadur}
\#
as long as $n  \gtrsim  p + t$, where $W_i  = \Sigma^{-1/2} X_i$.
\end{theorem}

Finally, we have the following Gaussian approximation result, which bounds the Kolmogorov distance between the distribution of the standardized statistic $\alpha \sqrt{n}\, a^\T(\hat\theta_\tau - \theta^*) / \varrho_{a } $ and the standard normal distribution uniformly over all (deterministic) vectors $a \in \RR^p$, where $\varrho_{a }^2  = a^\T \Sigma^{-1} \Omega \Sigma^{-1}a $ is the same as in Theorem~\ref{thm:CLT}.  A similar conclusion applies to the oracle robust estimate $\hat \theta^{{\rm ora}}_\tau$ \eqref{robust.oracle.ES}.  The following theorem shows that  the two-step robust estimator obtained via \eqref{qr.est} and \eqref{robust.ES.est} is asymptotically equivalent to the oracle Huberized estimator \eqref{robust.oracle.ES} (assuming $\beta^*$ were known).

\begin{theorem} \label{thm:ES.CLT}
Under the conditions of Theorem~\ref{thm:uniform.clt},  the robust $\alpha$-level ($\alpha \in (0, 1/2]$) ES  estimator $\hat \theta_\tau$ with $\tau \asymp   \overbar \sigma \sqrt{ n/(p+\log n )}$ satisfies
\#
\sup_{a\in \RR^p, \, t\in \RR} \, &  \big| \PP  \big(  \alpha \sqrt{n} \langle a,  \hat \theta_\tau - \theta^*  \rangle / \varrho_{a } \leq t   \big) - \Phi(t)   \big| \nn \\
	&  \lesssim   \frac{\alpha_3 }{ \underbar{$\sigma$}^3 }  \sqrt{\frac{p+\log n}{n}}  +   ( \overbar f  /\underbar{$f$}^2 \vee  \alpha_3^{1/3} )   \frac{ p \sqrt{\log n} + \sqrt{p} \log n }{ \underbar{$\sigma$} \sqrt{n}} .  \label{clt}
\# 
Moreover,  the oracle Huberized ES estimator $\hat \theta^{{\rm ora}}_\tau$ \eqref{robust.oracle.ES} with the same $\tau$ satisfies
\#
\sup_{a\in \RR^p, \, t\in \RR} \, &  \big| \PP  \big(  \alpha \sqrt{n} \langle a,  \hat \theta^{{\rm ora}}_\tau - \theta^*  \rangle / \varrho_{a } \leq t   \big) - \Phi(t)   \big|   \lesssim   \frac{\alpha_3 }{ \underbar{$\sigma$}^3 }  \sqrt{\frac{p+\log n}{n}} .	\label{oracle.be}
\#
\end{theorem}

The above Gaussian approximation result lays the theoretical foundation for the statistical inference problems of testing the linear hypothesis $H_0: a^\T \theta^* = c_0$ versus $H_1: a^\T \theta^* \neq c_0$ and constructing confidence intervals for $a^\T \theta^*$, where $a\in \RR^p$ and $c_0 \in \RR$ are predetermined. 
Given the joint quantile and ES regression estimates $(\hat \beta, \hat \theta_\tau)$, let $\hat \Omega_\gamma$ be the truncated estimator of $\Omega= \EE (  \omega^2 X X^\T )$ defined in \eqref{robust.omega} with $\gamma=\gamma(n,p)>0$ denoting a second robustification parameter. Then, we consider the robust test statistic 
$$
	T_a = \alpha \sqrt{n} ( a^\T \hat\theta_\tau - c_0) / \hat \varrho_{a, \gamma}
$$ 
for testing $H_0:a^\T \theta^* = c_0$, and the (approximate) 100$(1- \beta)$\% confidence interval
$ a^\T \hat \theta_\tau  \pm   z_{\beta/2}   \hat \varrho_{a, \gamma} /(\alpha \sqrt{n})$ for $a^\T \theta^*$, where $\hat \varrho_{a, \gamma}^2 :=  a^\T \hat \Sigma^{-1} \hat \Omega_\gamma  \hat \Sigma^{-1} a$ is a robust variance estimator and $z_{\beta/2}$ is the upper $(\beta/2)$-quantile of $\cN(0,1)$.

In view of Theorem~\ref{thm:ES.CLT}, the validity of the above normal-based confidence construction for $a^\T \theta^*$ (with a prespecified $a \in \RR^p$) depends on the consistency of the robust variance estimate $\hat \varrho_{a, \gamma}^2$. 
We will show in the proof of Theorem~\ref{thm:uniform.clt} that with probability at least $1-1/n$, $\| \Sigma^{-1/2} \hat \Sigma \Sigma^{-1/2} - {\rm I}_p \|_2 \lesssim \sqrt{(p+\log n)/n}$ and thus $\| \Sigma^{-1/2} \hat \Sigma \Sigma^{-1/2} - {\rm I}_p \|_2 = \cO_{\PP}( \sqrt{(p+\log n)/n})$.
The consistency of the truncated estimator $ \hat \Omega_\gamma$, on the other hand, is more subtle  in the increasing-$p$ regime as it involves both estimated regression coefficients $\hat \beta$ and $\hat \theta_\tau$. To account for the estimation errors $\| \hat \beta - \beta^* \|_\Sigma$ and $\| \hat \theta_\tau - \theta^* \|_\Sigma$, define for each pair of radius parameters $r_0, r_1>0$ the subset 
\#
	\Theta(r_0, r_1) =  \big\{ (\beta, \theta)\in \RR^p \times \RR^p : \| \beta-\beta^*\|_\Sigma\leq r_0, \, \alpha \| \theta - \theta^* \|_\Sigma \leq r_1 \big\} . \label{local.set}
\#
Conditioning on the event $\{ (\hat \beta, \hat \theta_\tau ) \in \Theta(r_0, r_1) \}$ for some $r_0$ and $r_1$ that determine the convergence rates of $\hat \beta$ and $\hat \theta_\tau$, respectively, we have 
\$
	\| \Sigma^{-1/2} ( \hat \Omega_\gamma - \Omega)   \Sigma^{-1/2}   \|_2 \leq \sup_{  (\beta, \theta ) \in \Theta(r_0, r_1) } \bigg\| \frac{1}{n} \sn \psi^2_\gamma( \omega_i(\beta, \theta)) W_i W_i^\T - \Sigma^{-1/2} \Omega \Sigma^{-1/2} \bigg\|_2 ,
\$
where $w_i(\beta, \theta) = (Y_i - X_i^\T \beta) \mathbbm{1}(Y_i \leq X_i^\T \beta) + \alpha X_i^\T(\beta - \theta)$ and $W_i = \Sigma^{-1/2} X_i$. The problem then boils down to controlling the supremum of $\| (1/n) \sn \psi^2_\gamma( \omega_i(\beta, \theta)) W_i W_i^\T -  \EE( \omega_i^2 W_i W_i^\T)   \|_2$ over $(\beta, \theta)$ in a local neighborhood of $(\beta^*, \theta^*)$.

\begin{theorem} \label{thm:var.consistency}
In addition to Conditions~\ref{cond:density}--\ref{cond:density2},  assume that  
\# \label{third.moment}
\EE_X   (  | \varepsilon_- |^3    ) \leq \alpha_3    ~\mbox{ almost surely over $X$}  ~~\mbox{ and }~~ \max_{1\leq i\leq n} \| W_i \|_2 \leq C_0 \sqrt{p}  
\#
for some constants $\alpha_3 , C_0 >0$. Conditioning on $\{ (\hat \beta, \hat \theta_\tau ) \in \Theta(r_0, r_1) \}$ for any predetermined $r_0, r_1>0$, it holds with probability at least $1-2/n$ that
\$
& \| \Sigma^{-1/2} ( \hat \Omega_\gamma - \Omega)   \Sigma^{-1/2} \|_2 \\
& \lesssim \max\{ \alpha_3^{1/2} , (\sqrt{p}\overbar r)^{3/2} \}  \sqrt{\gamma \frac{ p \log n}{n}} +  \gamma^2 \frac{   p  \log n}{n} + \frac{\alpha_3}{\gamma} + (  \overbar \sigma  + \overbar r ) \overbar r 
\$
as long as $n\gtrsim  p+\log n $, where $\overbar r = r_0+r_1$.
\end{theorem}

\begin{remark} \label{rmk4.3}
From Proposition~\ref{prop:qr} and Theorem~\ref{thm:huber.ES} we see that the regression estimates $\hat \beta$ and $\hat \theta_\tau$ with $\tau \asymp \overbar \sigma \sqrt{n/(p+\log n)}$ satisfy with probability at least $1-C n^{-1}$ that
\$
	\| \hat \beta - \beta^* \|_\Sigma  \leq r_0\asymp  \frac{1}{ \underbar{$f$}} \sqrt{\frac{p + \log n}{n}} ~\mbox{ and }~	\alpha \| \hat \theta_\tau - \theta^* \|_\Sigma  \leq r_1 \asymp   \overbar \sigma \sqrt{\frac{p+\log n}{n}} + \frac{\overbar f}{ \underbar{$f$}^2}\frac{p+\log n}{n} .
\$
The corresponding event $\{ (\hat \beta, \hat \theta_\tau ) \in \Theta(r_0, r_1) \}$ holds with high probability, and $\overbar r = r_0 + r_1= \cO( \sqrt{(p+\log n)/n})$. Furthermore, by taking $\gamma \asymp \{ \alpha_3 n / (p \log n) \}^{1/3}$ in Theorem~\ref{thm:var.consistency}, we conclude that up to multiplication constants depending on $(\alpha_3, \overbar \sigma,  \overbar f, \underbar{$f$})$, 
$$
	\| \Sigma^{-1/2} ( \hat \Omega_\gamma - \Omega)   \Sigma^{-1/2} \|_2 \lesssim  \bigg( \frac{p \log n}{n} \bigg)^{1/3}  
$$
with high probability as long as $n\gtrsim p^2$. This ensures the consistency of $\hat \varrho^2_{a, \gamma}$, i.e. $| \hat \varrho^2_{a, \gamma} / \varrho^2_a - 1| = o_{\PP}(1)$, under the constraint $p^2 = O(n)$ as $n\to \infty$.
\end{remark}

\begin{remark} \label{rmk4.4}
The bounded covariates assumption $\max_{1\leq i\leq n} \| W_i \|_2 \leq C_0 \sqrt{p}$ in \eqref{third.moment} is only imposed for technical convenience. In fact, under Condition~\ref{cond:covariate}, combining Theorem~2.1 in \cite{HKZ2012} with the union bound we have that with probability at least $1-1/n$, $\max_{1\leq i\leq n} \| W_i \|_2\leq C \upsilon_0 \sqrt{p+\log n}$ for some absolute constant $C>1$. 
We can modify the proof of Theorem~\ref{thm:var.consistency} to remove the bounded covariates assumption in \eqref{third.moment} by using a truncation argument,  namely that replacing $W_i$ by $W_i\mathbbm{1}( \| W_i \|_2 \leq C_0 \sqrt{p+\log n} )\}$ for each $1\leq i\leq n$ ($C_0 = C \upsilon_0$).
\end{remark}

\section{Nonparametric Expected Shortfall Regression}
\label{sec:nonparametric}

In this section, we consider nonparametric models for joint quantile and expected shortfall regression.  For a predetermined quantile level $\alpha\in(0,1)$, the goal is to estimate the unknown (conditional) quantile and expected shortfall functions $f^*_q(x) = Q_\alpha(Y| X=x)$ and $f^*_e(x) = {\rm ES}_\alpha(Y | X=x)$, with an emphasis on the latter.  By \eqref{ES.def1}, $f^*_q$ and $f^*_e$  can be identified as
 \$
  f^*_q  = \argmin_{f_q} \EE \rho_\alpha ( Y - f_q(X) ) ~~\mbox{ and }~~ f^*_e =  \argmin_{f_e } \EE \{ Y - f_e(X) \}^2 \mathbbm{1}_{\{ Y \leq f^*_q(X)  \}} .
 \$

Motivated by the two-step procedure developed under joint linear models,  in the following we propose a nonparametric ES estimator using the series regression method \citep{ES1990,A1991,N1997}.  Such a nonparametric estimate is carried out by regressing the dependent variable on an asymptotically growing number of approximating functions of the covariates, and therefore is closely related to the estimator define in \eqref{ES.est} under the so-called many regressors model \citep{BCCF2019}, that is, the dimension $p=p_n$ is allowed to grow with $n$.  The idea of series estimation is to first approximate $f^*_q$ and $f^*_e$  by their ``projections'' on the linear spans of $m_1$ and $m_2$ series/basis functions, respectively, and then fit the coefficients using the observed data.  Specifically,   we approximate functions $f^*_q$ and $f^*_e$ by linear forms $U(x)^\T \beta$ and $V(x)^\T \theta$, where
$$
	U(x)= (u_1(x), \ldots, u_{m_1}(x) )^\T ~~\mbox{ and }~~ V(x)= (v_1(x), \ldots,  v_{m_2}(x) )^\T
$$
are two vectors of series approximating functions of dimensions $m_1$ and $m_2$.  Here both $m_1$ and $m_2$ may increase with $n$.   We thus define the vectors of quantile and ES series approximation coefficients as
\#
	\beta^* \in \argmin_{\beta \in \RR^{m_1}} \EE \rho_\alpha (Y - U(X)^\T \beta) ~~\mbox{ and }~~
\theta^* \in \argmin_{\theta \in \RR^{m_2}} \EE  \{ Y - V(X)^\T \theta \}^2 \mathbbm{1}_{\{ Y \leq f^*_q(X)  \}} . \label{def:series.parameter}
\#

Given independent observations $(Y_i, X_i)$, $1\leq i\leq n$ from $(Y, X) \in \RR \times \cX$ with $\cX$ denoting a compact subset of $\RR^p$,   we write
$$
	U_i = U(X_i) \in \RR^{m_1} ~~\mbox{ and }~~ V_i = V_i(X_i) \in \RR^{m_2}. 
$$
Extending the two-step approach described in Section~\ref{sec:2.2}, we first define the (conditional) quantile series estimator of $f^*_q(x) = Q_\alpha(Y| X=x)$  \citep{BCCF2019}:
\#
	\hat f_q(x) = U(x)^\T \hat \beta  , \ \  x\in \cX , ~\mbox{ where }~ \hat \beta = \hat \beta_{m_1}  \in  \argmin_{\beta \in \RR^{m_1}} \frac{1}{n} \sn \rho_\alpha( Y_i - U_i^\T \beta) . \label{series.quant}
\#
With nonparametrically generated response variables $\hat Z_i := \{ Y_i - \hat f_q(X_i) \} \mathbbm{1} \{ Y_i \leq \hat f_q(X_i ) \} + \alpha \hat f_q(X_i)$, the second-stage ES series estimator is given by
\#
 \hat f_e(x) = V(x)^\T \hat \theta, \ \ x \in \cX , ~\mbox{ where }~ \hat \theta = \hat \theta_{m_2} \in \argmin_{\theta \in \RR^{m_2} } \frac{1}{n}  \sn (\hat Z_i - \alpha V_i^\T \theta)^2   . \label{series.es}
\#

Commonly used series functions with good approximation properties include B-splines, polynomials,
Fourier series and compactly supported wavelets.   We refer to \cite{N1997} and \cite{C2007} for a detailed description of these series functions.  In the context of quantile regression, \cite{C2007} established the consistency and rate of convergence at a single quantile index.  More recently,  \cite{BCCF2019} developed large sample theory for quantile series coefficient process, including convergence rate and uniform strong approximations.  The choice of the parameter $m_1$, also known as the order of the series estimator, is crucial for establishing the balance between bias and variance.

Note that the quantile series estimator $\hat f_q$ in \eqref{series.quant} has been well studied by \cite{BCCF2019}.  Because the number of regressors increases with the sample size,  conventional
central limit theorems are no longer applicable to capture the joint asymptotic normality of the regression coefficients.  The growing dimensionality is the primary source of technical complication.  Our theoretical analysis under the joint linear model \eqref{joint.reg.model}, which leads to novel non-asymptotic high probability bounds,  can be used as a starting point for studying the two-step nonparametric ES series estimator  $\hat f_e$ defined in \eqref{series.es}.   Of particular interest is to develop a uniform inference procedure for the conditional ES function $f^*_e$ and its theory.
That is,  at a given confidence level $1-\gamma$,  we aim to construct a pair of functional estimates $[\hat f^L_e, \hat f^U_e]$ from $\{ (Y_i, X_i)\}_{i=1}^n$ such that
$$
	\PP \big\{ \hat f^L_e(x) \leq f^*_e(x) \leq    \hat f^U_e(x) ~\mbox{ for all }  x \in \cX \big\} \to 1 - \gamma , ~\mbox{ as }~ n \to \infty.
$$
Since a significant amount of additional work is still needed, including explicit characterizations of the ES series approximation error and the impact of first-stage nonparametric QR estimation error, we leave a rigorous theoretical investigation of $\hat f_e$ to future work.
Although we have only focused on series methods, there are other nonparametric techniques which offer superior empirical and theoretical performance. Among those, deep neural networks have stood out as a promising tool for nonparametric estimation, from least squares,  logistic to quantile regressions \citep{SH2020,FLM2021,Shen2021}.  It is practically useful to construct deep learning
implementations of two-step estimators,  and statistically important to deliver valid inference on finite-dimensional parameters following first-step estimation (of both quantile and ES functions) using deep learning.  A detailed investigation of these problems is beyond the present scope but of future interest.

\section{Numerical Studies and Real Data Examples}
\label{sec:6}

\subsection{Monte Carlo Experiments}
\label{sec:numerical}

In this section, we assess the numerical performance of the proposed method for fitting expected shortfall regression. For its \texttt{R} implementation, we first obtain a QR estimate via the \texttt{quantreg} library \citep{K2022}, and in step two use the \texttt{adaHuber} library \citep{PZ2022} to solve \eqref{robust.ES.est} with the robustification parameter selected adaptively as described in Section~\ref{sec:2.3}.  

\begin{example}[\texttt{R} code for fitting $\alpha$-level ES regression to data ${\rm x}\in \RR^{n\times p}$ and ${\rm y}\in \RR^n$]  ~\\
\noindent
\texttt{library(quantreg) \\
library(adaHuber) \\
qr\_fit ~<- rq(y$\sim$x, tau=alpha, method=`pfn') \\ 
fit\_q  ~~<- x  \%*\%  qr\_fit\textdollar coef[-1] +  qr\_fit\textdollar coef[1]    \\ 
ynew \,\,\,\,\,~<- (y -   fit\_q) * (y <= fit\_q) / alpha +  fit\_q  \\
es\_fit ~<- adaHuber.reg(x, ynew, method=`adaptive') \\
coef\_e ~<- es\_fit\textdollar coef
}
\end{example} 
\noindent
We compare the proposed two-step adaptive Huber ES estimator (\texttt{2S-AH}) to several competitors: (i) the joint regression estimate (\texttt{joint}) via FZ loss minimization, implemented by the \texttt{R} library \texttt{esreg} with the default option \citep{BD2019}; (ii) the two-step least squares estimator \eqref{two-step.ES.est1} (\texttt{2S-LS}), and (iii) the oracle two-step ``estimator'' (\texttt{2S-oracle}).  
Recall that the two-step procedure first obtains a QR estimator $\hat{\beta}$ via either standard \citep{KB1978} or smoothed QR regression \citep{HPTZ2021}, and subsequently computes the ES estimator based on fitted quantile thresholds $\{ X_i^\T \hat{\beta} \}_{i=1}^n$.  The oracle method refers to the two-step ES estimate based on the true quantile thresholds $\{ X_i^\T \beta^* \}_{i=1}^n$. %

In our simulation studies,  we first generate  $\gamma^*= (\gamma_1^*, \ldots, \gamma_p^*)^\T $ and $\eta^*=( \eta_1^*, \ldots, \eta_p^*)^\T $ independently, where $\gamma_j^*$'s are independent Rademacher random variables and $\eta_j^*\sim_{{\rm i.i.d.}}  0.5 \cdot  {\rm Bernoulli}(1/2)$. Data are then generated from the heteroscedastic model 
\begin{equation}
\label{eq:sim:model}
	Y_i = X_i^\T  \gamma^*  + X_i^\T \eta^*   \cdot \varepsilon_i, 
\end{equation}
where $X_i = (X_{i1}, \ldots, X_{ip} )^\T$ with $X_{ij} \sim_{{\rm i.i.d.}} \mathrm{Unif}(0,1.5)$, and the random noise $\varepsilon_i$ follows one of the following two distributions: (i) standard normal distribution, and (ii) $t$-distribution with $\nu >2$ degrees of freedom ($t_{\nu}$). Given $\gamma^*$ and $\eta^*$, the true quantile and expected shortfall regression coefficients are
\[
\beta^* =  \gamma^* + \eta^*  \cdot Q_{\alpha}(\varepsilon) \qquad \mathrm{and}\qquad \theta^* = \gamma^* + \eta^* \cdot \mathrm{ES}_{\alpha}(\varepsilon),
\]
where $Q_{\alpha}(\varepsilon)$  and $\mathrm{ES}_{\alpha}(\varepsilon)$ are the $\alpha$-level quantile and expected shortfall of $\varepsilon$, respectively.

\begin{figure}[!htp]
  \centering
  \subfigure[Normal model with $(p,n)=(5, 5000)$]{\includegraphics[height=0.22\textheight, width=0.48\textwidth]{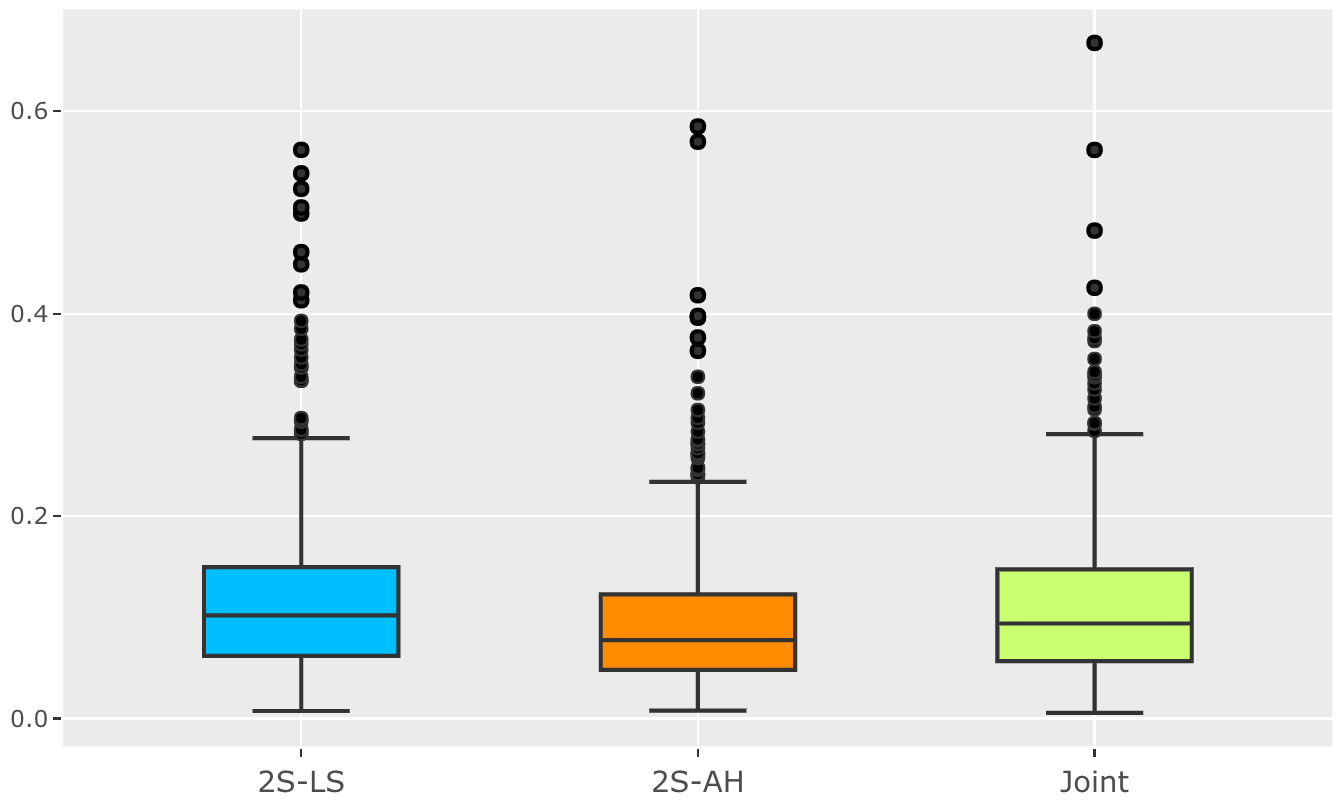}} 
  ~~\subfigure[$t_3$ model with $(p,n)=(5, 10000)$]{\includegraphics[height=0.22\textheight, width=0.48\textwidth]{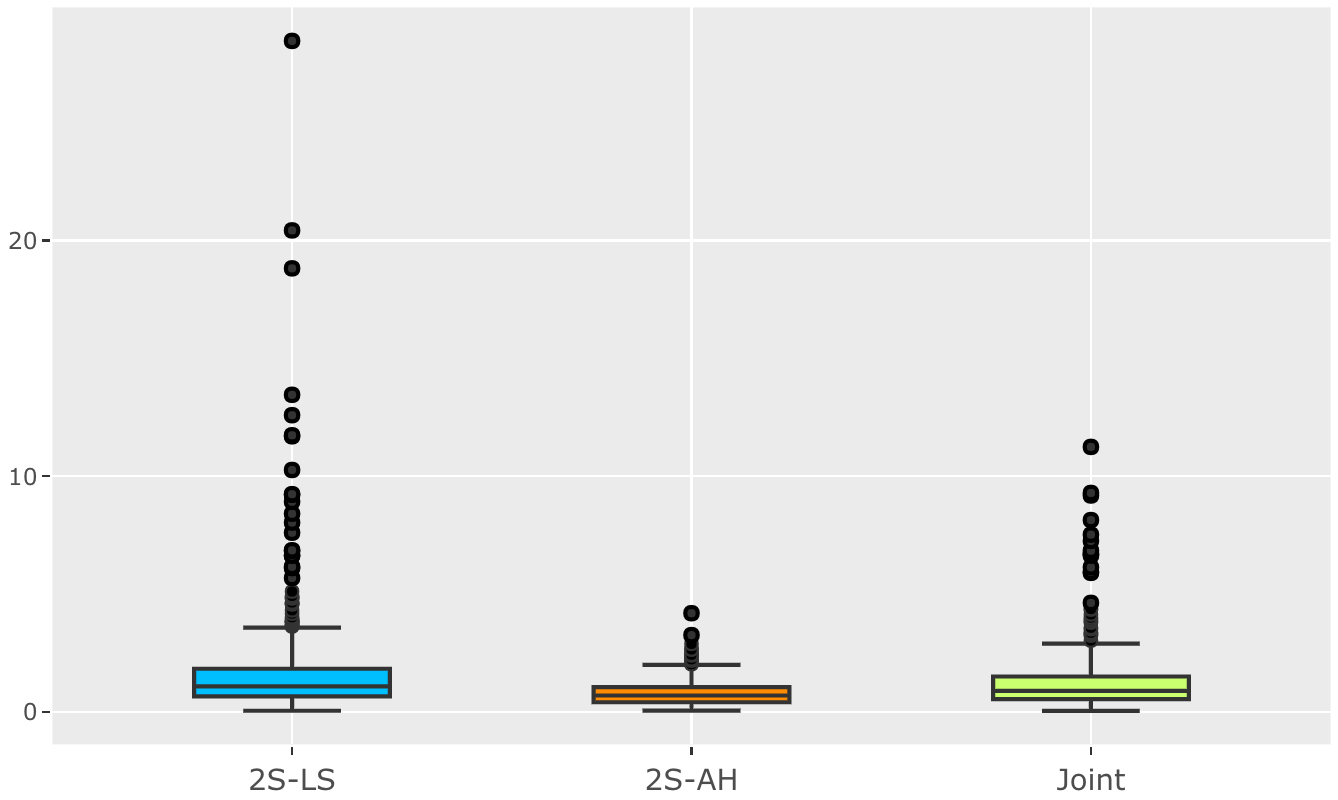}} 
\caption{Boxplots of squared total $\ell_2$-errors (including the intercept when its true value is 2), based on 500 replications,  for three ES regression estimators (\texttt{2S-LS}, \texttt{2S-AH} and \texttt{joint}) at quantile level $\alpha = 0.01$.   The mean squared errors of these three estimators are 0.1219, 0.0983 and 0.1119 in the normal model, and 1.7401, 0.8017 and 1.2542  in the $t_3$ model.}
  \label{fig:extreme}
\end{figure}

\begin{figure}[!htp]
  \centering
  \subfigure[Normal noise, $\alpha=0.05$.]{\includegraphics[width=0.32\textwidth]{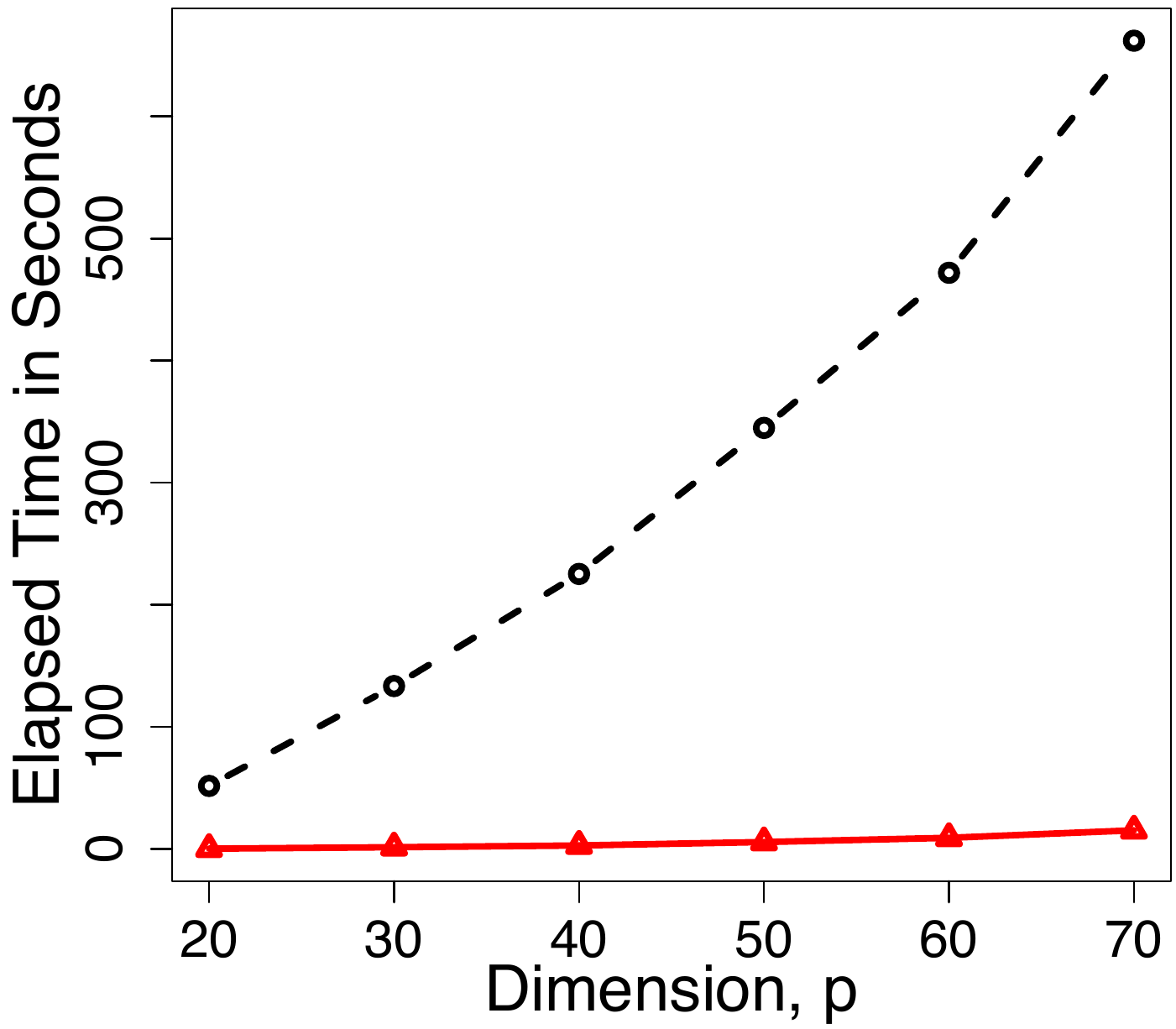}} 
    \subfigure[Normal noise, $\alpha=0.1$.]{\includegraphics[width=0.32\textwidth]{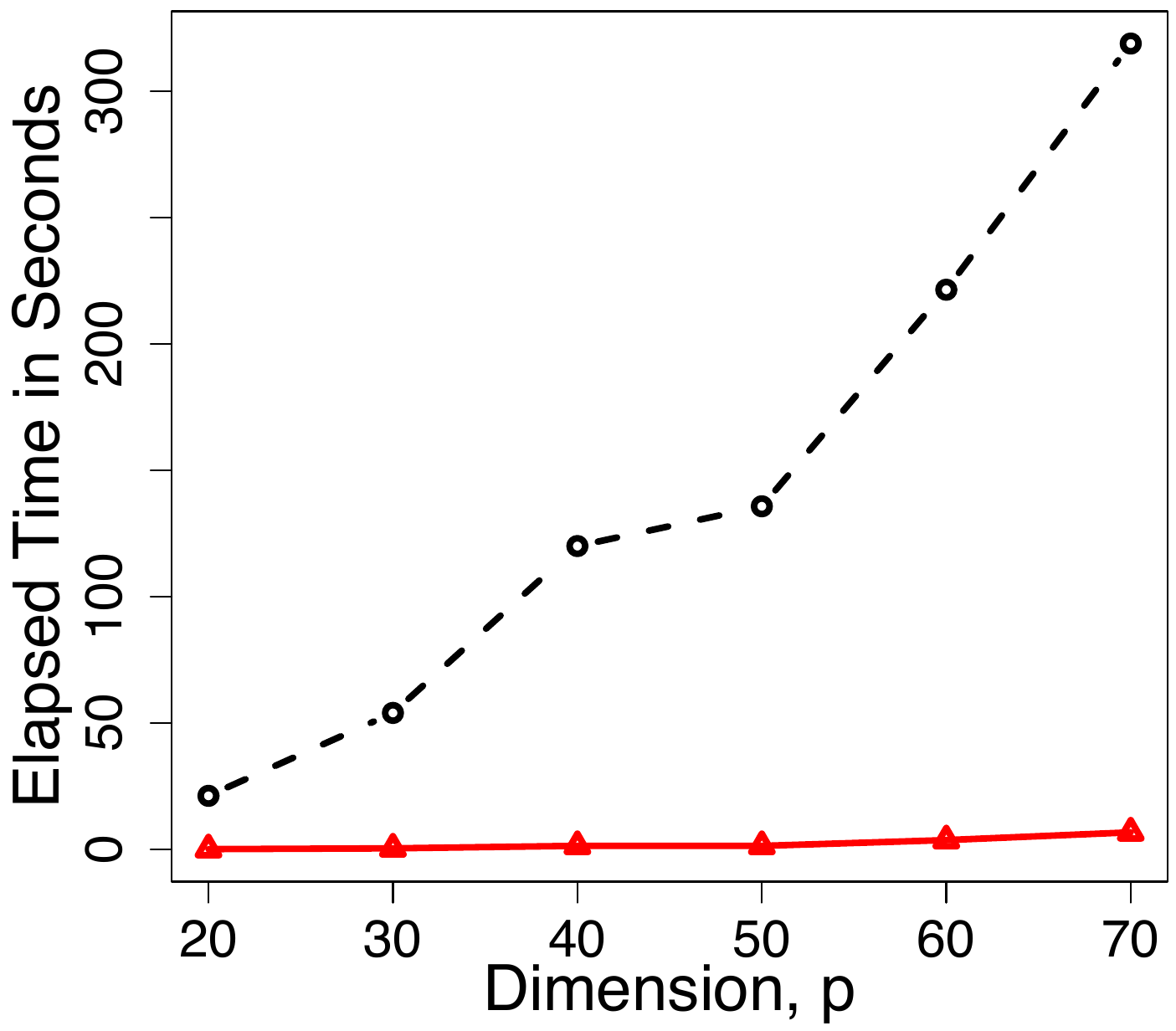}} 
      \subfigure[Normal noise, $\alpha=0.2$.]{\includegraphics[width=0.32\textwidth]{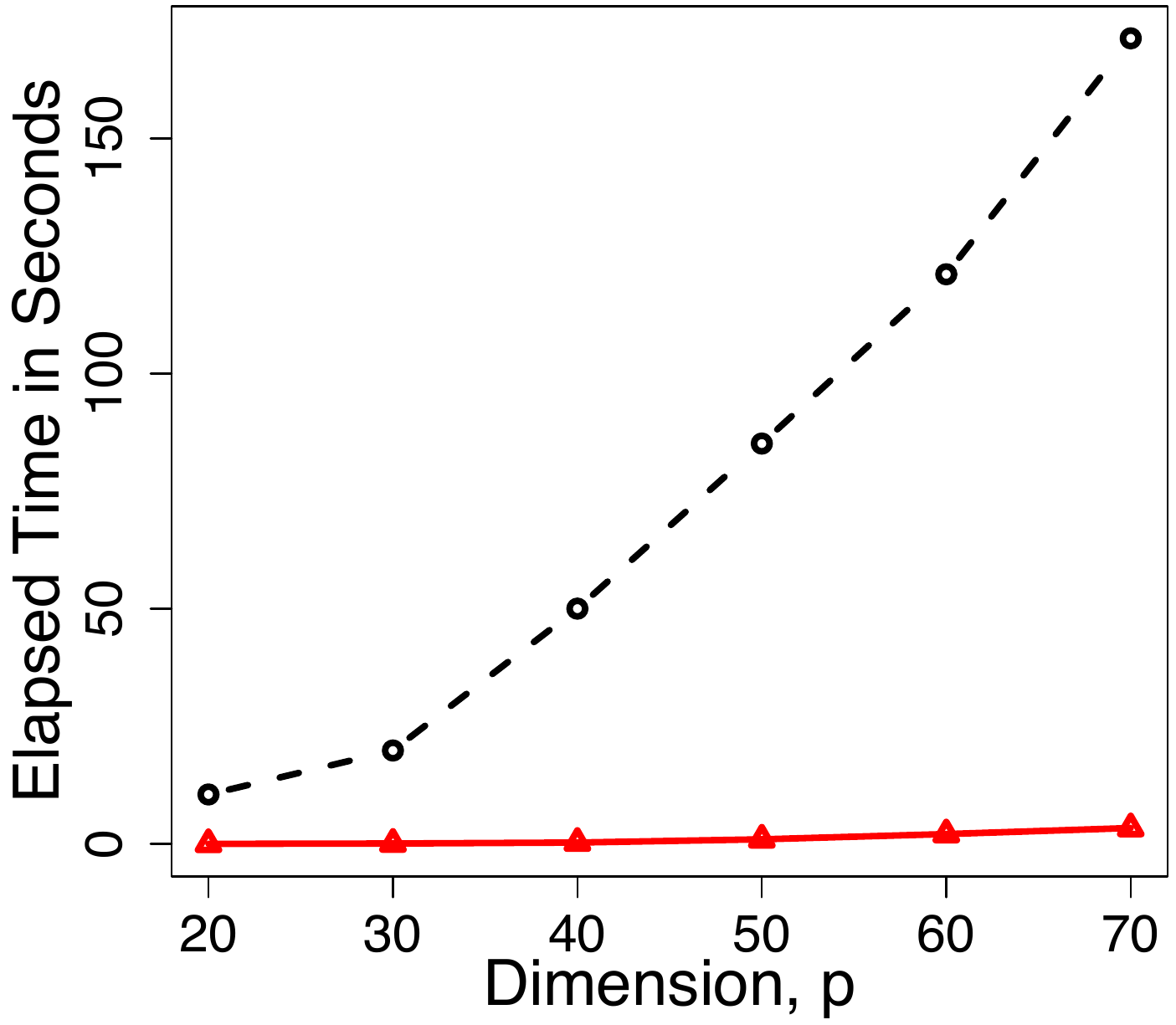}} 
        \subfigure[$t_{2.5}$ noise, $\alpha=0.05$.]{\includegraphics[width=0.32\textwidth]{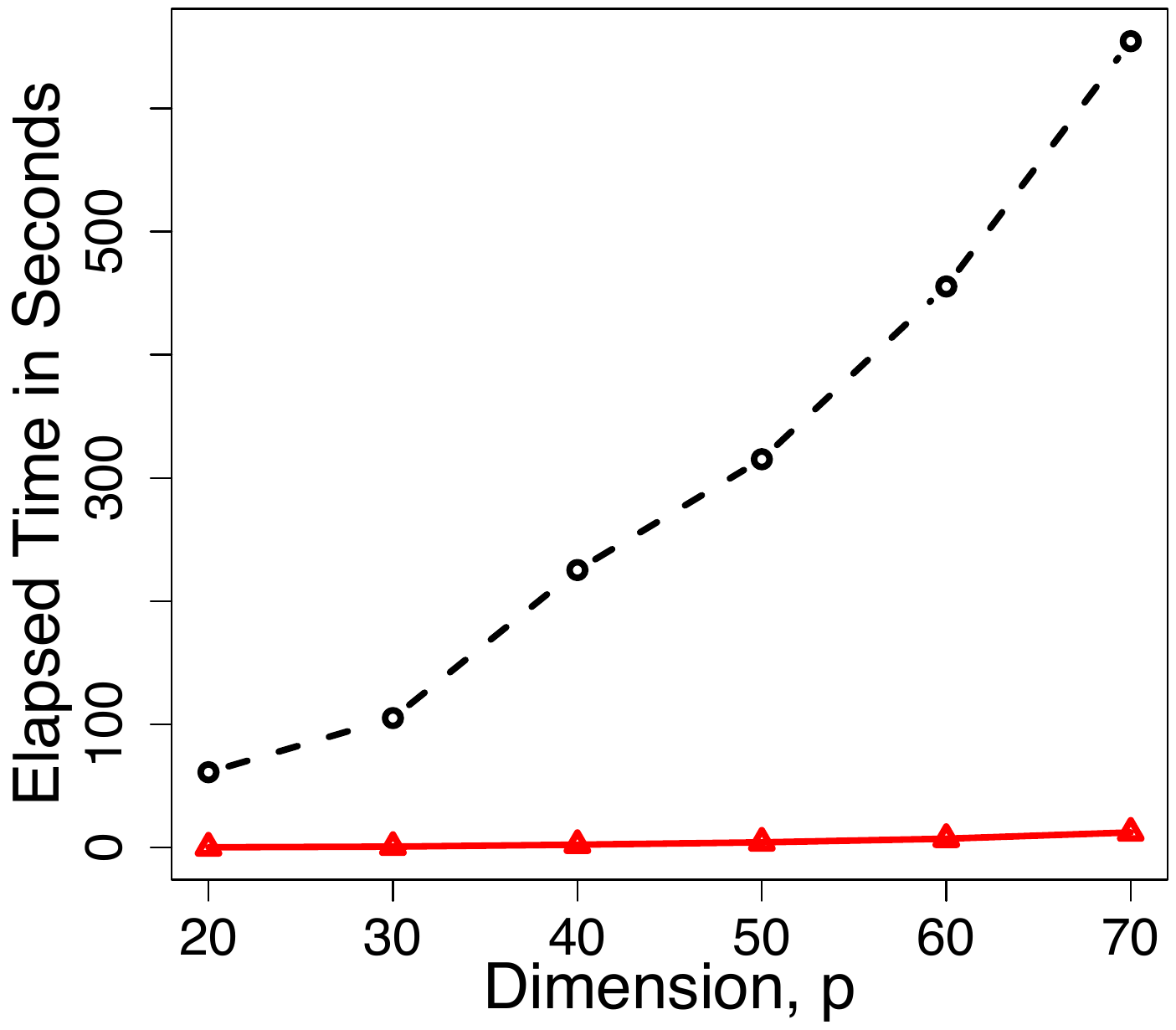}} 
          \subfigure[$t_{2.5}$ noise, $\alpha=0.1$.]{\includegraphics[width=0.32\textwidth]{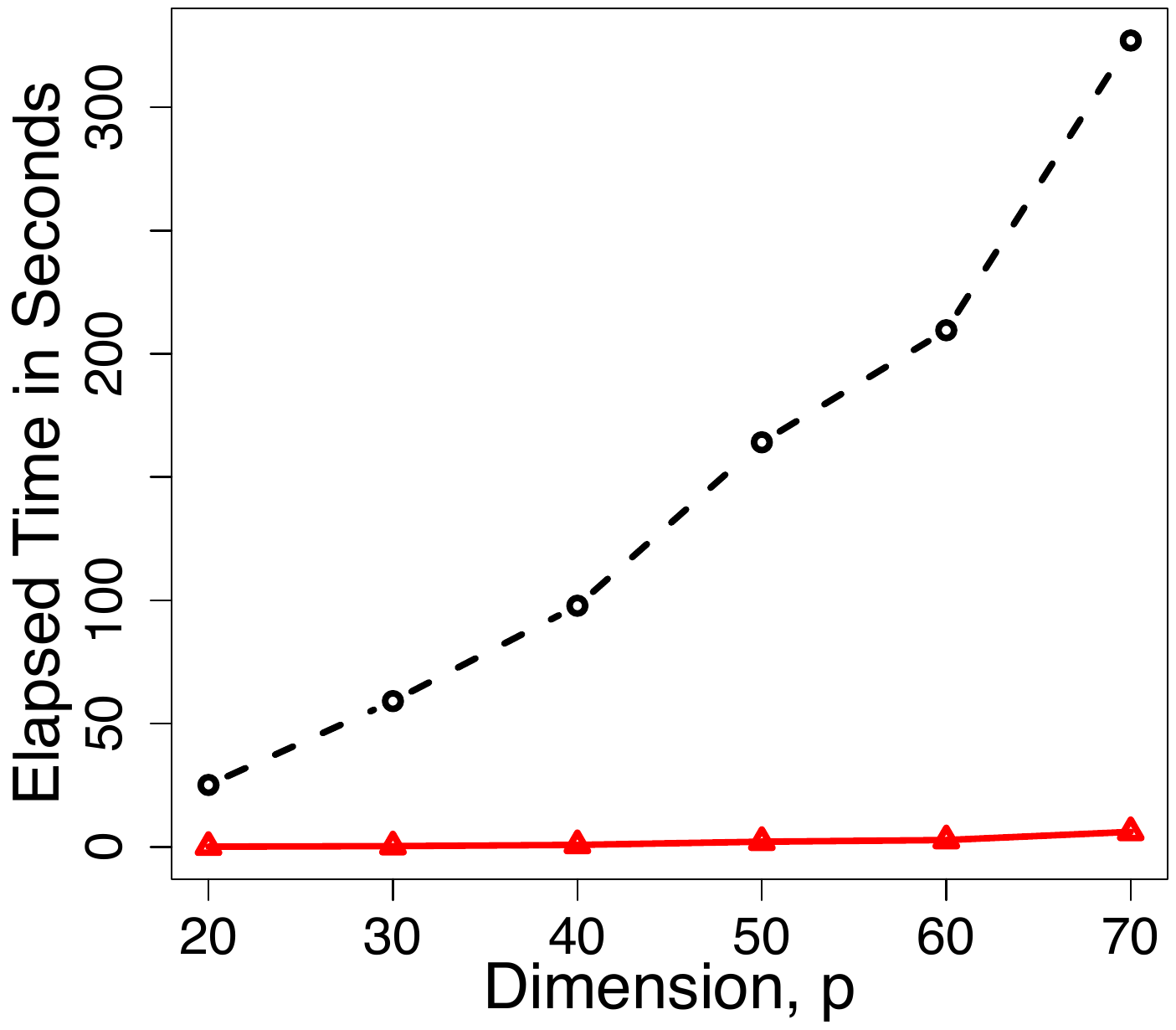}} 
            \subfigure[$t_{2.5}$ noise, $\alpha=0.2$.]{\includegraphics[width=0.32\textwidth]{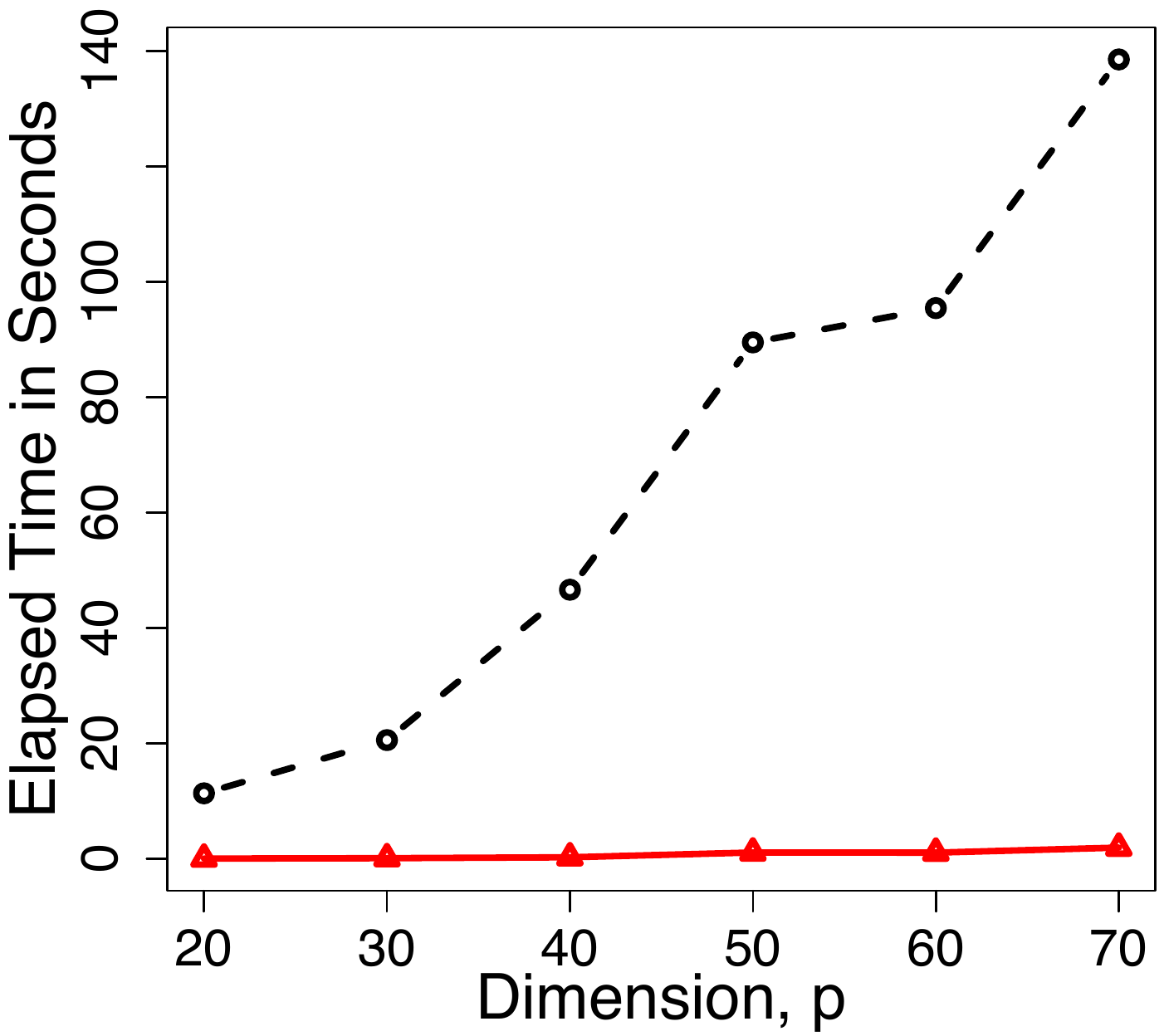}} 
\caption{Average elapsed time (in seconds) over 50 replications for the proposed method implemented by a combination of \texttt{quantreg} and \texttt{adaHuber} and the joint regression approach implemented by \texttt{esreg} under $\cN(0,1)$ and $t_{2.5}$ error models when $\alpha \in  \{0.05,0.1,0.2\}$.  The sample size is set to be $n =  \lceil{50 p/\alpha}\rceil$. The red solid and black dashed lines correspond to the proposed method and the joint regression approach, respectively.}
  \label{fig:timing}
\end{figure}

We first set the dimension $p=20$ and sample size $n =  \lceil{50 p/\alpha}\rceil $, where the quantile level $\alpha$ takes values in $\{0.05,0.1,0.2\}$. Simulation results on the relative $\ell_2$-error $ \|\hat{\theta}-\theta^*\|_2/\|\theta^*\|_2$, averaged over 200 replications, are reported in Tables~\ref{table:normal} and~\ref{table:t} under the $\cN(0,1)$ and $t_{2.5}$ noise model, respectively.  All four methods have very similar performance across different quantile levels in the normal model, while in the presence of heavy-tailed errors, the proposed robust method achieves  consistently more favorable performance.   This demonstrates that the use of adaptive Huber regression (in stage two) gains robustness against heavy-tailed errors without compromising statistical efficiency when the error distribution is light-tailed.  
In a more extreme setting where $\alpha=0.01$, Figure~\ref{fig:extreme} shows the boxplots of squared $\ell_2$-errors for three ES estimates (\texttt{2S-LS}, \texttt{2S-AH} and \texttt{joint}) under the normal and $t_3$ models.   Although the \texttt{2S-LS} estimator is easy-to-compute, it is more sensitive to heavy tailedness than the the joint estimator obtained via FZ loss minimization.
We further compare the proposed method with the joint regression approach in terms of computational efficiency. The computational time  in seconds, averaged over 50 independent replications, for the two methods with growing $(n, p)$ subject to $n =  \lceil{50 p/\alpha}\rceil $ ($\alpha \in \{0.05, 0.1, 0.2\}$) are reported in Figure~\ref{fig:timing}. These numerical results show evidence that our \texttt{R} implementation of the robust two-step method  can be faster than the \texttt{esreg} library for the joint regression approach by several orders of magnitude.

To shed some light on the drastic difference in numerical efficiency between the two methods,  note that the joint regression approach \citep{DB2019} relies on the Nelder-Mead simplex method, which is sensitive to the starting values and not guaranteed to converge to a local minimum. The convergence of the Nelder-Mead method is already very slow  for large-scale problems because it is a direct search method based on function comparison.  And due to its sensitivity to starting values, \cite{DB2019} proposed to re-optimize the model (several times) with the perturbed parameter estimates as new starting values. This explains, to some extent, the fast increase in runtime of \texttt{esreg} as both $n$ and $p$ grow.
The function in \texttt{quantreg} that fits linear QR is coded in \texttt{fortran}, and thus is very fast in larger problems. The computation of adaptive Huber regression is based on the Barzilai-Borwein gradient descent  method \citep{BB1988}, implemented via RcppArmadillo \citep{ES2014} in \texttt{adaHuber}.

Next, we construct entrywise (approximate) 95\% confidence intervals (CIs) for the expected shortfall regression parameter $\theta^*$.  The CI for the two-step estimator is based on \eqref{eq:twostep:ci} (non-robust) and \eqref{eq:ci} (robust), and we use the default option in the \texttt{esreg} package to implement \cite{DB2019}'s method. To evaluate the accuracy and reliability of the CIs, we compute the empirical coverage probability and interval width based on 500 independent replications, then averaged over the $p$ slope coefficients.  
Results for $p=20$ and $n =  \lceil{50 p/\alpha}\rceil$ ($\alpha \in  \{0.05,0.1,0.2\}$)  are reported in Tables~\ref{table:normal:inf} and \ref{table:t:inf}. Once again, all three methods perform similarly under normal errors, while the robust approach gives the narrowest CIs while maintaining the desired coverage level under $t_{2.5}$ errors.  Together, the results in Tables~\ref{table:t} and \ref{table:t:inf} demonstrate the robustness of the proposed method, as indicated by the theoretical investigations in Section~\ref{sec:3.2}.

\begin{table}[!htp]
\footnotesize
\begin{center}
\caption{Mean relative $\ell_2$-error $\|\hat{\theta}-\theta^*\|_2/ \|\theta^*\|_2$ (and standard error), averaged over 200 replications, when $\varepsilon_i \sim t_{2.5}$, $p = 20$, $n =  \lceil{50 p/\alpha}\rceil $ and $\alpha=\{0.05,0.1,0.2\}$.}
\begin{tabular}{l |     c |  c|c}
  \hline
&   \multicolumn{3}{c}{$t_{2.5}$ noise} \\ \hline
Method& $\alpha= $ 0.05& $\alpha= $ 0.1  &$\alpha= $ 0.2 \\ 
\hline
\texttt{2S-AH}  & 0.484 (0.008)  & 0.470 (0.009)& 0.429 (0.008)\\
\texttt{2S-LS}  & 0.612 (0.013)& 0.606 (0.016) & 0.532 (0.013)\\
\texttt{joint}  &0.581 (0.012) &0.567 (0.014) & 0.511 (0.013)\\
\texttt{2S-oracle}  & 0.612 (0.013)& 0.607 (0.016)& 0.532 (0.013)\\
\hline
\hline
\end{tabular}
\label{table:t}
\end{center}
\end{table}

\begin{table}[!htp]
\footnotesize
\begin{center}
\caption{Empirical coverage probability and mean width (based on 500 replications) of 95\% confidence intervals averaged over $p=20$ variables when $n =  \lceil{50 p/\alpha}\rceil $,  $\alpha=\{0.05,0.1,0.2\}$ and $\varepsilon_i \sim  t_{2.5}$.}
\begin{tabular}{c|     c   c|cc| cc}
  \hline
$t_{2.5}$ &   \multicolumn{2}{c}{$\alpha=0.05$} & \multicolumn{2}{c}{$\alpha=0.1$} & \multicolumn{2}{c}{$\alpha=0.2$} \\ \hline
Method & Coverage & Width& Coverage & Width& Coverage & Width    \\ \hline
\texttt{2S-AH}  & 0.947   & 3.633& 0.946 & 2.790 & 0.948&2.243\\
\texttt{joint}  &0.959  &5.771  & 0.959 &3.571& 0.954 & 2.872\\
\texttt{2S-LS}  & 0.952& 4.521  & 0.950& 3.397 & 0.953&2.687 \\
\hline
\end{tabular}
\label{table:t:inf}
\end{center}
\end{table}

\begin{table}[!htp]
\footnotesize
\begin{center}
\caption{Empirical coverage probability and mean width (based on 500 replications) of 95\% confidence intervals averaged over $p=20$ variables when $n =  \lceil{50 p/\alpha}\rceil $, $\alpha=\{0.05,0.1,0.2\}$ and $\varepsilon_i \sim \cN(0, 1)$.
}
\begin{tabular}{c|     c   c|cc| cc}
  \hline
$\cN(0, 1)$  &   \multicolumn{2}{c}{$\alpha=0.05$} & \multicolumn{2}{c}{$\alpha=0.1$} & \multicolumn{2}{c}{$\alpha=0.2$} \\ \hline
Method & Coverage & Width& Coverage & Width& Coverage & Width    \\ \hline
\texttt{2S-AH}  & 0.950  & 0.595& 0.949& 0.660 & 0.948 &0.744\\
\texttt{joint}  &0.946 &0.584 & 0.944 &0.651 & 0.942 & 0.740\\
\texttt{2S-LS}  & 0.950& 0.595  & 0.949 & 0.661 & 0.948&0.745\\
\hline
\end{tabular}
\label{table:normal:inf}
\end{center}
\end{table}

\begin{table}[!htp]
\footnotesize
\begin{center}
\caption{Mean relative $\ell_2$-error $ \|\hat{\theta}-\theta^*\|_2/ \|\theta^*\|_2$ (and standard error), averaged over 200 replications, when $\varepsilon_i \sim \cN(0,1)$, $p = 20$, $n =  \lceil{50 p/\alpha}\rceil $ and $\alpha=\{0.05,0.1,0.2\}$. }
\begin{tabular}{ l|     c |  c|c}
  \hline
&   \multicolumn{3}{c}{$\cN(0,1)$ noise} \\ \hline
Method& $\alpha= $ 0.05& $\alpha= $ 0.1  &$\alpha= $ 0.2\\ 
\hline
\texttt{2S-AH}  & 0.130 (0.003)  & 0.150 (0.003)& 0.171 (0.004)\\
\texttt{2S-LS}  & 0.130 (0.003)& 0.150 (0.003) & 0.171 (0.004)\\
\texttt{joint}  &0.130 (0.003) &0.151 (0.003) & 0.177 (0.004)\\
\texttt{2S-oracle}  & 0.129 (0.003)& 0.149 (0.003)& 0.171 (0.004)\\
\hline
\hline
\end{tabular}
\label{table:normal}
\end{center}
\end{table}

\subsection{Data Application I: Health Disparity}
\label{sec:data2}
Iron deficiency is one of the most common nutritional deficiency worldwide and is one of the leading cause of anemia \citep{camaschella2015iron}.  Being able to detect iron deficiency is essential in medical care for patients with inflammation, infection, or chronic disease. It is also important in preventive care since iron deficiency tends to present signs of a more serious illness such as gastrointestinal malignancy \citep{rockey1993evaluation}.  
One measure of iron deficiency that has proven to be useful is the soluble transferrin receptor (sTRP), a carrier protein for transferrin \citep{Mast1998}.  A high value of sTRP indicates iron deficiency.

The scientific goal here is to assess whether there is any disparity in sTRP levels among four different ethnic groups: Asian,  Black,  Mexican American, and White.  To this end, we analyze a dataset obtained from the National Health and Nutrition Examination Survey from 2017 to 2020 (pre-covid).    In this dataset,  the response variable sTRP was measured for female participants who range in ages from 20 to 49 years. The covariates of interest are three dummy variables that correspond to Asian,  Mexican American and Black, using White as the baseline.  We adjust for the demographic variables such as age, education level, and health diet throughout our analysis. For simplicity, we remove all participants with missing values on the covariates and the final dataset consists of $n=1689$ observations and $p=7$ covariates.

\begin{figure}[!htp]
  \centering
\includegraphics[scale=0.6]{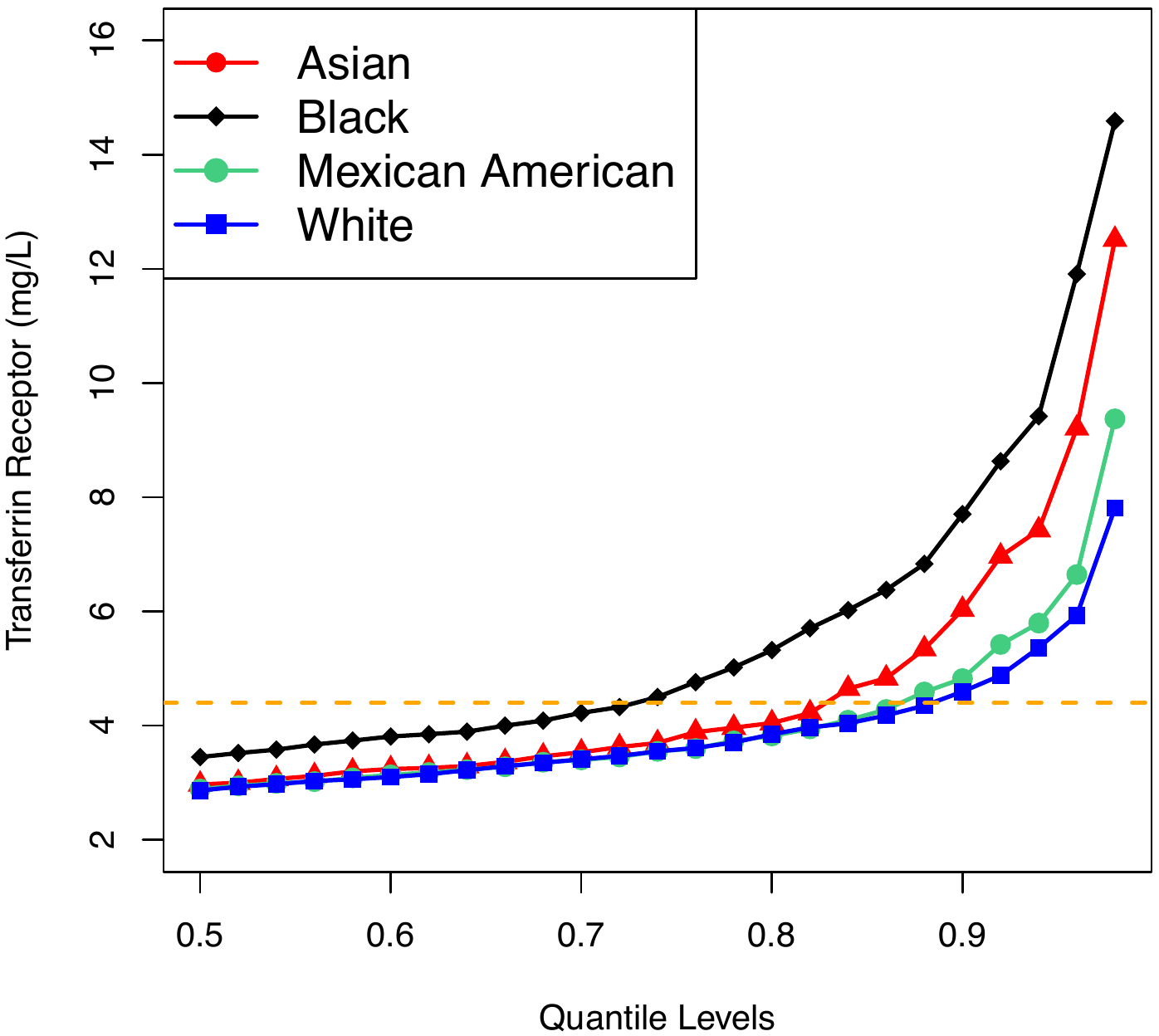}
\caption{The soluble transferrin receptor levels (mg/L) versus quantile levels (ranging from 0.5 to 0.99) for the female population in four different ethnic groups: Asian,  Black, Mexican American and White. The orange horizontal dash line indicates the upper bound of the normal range  (1.9--4.4 mg/L) for transferrin receptor among females.}
  \label{fig:disparity}
\end{figure}

As an exploratory analysis,  in Figure~\ref{fig:disparity} we plot the quantile curves of  sTRP measurements at levels from  50\% to 99\% for each of the four different ethnic groups. In this dataset, the sTRP values range from 1.24 to 35.1 mg/L.  We note that the normal range for females is between 1.9 to 4.4 mg/L \citep{Ketal2007},  and values that are much higher than 4.4 mg/L indicate severe iron deficiency.  We see from Figure~\ref{fig:disparity} that majority of the population have sTRP levels within the normal range.  However, there are large disparities between Black and the other three ethnic groups,  reflected in higher quantiles of the marginal distributions of sTRP.

To quantify the statistical significance of the aforementioned disparity, we fit robust expected shortfall regression at $\alpha= 0.75$ (upper tail),  with the robustification parameter tuned by the  procedure described in Section~\ref{sec:2.3}.  
This is equivalent to fitting the proposed \texttt{2S-AH} method at level $1-\alpha$ (see Section~\ref{sec:resr}) after flipping the signs of both the response and the covariates.   
We also implement the  standard quantile regression at level $\alpha$. 

Table~\ref{tab:disparity} reports the estimated coefficients and the associated 95\% confidence intervals for the three indicator covariates on the ethnic groups Asian, Mexican American and Black, using White as a baseline.  We see that both the quantile and robust expected shortfall regression methods are able to detect a health disparity between Black and White.   Specifically,  the estimated robust ES regression coefficient and 95\% CI (in the parenthesis) is 3.03 (1.88, 4.19)  versus its QR counterparts 0.86 (0.37, 1.35). 
With the use of quantile regression (at level 0.75),  we do not observe a statistically significant health disparity between Asian and White.
In contrast, \texttt{2S-AH} detects health disparity between Asian and White with estimated coefficient 2.34 (0.59, 4.09).  
We also see that the quantile regression detects health disparity between Mexican American and White, but the effect size is close to zero.  
In summary, ES regression complements QR, and can be more effective, as a tool to detect health disparity especially when it only occurs in the tail of the conditional distribution.

\begin{table}[!htp]
\footnotesize
\begin{center}
\caption{The estimated regression coefficients (and 95\% confidence intervals) for three dummy variables: Asian,  Black and Mexican American, using White as baseline.   Results of the upper-tail robust ES regression method  \texttt{2S-AH} and standard quantile regression at quantile level $\alpha =0.75$  are reported.  
}
\begin{tabular}{c|     c  | c|c}
  \hline
&   \multicolumn{1}{c}{Asian} & \multicolumn{1}{c}{Black} & \multicolumn{1}{c}{Mexican American} \\ \hline
Quantile Regression  &  0.31 (-0.02, 0.64)    &  0.86 (0.37, 1.35)&  -0.22 (-0.42, -0.01) \\
ES Regression (\texttt{2S-AH})  & 2.34 (0.59, 4.09)    & 3.03 (1.88, 4.19)  & 0.13 (-0.76, 1.03) \\
\hline
\end{tabular}
\label{tab:disparity}
\end{center}
\end{table}

\subsection{Data Application II: Job Training Partnership Act}
\label{sec:data}
We consider the Job Training Partnership Act (JTPA) study, a publicly-funded training program that provides training for adults with the goal of improving their earnings.    Specifically, we focus on Title II subprogram of the JPTA study that is mainly offered to adults with barriers to employment and out-of-schools youths.  
This dataset was previously analyzed in \citet{BOBCDLB1997}. It consists of 30-months accumulated earnings for 6102 females and 5102 males, with 16 covariates that are related to the demographics of the individuals such as age, race,  and the indicator variable that indicates whether the individual received JPTA training. 
After removing individuals with zero income, there are 4576 males and 5296 females. 
Our goal is to assess the effect of JPTA training on participants' earnings with an emphasis on the low income population that are employed, for both male and female subgroups.

To this end, we fit an expected shortfall regression model using the proposed robust method with $\alpha = \{0.05,0.1, 0.2\}$. The robustification parameter $\tau$ is selected automatically via the procedure described in Section~\ref{sec:2.3}.  Specifically, we regress the 30-months accumulated earnings on 
the JPTA training to assess the effect of JPTA training on low-income individuals, adjusting for whether individuals completed high school, race, Hispanic/non-Hispanic, marital status, working less than 13 weeks in the past year, and age. We report the estimated regression coefficient for the binary variable JPTA training and its associated 95\% confidence intervals.   
The results are summarized in Table~\ref{tab:JPTA}. 


From Table~\ref{tab:JPTA}, we see that 95\% confidence intervals for the robust method do not contain zero for all $\alpha \in \{0.05,0.1,0.2\}$.  This indicates that the JPTA training is statistically effective to improve earnings for the low-income population.  Specifically, for the male subpopulation, the estimated ES-effects of JPTA training are 283, 552, and 1093 dollars at levels $0.05$, $0.1$, and $0.2$, respectively. 
To further assess whether the estimated effects are scientifically meaningful, we compute the average 30-months accumulated earnings below the quantile levels $0.05,0.1$, and $0.2$ for the male subgroup, which are  214, 566, and 1496, respectively. We find that the JPTA training doubles the average income for individuals with income below the quantile levels 0.05 and 0.1, and becomes less effective for individuals with higher income.  Similar findings  are also observed for the female subgroup.




\begin{table}[!htp]
\footnotesize
\begin{center}
\caption{The estimated regression coefficient of the binary predictor JPTA training (and its 95\% confidence interval) for the proposed robust method and the standard quantile regression at quantile level $\alpha \in \{0.05,0.1,0.2\}$.  Results are rounded to the closest integer.   
}
\begin{tabular}{c|     c  | c|c}
  \hline
Male Subgroup &   \multicolumn{1}{c}{$\alpha=0.05$} & \multicolumn{1}{c}{$\alpha=0.1$} & \multicolumn{1}{c}{$\alpha=0.2$} \\ \hline
Quantile Regression & 465 (255, 675)  & 882 (603, 1161)& 2031 (1431, 2603)\\
ES Regression  (\texttt{2S-AH})  & 283 (149, 418)  & 552 (333, 771)& 1093 (641, 1546)\\
\hline
\hline
Female Subgroup &   \multicolumn{1}{c}{$\alpha=0.05$} & \multicolumn{1}{c}{$\alpha=0.1$} & \multicolumn{1}{c}{$\alpha=0.2$} \\ \hline
Quantile Regression  & 202 (76, 328)  & 480 (307, 653)& 1086 (719, 1452)\\
ES Regression  (\texttt{2S-AH}) & 123 (41, 205)  & 300 (146, 453)& 672 (385, 958)\\
\hline
\end{tabular}
\label{tab:JPTA}
\end{center}
\end{table}

\section{Conclusion and Discussions}
\label{sec:7}

This paper considers expected shortfall regression under a joint quantile and ES model recently proposed in \cite{DB2019} and \cite{PZC2019}. The existing approach is based on a joint $M$-estimator, defined as the global minimum of any member of a class of strictly consistent joint loss functions \citep{FZ2016} over some compact set. Since the loss function is non-differentiable and non-convex,  the  computation of such a joint $M$-estimator is intrinsically difficult especially when the dimensionality is large.  To circumvent the aforementioned challenge, \cite{B2020} proposed a two-step procedure for estimating the joint quantile and ES model based on Neyman-orthogonalization:  the first step involves fitting the quantile regression, and the second step employs the Neyman-orthogonal scores to estimate the ES parameters.  Due to the use of least squares method in the second step, the resulting estimator is sensitive to heavy-tailed error distributions.  

To address the robustness and computation concerns simultaneously, we propose a robust two-step method that applies adaptive Huber regression \citep{ZBFL2018} in the second step. The key is the use of a diverging robustification parameter for bias-robustness tradeoff, tuned by a convenient data-driven mechanism. The proposed method can be efficiently implemented by a combination of \texttt{R} packages \texttt{quantreg}/\texttt{conquer} and \texttt{adaHuber}. We establish a finite-sample theoretical framework for this two-step method, including deviation bound, Bahadur representation and (uniform) Gaussian approximations, in which the dimension of the model, $p$,  may depend on and increase with the sample size, $n$. Robust confidence intervals/sets are also constructed. Numerical experiments further demonstrate that the proposed robust ES regression approach achieves satisfying statistical performance, high degree of robustness (against heavy-tailed data) and superior computational efficiency and stability.  Through a real data application to the Job Training Partnership Act study, we show that ES regression complements QR as a useful tool to explore heterogeneous covariate effects on the average tail behavior of the outcome.

 Although we restrict attention to (joint) linear models in this work, our non-asymptotic theory and the underpinning techniques pave the way for analyzing (i) series/projection estimators under joint nonparametric quantile-ES models, and (ii) penalized estimators under high-dimensional sparse quantile-ES models. We leave these extensions in future research.  One limitation in our data analysis for the JTPA study is that we do not account for potential selection bias. Specifically, as pointed out by \citet{AAI2002}, out of all subjects that were assigned to participate in the training program, only approximate 60\% of them  (compliers) actually committed to the training program.   These individuals may simply have higher motivation in improving their earnings, and thus, the training status is likely positively correlated with potential income earnings. Generalizing the proposed method to estimating complier expected shortfall treatment effect, using an instrumental variable approach previously considered in \citet{AAI2002}, is another direction for future research.

The ES regression methods considered in this paper are suited for a fixed quantile level $\alpha \in (0, 1)$, independent of the sample size. For extreme quantiles satisfying $\alpha=\alpha_n \to$ 0 or 1 as $n\to \infty$, both the FZ loss minimization method (see \eqref{def:rho} and \eqref{def:M-est}) and two-step procedures perform poorly because observations become scarce at that level, i.e., $\alpha n$ is not large enough. In fact, if dimension $p$ is fixed, Theorem~\ref{thm:ES} and Theorem~\ref{thm:huber.ES} imply that the two-step ES regression estimates, robust and non-robust, are consistent if $\alpha_n^2 n \to \infty$ as $n\to \infty$.
In the case where $\alpha_n^2 n = O(1)$, these methods are no longer useful and one may need to resort to extreme value theory  \citep{HF2006,WLH2012}, which provides the statistical tools for a feasible extrapolation into the tail of the variable of interest. A more detailed discussion on modeling the extremes are deferred to Section~B of the Appendix.

 
 \appendix

\section{Expected Shortfall Regression without Crossing}
\label{sec:nc}

Recall the joint loss function $S(\beta, \theta; Y, X)$ and the score function $S_0(\beta, \theta; Y, X)$ given in \eqref{def:rho}.   Under model \eqref{joint.reg.model},  both the joint $M$-estimator \citep{DB2019} and the two-step procedure \citep{B2020}  rely on the moment conditions 
\$
 \PP_X  ( Y \leq X^\T \beta^* )  = \alpha  ~\mbox{ and }~  \EE_X   ( Y - X^\T \beta^*)  \mathbbm{1}( Y \leq X^\T \beta^* ) = \alpha X^\T(\theta^* - \beta^*) .
\$
The latter follows from the fact that 
$$
 \EE_X\{ Y  | Y\leq  Q_\alpha   \} = \frac{\EE_X  \{Y \mathbbm{1}( Y\leq Q_\alpha  )\} }{\alpha} = \frac{\EE_X ( Y - Q_\alpha  ) \mathbbm{1} ( Y\leq Q_\alpha ) }{\alpha} + Q_\alpha  ,
$$
where $Q_\alpha= Q_\alpha(Y|X)$ is the conditional $\alpha$-quantile of $Y$ given $X$.  By definition, the quantile and expected shortfall satisfy a monotonicity condition. At the population level, it holds under model \eqref{joint.reg.model} that
\#
	X_i^\T \theta^* = \ES_\alpha (Y_i |X_i) \leq Q_\alpha(Y_i|X_i) = X_i^\T \beta^*, \ \ i = 1,\ldots, n. \label{inequality.constraints}
\#
This requirement, however, is not necessarily satisfied by any of the estimators described in Sections~\ref{sec:2.2} and \ref{sec:2.3}. Even when both models (for quantile and ES) are correctly specified, for a dataset of modest size, the variability of the estimates may be large enough to upset the above inequality constraints enjoyed by their population counterparts. A large dataset may also contain scarce observations with fitted ES larger than fitted quantiles.

To obtain non-crossing conditional quantile and ES regression estimates, in the following we propose a constrained two-step method and its robust counterpart by directly incorporating the constraints in \eqref{inequality.constraints} to the optimization programs.
Let $\hat \beta \in \RR^p$ be the QR estimator of $\beta^*$, and set 
$$
  \hat Z_i = (Y_i -  X_i^\T \hat \beta ) \mathbbm{1}(Y_i \leq  X_i^\T \hat \beta ) + \alpha  X_i^\T \hat \beta , \ \ i = 1,\ldots, n.
$$
In the second step,  we define the non-crossing robust ES estimator $\hat \theta^{{\rm nc}}_\tau$ as a solution to the  constrained Huber loss minimization problem 
\# \label{constrained.es}
\begin{split}
	{\rm minimize}_{\theta \in \RR^p}  & ~~~~ \frac{1}{n} \sn \ell_\tau (\hat Z_i - \alpha X_i^\T \theta)    \\
  {\rm subject~to} &~ ~~~ X_i^\T \theta \leq  X_i^\T \hat \beta ,  \, i =1 ,\ldots , n,
  \end{split}
\#
where $\ell_\tau$ ($\tau>0$) is the Huber loss. For simplicity, we denote $\hat \theta^{{\rm nc}}= \hat \theta^{{\rm nc}}_\infty$ as the two-step ES estimator without crossing.

For any given $\beta \in \RR^p$,  define the subset $\cD_n(\beta) \subseteq \RR^{ p}$  as
$$
	\cD_n(\beta) = \{ \theta  \in \RR^p  :   X_i^\T \theta \leq X_i^\T \beta  , \, i=1 , \ldots , n \},
$$
which is a polyhedron and thus is convex. Under the joint model \eqref{joint.reg.model}, the true ES regression coefficient $\theta^*$ lies in the interior of $\cD_n(\beta^*)$, that is, $ X_i^\T \theta^* <  X_i^\T \beta^* $ for all $i=1,\ldots , n$.  Let $\hat \beta$ and $\hat \theta$ be the quantile and ES regression estimates given in \eqref{qr.est} and \eqref{ES.est}, respectively. By Proposition~\ref{prop:qr}, Theorem~\ref{thm:ES} and Remark~\ref{rmk4.4}, we have
\$
\max_{1\leq i\leq n} | X_i^\T(\hat \beta - \beta^* ) | \leq \| \hat \beta - \beta^* \|_\Sigma \cdot \max_{1\leq i\leq n} \| \Sigma^{-1/2} X_i \|_2  = \cO_{\PP}\bigg(\frac{p + \log n}{\sqrt{n}} \bigg) , \\
\max_{1\leq i\leq n} | X_i^\T(\hat \theta - \theta^* ) | \leq \| \hat \theta - \theta^* \|_\Sigma \cdot \max_{1\leq i\leq n} \| \Sigma^{-1/2} X_i \|_2  = \cO_{\PP}\bigg(\frac{p + \log n}{\alpha \sqrt{n}} \bigg) .
\$
Provided $\delta := \min_{1\leq i\leq n} X^\T_i(\beta^* - \theta^*)$ is bounded away from zero,  and under the condition $p^2 = o(n)$ as $n\to\infty$,   we see that with probability approaching one  the ES estimate $\hat \theta = \argmin_\theta  \sn \ell_\tau (\hat Z_i - \alpha X_i^\T \theta) $ maintains the proper ordering at each of the data points, that is,  $\hat \theta \in \cD_n(\hat \beta)$. This further implies $\PP\{ \hat \theta = \hat \theta^{{\rm nc}} \} \to 1$ as $n \to \infty$. The same conclusion also applies to their robust counterparts. Therefore, the non-crossing estimator obtained via \eqref{constrained.es} is asymptotically equivalent to the vanilla ES estimator.

Note that problem \eqref{constrained.es} with $\tau=\infty$ can be cast as a linearly constrained quadratic program ${\rm QP}(C, d, A, b)$ that is of the form
\#
 {\rm minimize}_{\theta \in \RR^p}   ~ \frac{1}{2}   \theta^\T C \theta + d^\T \theta  ~~~{\rm ~subject~to}~~ A \theta \leq b,  \label{cstr.qp}
\#
where $C =  (\alpha^2/n) \sn X_i X_i^\T \in \RR^{p \times p}$, $d = -(\alpha/n) \sn \hat Z_i X_i \in \RR^p$, $A= (X_1, \ldots, X_n)^\T \in \RR^{n\times p}$ and $b= ( X_1^\T \hat \beta, \ldots,  X_n^\T \hat \beta)^\T \in \RR^n$.  This can be solved efficiently by the dual method \citep{GI1983} implemented in the \texttt{quadprog} package. To compute $\hat \theta^{{\rm nc}}_\tau$ for any given $\tau>0$, in principle one can use \texttt{CVXR} \citep{FNB2020}, an R library for disciplined convex optimization, to solve the inequality constrained optimization problem \eqref{constrained.es} via generic solvers.
In fact, \texttt{CVXR} applies to any user-specified convex objective function, and therefore offers a high degree of flexibility.
For the Huber loss in particular, such a generic toolbox is blind to the problem structure and  tend to be much slower than standard quadratic programming solvers when the sample size and/or dimension are large. 
In the following, we take advantage of the special structure of the Huber loss and propose a tailored combination of the iteratively reweighted least squares (IRLS) algorithm and quadratic programming.  For any $\tau>0$,  it can be shown that   the linearly constrained Huber loss minimization problem \eqref{constrained.es} is equivalent to   (see Exercise 4.5 in \cite{BV2004})
\$
  \begin{split}
	{\rm minimize}_{\theta\in \RR^p ,    w_1,\ldots, w_n \in \RR}  & ~~~~ \sn \bigg\{  \frac{(\hat Z_i - \alpha X_i^\T \theta)^2 }{w_i + 1} + \tau^2 w_i \bigg\}    \\
  {\rm subject~to} &~ ~~~  X_i^\T \theta \leq  X_i^\T \hat \beta ,  w_i \geq 0 , i =1,\ldots , n.
  \end{split}  
\$
This problem can be interpreted as a (constrained) weighted least squares problem, which naturally motivates an iteratively reweighted least squares algorithm as follows.
Starting  at iteration 0 with an initial estimate $\theta^0$, say $\theta^0 =  \hat \theta^{{\rm nc}}$,  we repeat the following two steps until convergence.
\begin{itemize}
\item[(i)] At iteration $t=0, 1,  2, \ldots$,  compute the residuals $\{ \omega^t_i = \hat Z_i - \alpha X_i^\T \theta^t \}_{i=1}^n$.  Let $\tau^t>0$ be the solution to the equation $ \sn ( | \omega^t_i | \wedge \tau )^2/\tau^2 = p + \log n$.

\item[(ii)] Compute the weight of the $i$th residual as $w_i^t= (  | \omega_i^t | / \tau^t  - 1  ) \mathbbm{1} (  | \omega_i^t |  > \tau^t )$. Then use \texttt{quadprog} to solve the constrained weighted least squares problem
\$
 {\rm minimize}_{\theta \in \RR^p}   \frac{1}{n} \sn \frac{(\hat Z_i - \alpha X_i^\T \theta)^2}{1 + w_i^t}    ~~{\rm subject~to}  ~ A \theta \leq b
\$
to obtain $\theta^{t+1}$, where the matrix $A$ and vector $b$ are as in \eqref{cstr.qp}.
Note that this is a linearly constrained quadratic program, denoted by  ${\rm QP}(C^t,  d^t,  A, b)$ with
$$
C^t = \frac{\alpha^2}{n} \sn \frac{1}{1+w_i^t} X_i X_i^\T ~~\mbox{ and }~ \quad d^t = - \frac{\alpha}{n} \sn \frac{\hat Z_i}{1+w_i^t}X_i  .
$$ 
\end{itemize}

\begin{figure}[!htp]
  \centering
  \subfigure[Normal model with $(n,p)=(5000, 10)$]{\includegraphics[height=0.2\textheight, width=0.47\textwidth]{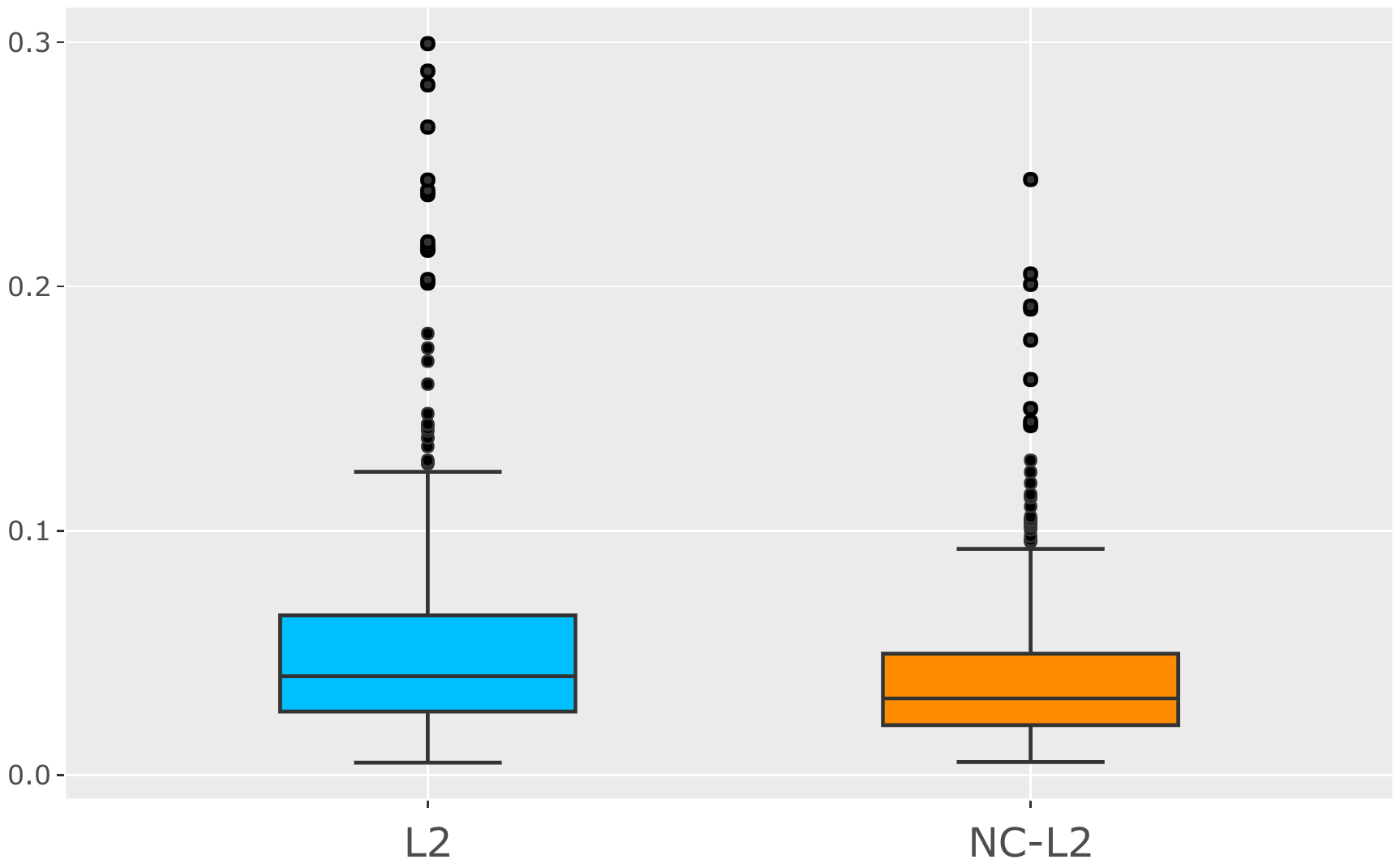}}~~~~
   \subfigure[Normal model with $(n,p)=(5000, 10)$]{\includegraphics[height=0.2\textheight, width=0.47\textwidth]{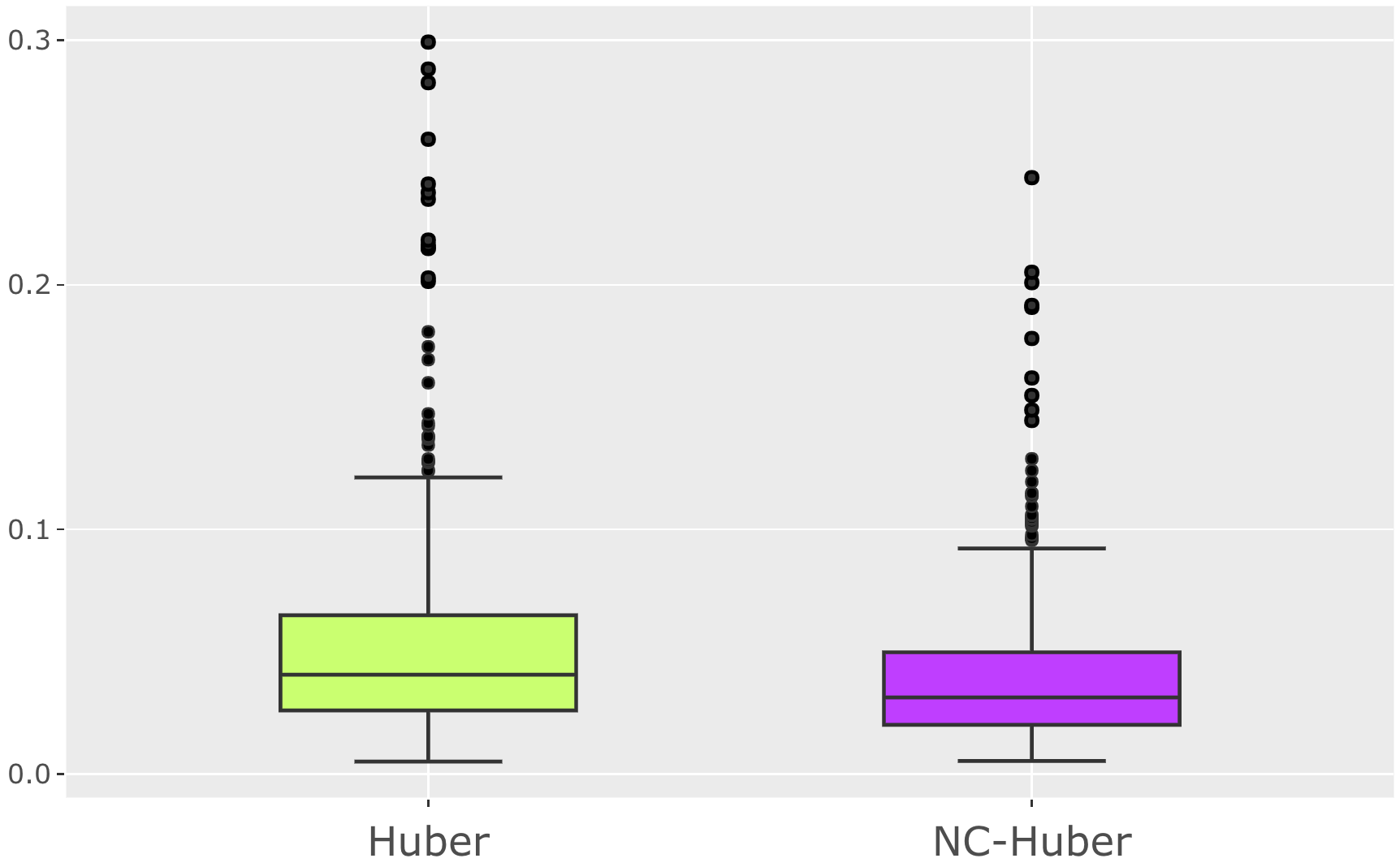}}
     \subfigure[$t_{2.5}$ model with $(n,p)=(10000, 10)$]{\includegraphics[height=0.2\textheight, width=0.47\textwidth]{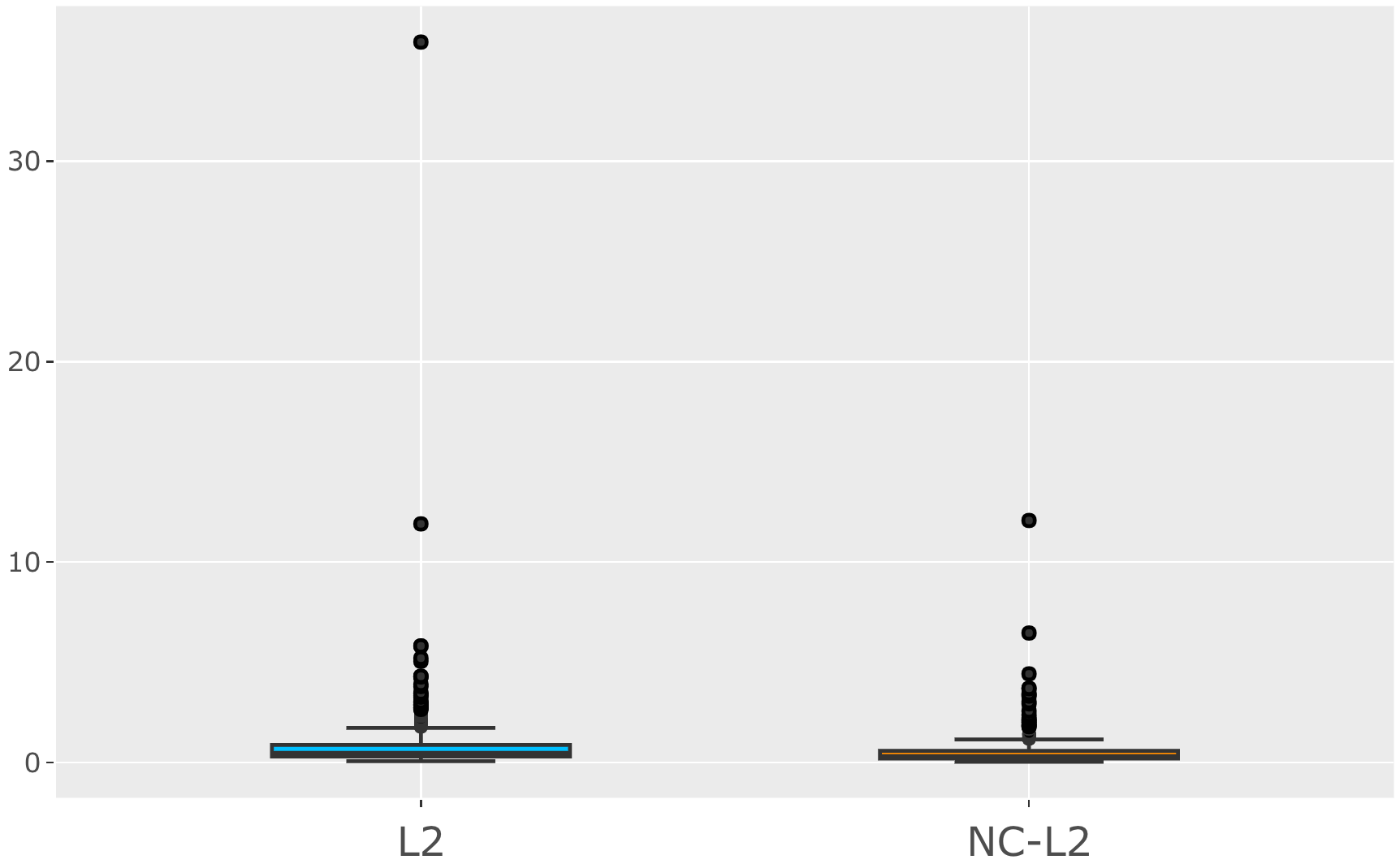}}~~~~
   \subfigure[$t_{2.5}$ model with $(n,p)=(10000, 10)$]{\includegraphics[height=0.2\textheight, width=0.47\textwidth]{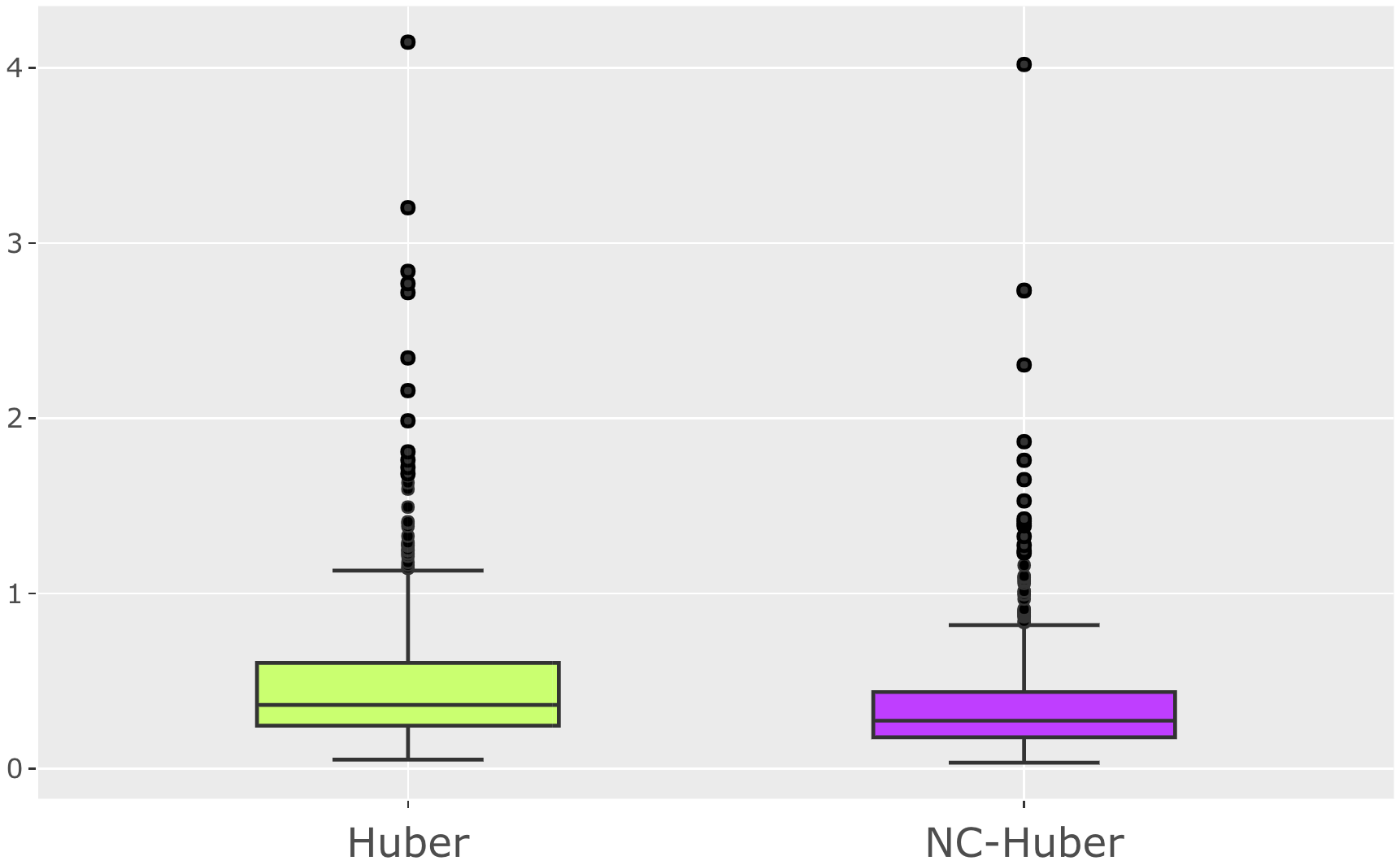}}  
\caption{Boxplots of squared $\ell_2$-error (based on 500 replications) for twp-step ES estimates and their non-crossing counterparts at quantile level $\alpha=0.1$. Under the normal model, the MSEs of the four methods ($L_2$, NC-$L_2$, Huber and NC-Huber) are 0.0534, 0.0407, 0.0532 and 0.0408; under the $t_{2.5}$ model, the MSEs of the four methods are 0.8587, 0.5277, 0.4951 and 0.3794. }
  \label{fig:all}
\end{figure}

\begin{figure}[!htp]
  \centering
  \subfigure[Normal model with $(n,p)=(5000, 10)$]{\includegraphics[height=0.21\textheight, width=0.47\textwidth]{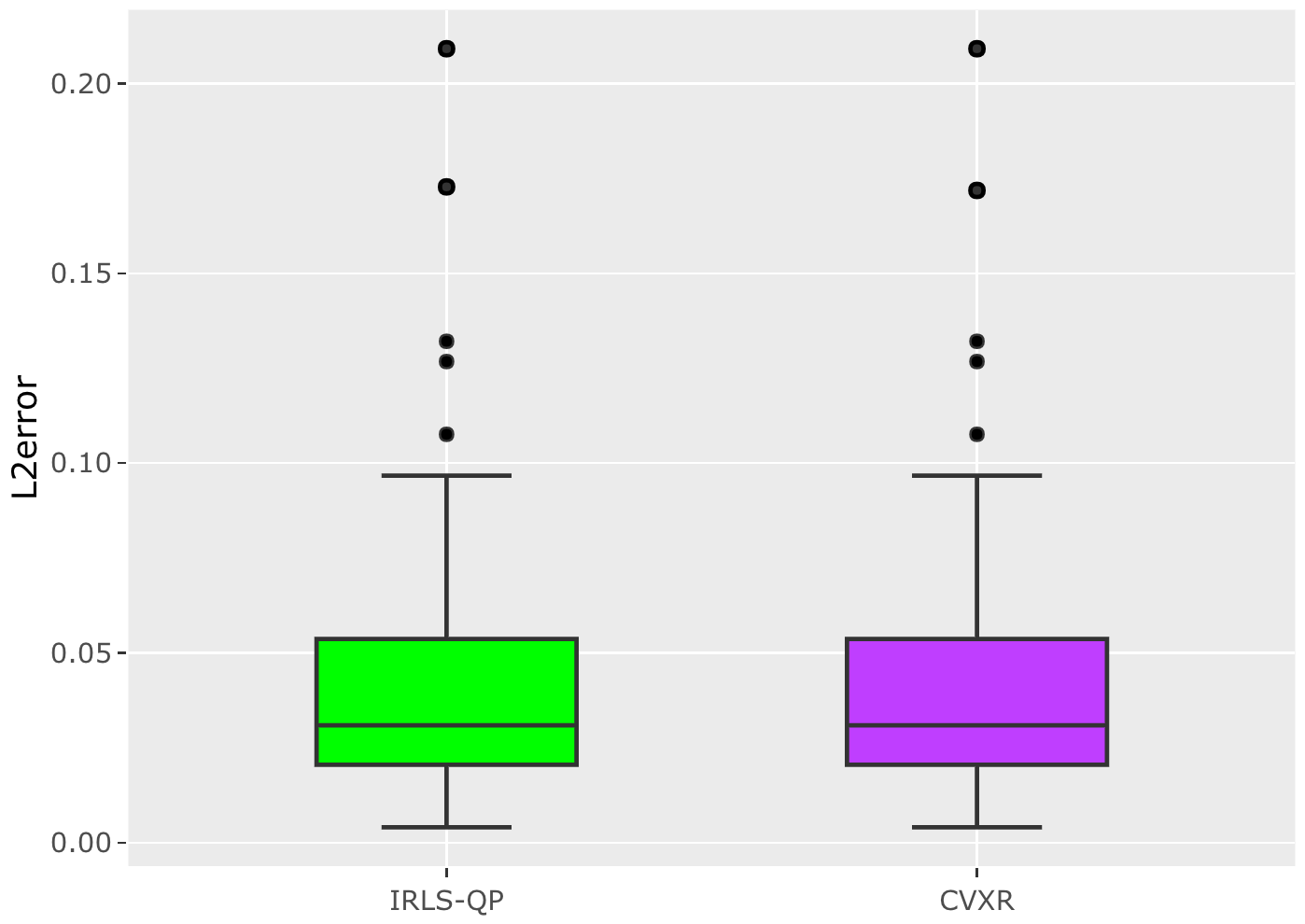}}~~~~
   \subfigure[Normal model with $(n,p)=(5000, 10)$]{\includegraphics[height=0.21\textheight, width=0.47\textwidth]{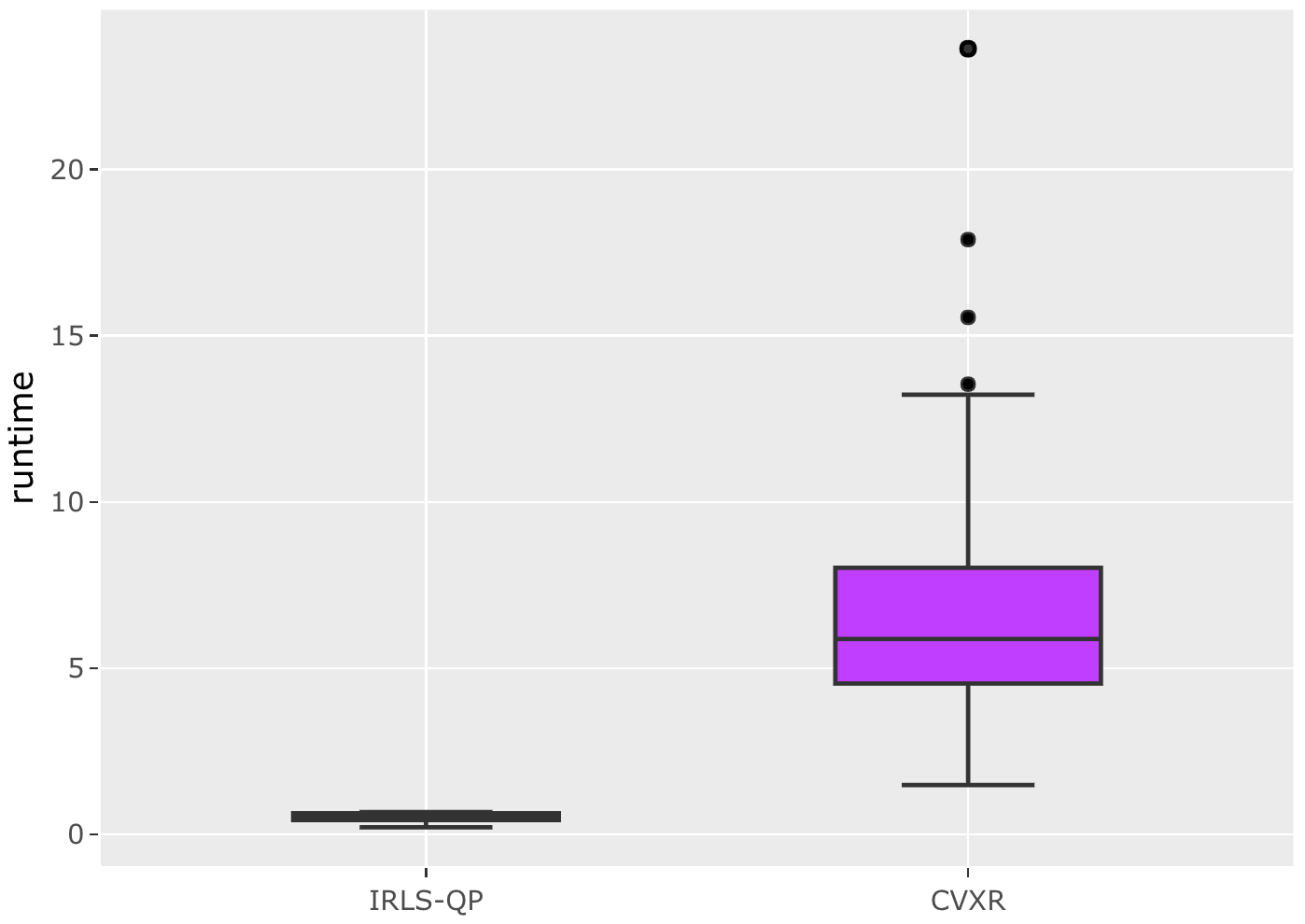}}
     \subfigure[$t_{2.5}$ model with $(n,p)=(8000, 10)$]{\includegraphics[height=0.21\textheight, width=0.47\textwidth]{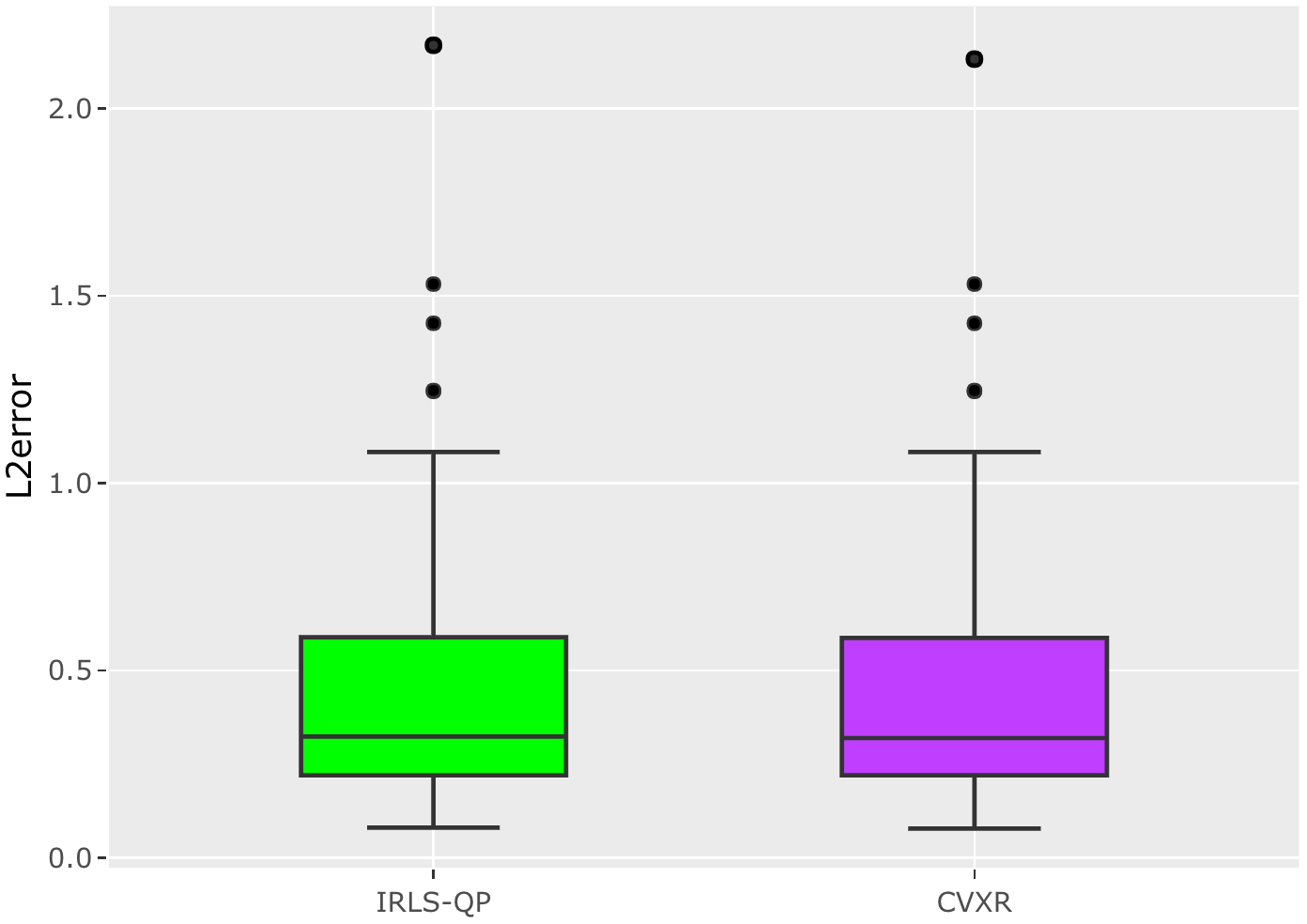}}~~~~
   \subfigure[$t_{2.5}$ model with $(n,p)=(8000, 10)$]{\includegraphics[height=0.21\textheight, width=0.47\textwidth]{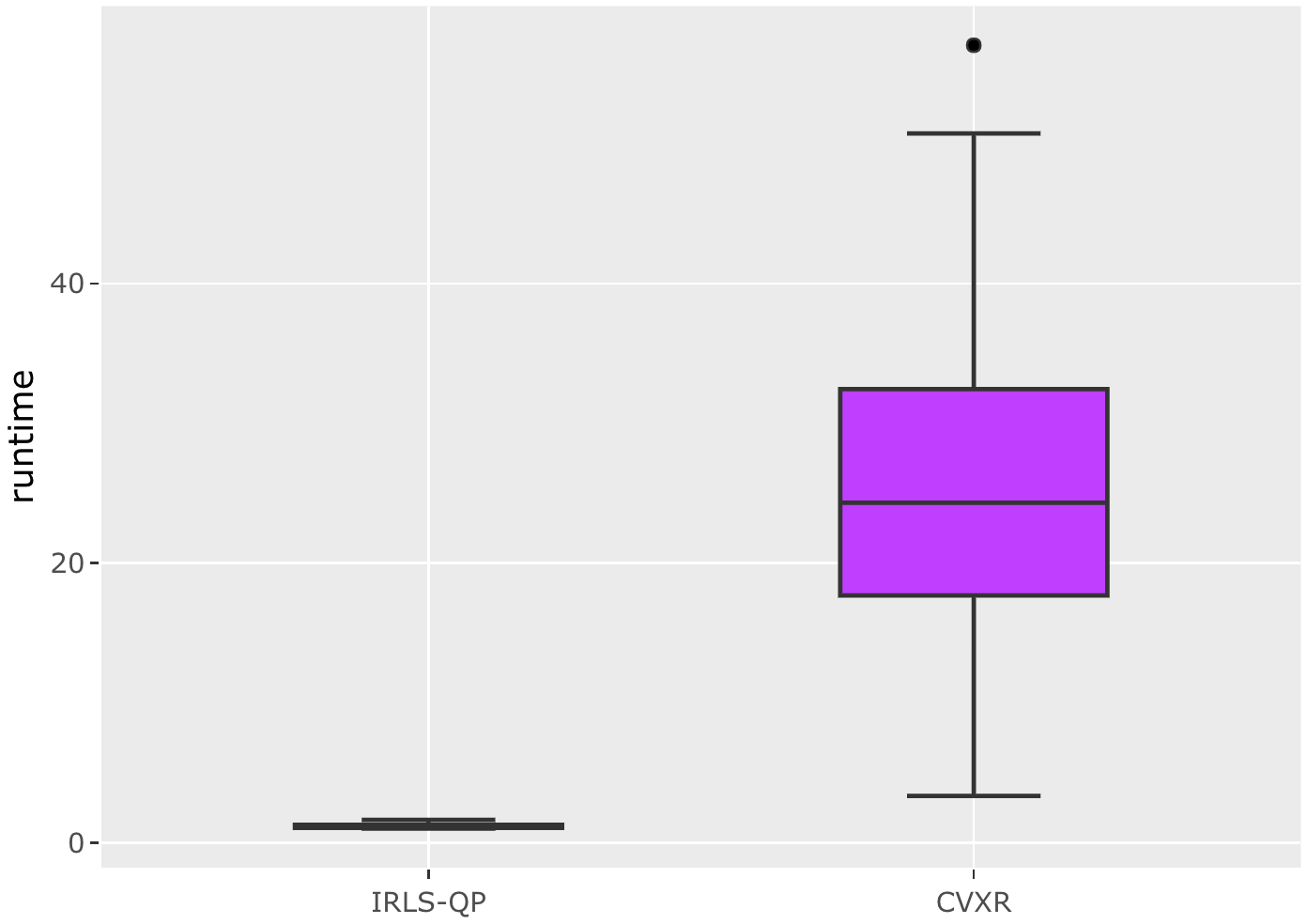}}  
\caption{Boxplots of  squared $\ell_2$-errors (panels (a) \& (c)) and computational time in seconds (panels (b) \& (d)),  based on 100 replications,  for non-crossing two-step robust ES estimation (at quantile level $\alpha=0.1$) implemented by the IRLS-QP algorithm and \texttt{CVXR} library.}
  \label{fig:nc}
\end{figure}

We end this section with an additional simulation study to  demonstrate the  effectiveness of non-crossing ES estimation in finite samples.  We generate random samples $\{ (Y_i, X_i) \}_{i=1}^n$ from the heteroscedastic error model $Y_i = 2 + X_i^\T \gamma^* + X_i^\T \eta^*  \cdot  \varepsilon_i $, where $X_i \in \RR^p$ consists of independent $\mathrm{Unif}(0, 2)$ components,  $\eta^* = (0.5, 0.5, 0, \ldots, 0)^\T \in \RR^p$ and $\varepsilon_i$  follows either the standard normal distribution or the $t_{2.5}$-distribution.  In the normal error case,  we fix some $\gamma^* \in \RR^p$,  generated from the uniform distribution on the unit sphere $\mathbb{S}^{p-1}$;  in the $t_{2.5}$-error case,  we set  $\gamma^*\in \sqrt{5}\, \mathbb{S}^{p-1}$. Figure~\ref{fig:all} shows the boxplots of squared $\ell_2$-errors for four different two-step ES regression estimates ($L_2$, non-crossing $L_2$, Huber and non-crossing Huber) at quantile level $\alpha=0.1$ under the $\cN(0,1)$ and $t_{2.5}$-error models. The non-crossing estimates, computed by the IRLS-QP algorithm, achieve consistently more favorable performance. For algorithmic comparisons, from Figure~\ref{fig:nc} we see that the proposed IRLS-QP algorithm shows significant improvement over \texttt{CVXR} implementation in computational efficiency while achieves the same level of statistical accuracy. The average runtime in seconds is $0.41$  versus $7.38$ in the normal model with $(n, p) = (5000, 10)$, and $1.19$  versus $25.94$ in the $t_{2.5}$ model with $(n, p )= (8000, 10)$. The reference machine for the above experiments is an iMac with a 3.7 GHz 6-Core Intel Core i5 processor and 32 GB of RAM.

\section{Modeling Extremes}
\label{sec:extremesupp}

A commonly used approach for the estimation of extreme conditional quantiles models extremes by fitting a fully parametric model,  such as generalized extreme value distribution or generalized Pareto distribution (GPD),  where the location, shape and scale parameters are allowed to depend on covariates either parametrically or nonparametrically. For example, the CDF of the GPD with positive shape and scale parameters $\xi$ and $\sigma$ is $G_{\xi, \sigma}(x) = 1- (1+\xi x / \sigma)^{-1/\xi}$ for $x\geq 0$; see, e.g. Definition 7.16 in \cite{MFE2015}. 
Let $Y$ represent some loss random variable, and assume that for some high threshold $u$, its excess distribution over $u$, denoted by
\$
	F_u(x) = \PP(Y-u \leq x  \, | Y > u) ~~\mbox{ for }~ x \geq 0 ,
\$
satisfies $F_u(x) = G_{\xi, \sigma}(x)$ for some $\xi >0$ and $\sigma>0$. Following (7.17)--(7.19) in \cite{MFE2015}, we have that for $\alpha \geq \alpha_0 = F(u)$ and $\xi<1$, the $\alpha$-quantile and upper $\alpha$-ES of $Y$ can be written as
\# \label{extreme.model}
	\begin{cases}  \vspace{.2cm}
	Q_\alpha(Y)  =  Q_{\alpha_0}(Y) + \frac{\sigma}{\xi}  \big\{ \big( \frac{1-\alpha_0}{1-\alpha } \big)^{ \xi} - 1 \big\} , \\
 {\rm ES}^+_\alpha(Y)  = \frac{1}{1-\alpha}  \int_\alpha^1 Q_u(Y) {\rm d}u  =  \frac{Q_\alpha(Y)}{1-\xi} + \frac{\sigma - \xi  Q_{\alpha_0}(Y) }{1- \xi},
 \end{cases}
\#
where the threshold $u$ is chosen as an intermediate quantile $Q_{\alpha_0}(Y)$ at level $\alpha_0 \in (0, 1)$ close to 1 but fixed. It follows immediately that $\lim_{\alpha \to 1} {\rm ES}^+_\alpha(Y) / Q_\alpha(Y) = (1-\xi)^{-1}$. In the presence of covariates,  we have a conditional version of \eqref{extreme.model}:
\$
	\begin{cases} \vspace{.2cm}
	Q_\alpha(Y|X=x)  =  Q_{\alpha_0}(Y|X=x) + \frac{\sigma(x)}{\xi(x)}  \big\{ \big( \frac{1-\alpha_0}{1-\alpha } \big)^{ \xi(x)} - 1 \big\} , \\ 
 {\rm ES}^+_\alpha(Y|X=x)   =  \frac{Q_\alpha(Y|X=x)}{1-\xi(x)} + \frac{\sigma - \xi(x)  Q_{\alpha_0}(Y|X=x) }{1- \xi(x)} .
 \end{cases}
\$
This shows that the estimation of ${\rm ES}^+_\alpha(Y|X=x)$ essentially depends on that of the extreme quantile $Q_\alpha(Y|X=x)$,  which further requires estimates of the intermediate quantile $Q_{\alpha_0}(Y|X=x)$ and of the conditional GPD parameters $\xi(x)$ and $\sigma(x)$. For a more complete review of extreme quantile regression that dates back to \cite{C2005}, we refer to \cite{VDCE2021} and the references therein.

\section{Expected Shortfall Autoregression: A Numerical Study}
 
In this section,  we conduct additional empirical investigations in serial dependent settings  where the covariates include  lagged values of the response.  Let $\{ U_t\}_{t\geq 1}$ be a sequence of i.i.d.  $\mathrm{Unif}(0,1)$ random variables, and  $\{Z_t\}_{t\geq 1}$ a sequence  of $p$-vectors of covariates that are independent of $\{ U_t\}$.
Motivated by \cite{KX2006},  we consider the following data generating mechanism 
\#
	Y_t = \beta_0(U_t)  +  \sum_{j=1}^{q} \beta_j(U_t) Y_{t-j} + Z_{t-1}^\T \gamma(U_t), \label{qar.model}
\#
where  $\theta_j: [0, 1] \to \RR$ ($0\leq j\leq q$)  and $\gamma: [0, 1]\to \RR^p$ ($p\geq 1$) are unknown functions. Provided the right-hand side of \eqref{qar.model} is increasing in $U_t$,   for any $\alpha \in (0, 1)$ it holds
\#
	Q_\alpha(Y_t | \cF_{t-1} ) =  \beta_0(\alpha)+  \sum_{j=1}^{q} \beta_j(\alpha) Y_{t-j}  + Z_{t-1}^\T  \gamma(\alpha)  ,  \label{q.ar}
\#
where $\cF_t$ is the $\sigma$-field generated by $\{ (Y_s, Z_s)\}_{s\leq t}$. Combining \eqref{q.ar} with \eqref{ES.def2},   we further obtain the conditional expected shortfall of $Y_t$ given $\cF_{t-1}$ as
\#
	\ES_\alpha  (Y_t | \cF_{t-1} ) =  \theta_0(\alpha)+  \sum_{j=1}^{q} \theta_j(\alpha) Y_{t-j}  + Z_{t-1}^\T  \eta(\alpha)  , \label{es.ar}
\#
where $\theta_j (\alpha) = \alpha^{-1}\int_0^\alpha \beta_j(u) {\rm d}u $, $0\leq j\leq q$ and $\eta(\alpha) = \alpha^{-1} \int_0^\alpha \gamma(u) {\rm d}u$.

Write $X_t = (1,  Y_{t-1} , \ldots, Y_{t-q} , Z_{t-1}^\T)^\T \in \RR^{p+q+1}$.  The above models can be expressed in a more compact form
 \#
 Q_\alpha(Y_t | \cF_{t-1} )= X_t^\T \beta_\alpha^* , \quad \ES_\alpha  (Y_t | \cF_{t-1} ) = X_t^\T \theta_\alpha^* , \label{joint.ar}
 \#
 where $\beta_\alpha^* = (\beta_0(\alpha) , \beta_1(\alpha), \ldots, \beta_q(\alpha), \gamma(\alpha)^\T )^\T$ and  $\theta_\alpha^* =  (\theta_0(\alpha) , \theta_1(\alpha), \ldots, \theta_q(\alpha), \eta(\alpha)^\T )^\T$.  In particular,  model \eqref{q.ar} with $\gamma \equiv 0$ is named the QAR$(q)$ model by  \cite{KX2006}.  We thus refer to \eqref{es.ar} with $\eta\equiv 0$ as the  ESAR$(q )$ model.
 
To estimate $\beta(\alpha)  : =  ( \beta_0(\alpha), \beta_1(\alpha), \ldots, \beta_q(\alpha) )^\T$ at each $\alpha \in (0 ,1)$ under the QAR$(q)$ model,   \cite{KX2006}  used the  standard quantile regression estimate 
$$
	\hat \beta(\alpha) \in \argmin_{ b \in \RR^{q+1} } \sum_{t=1}^T \rho_\alpha( Y_t - X_t^\T b)  ~\mbox{ with }~ X_t = (1, Y_{t-1}, \ldots, Y_{t-q})^\T 
$$
and proved its asymptotic normality under certain distributional assumptions.  For  the estimation and inference of 
$ 
	\theta(\alpha) : = ( \theta_0(\alpha),  \theta_1(\alpha), \ldots, \theta_q(\alpha) )^\T
$ 
under the ESAR$(q)$ model \eqref{es.ar} with $\eta \equiv 0$,  it is natural to consider one of the following two-step estimates:
\#
\begin{cases}
	\hat \theta_{{\rm FZ}} (\alpha)   \in  \argmin_{\theta\in \RR^{q+1}} \frac{1}{T} \sum_{t=1}^{T} S(\hat \beta (\alpha) , \theta; Y_t, X_t)   ,   \\
	\hat \theta_{{\rm LS}} (\alpha)  = \argmin_{\theta\in \RR^{q+1}} \frac{1}{T} \sum_{t=1}^{T} S^2_0(\hat \beta (\alpha) , \theta; Y_t, X_t) ,   \\
	\hat \theta_{{\rm AH}} (\alpha)  = \argmin_{\theta\in \RR^{q+1}} \frac{1}{T} \sum_{t=1}^{T} \ell_\tau(S_0(\hat \beta (\alpha) , \theta; Y_t, X_t) ) ,   
	\end{cases} \label{ar.est}
\#
where $S(\beta, \theta; Y , X)$ and $S_0(\beta, \theta; Y , X)$ are defined in \eqref{def:rho}, and $\ell_\tau$ denotes the Huber loss.

 In the following, we report a Monte Carlo experiment conducted to examine the performance of three ES estimation procedures under  serial dependence,  which are  the joint regression method via FZ loss minimization \citep{DB2019, PZC2019},  two-step least squares method and two-step adaptive Huber method.  The data $\{ (Y_t , Z_t ) \}_{t=1}^{T+20}$ in this experiment were generated from model \eqref{qar.model} with $q=p=1$,  $Z_t   \sim_{{\rm i.i.d.}} {\rm Unif}(0,1)$ and
$$
	\beta_0(u) = F^{-1}(u), \quad  \beta_1(u) = a_0 + a_1 u , \quad \gamma(u) = b_0 + b_1 u  ~\mbox{ for }  u \in (0, 1),  
$$
where $F$ is the CDF of either the standard normal distribution (normal model) or $t_\nu$-distribution with $\nu$ degrees of freedom ($t_\nu$ model). Here we require $a_0, a_1, b_1 >0$ and $a_0 + a_1 \leq 1$.
 The  first 20 observations were discarded to allow for a reasonably long burn-in period.   At quantile level $\alpha \in (0, 1)$,  the correspondent ES regression coefficients in \eqref{es.ar} are 
\#
\theta_0(\alpha) = \frac{1}{\alpha} \int_0^\alpha F^{-1}(u) {\rm d}u , \quad   \theta_1(\alpha) = a_0 +  0.5 a_1 \alpha  , \quad \gamma(\alpha) = b_0 +   0.5  b_1 \alpha . \label{sim.coef}
\# 
Specifically, we set model parameters $(a_0, a_1) = (0.5, 0.5)$, $(b_0, b_1) = (0.95, 0.5)$, $\nu = 3.5$ and consider the quantile level $\alpha = 0.05$.   The sample size $T$ is $1000$ in the normal model and $1500$ in the $t_{3.5}$ model.   
 
\begin{figure}[!htp]
  \centering
  \subfigure[Normal model]{\includegraphics[height=0.2\textheight, width=0.48\textwidth]{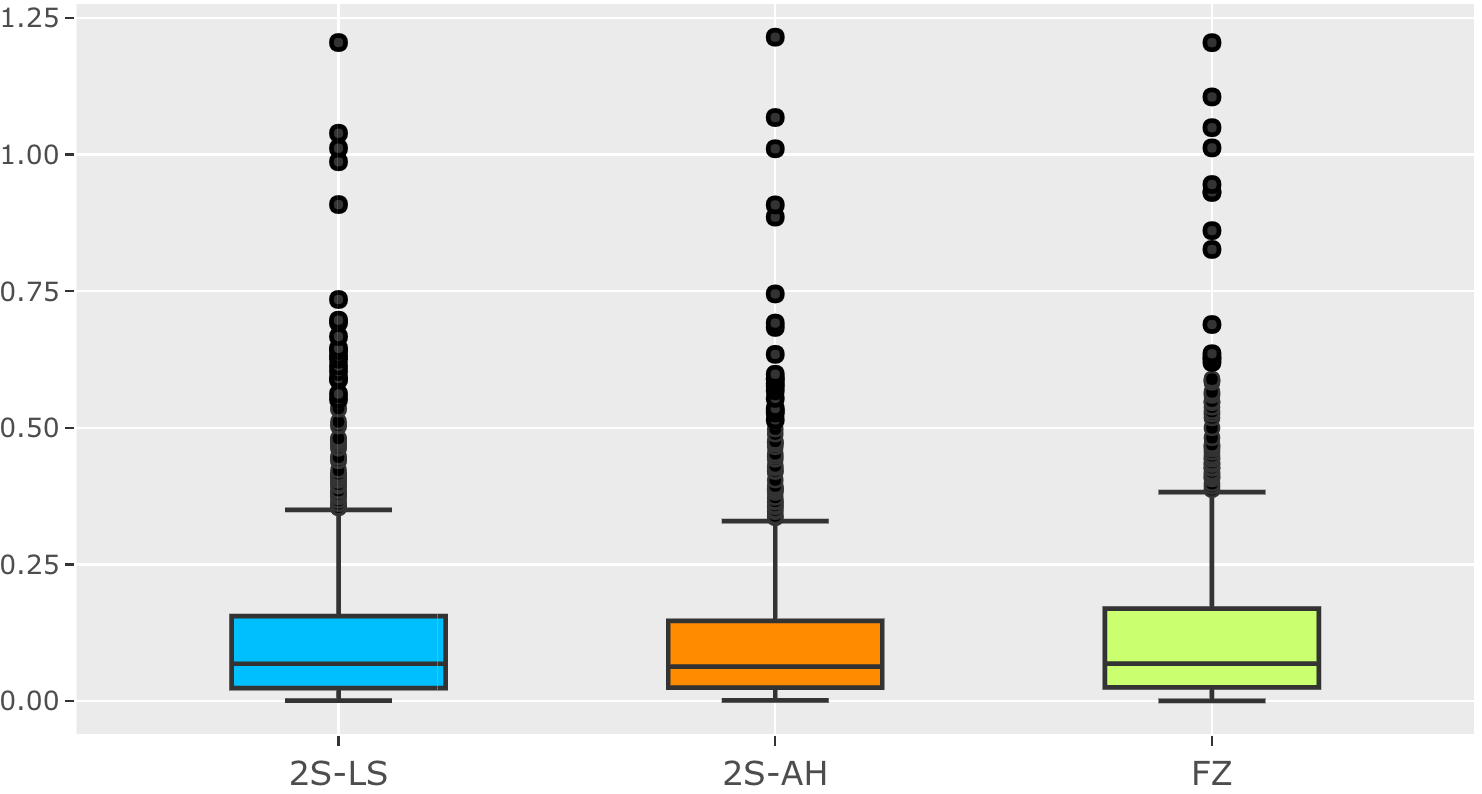}}~~~~
   \subfigure[$t_{3.5}$ model]{\includegraphics[height=0.2\textheight, width=0.48\textwidth]{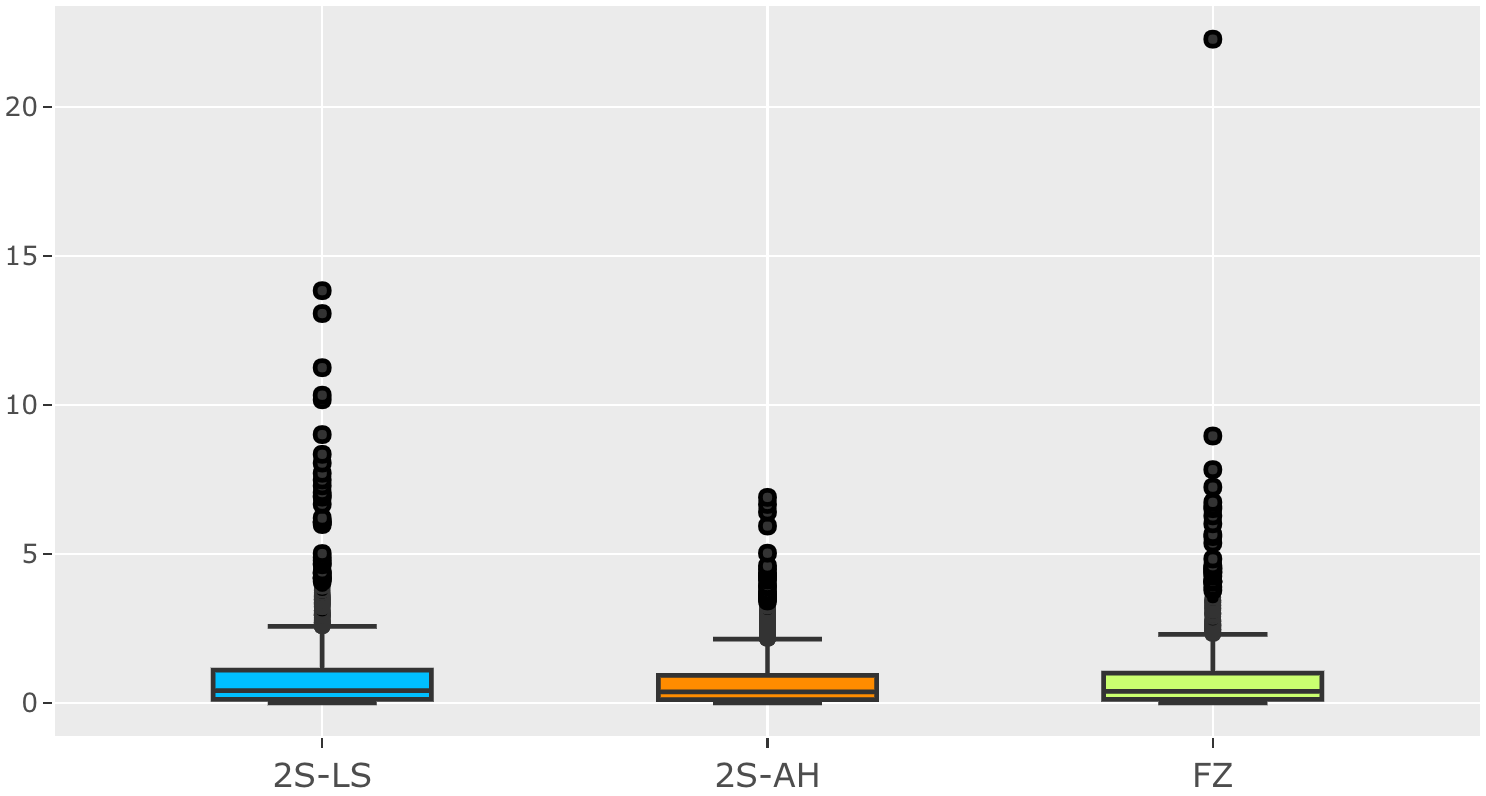}} 
\caption{Boxplots of squared $\ell_2$-errors,  based on 1000 replications, for three ES regression estimates at quantile level $\alpha=0.05$. The data are generated from model \eqref{qar.model} with $T=1000$ in the normal model and $T=1500$ in the $t_{3.5}$ model.}
  \label{fig:qar}
\end{figure}

Figure~\ref{fig:qar} shows the boxplots of squared $\ell_2$-errors (over 1000 replications) for three ES regression estimates,  two-step least squares (2S-LS),  two-step adaptive Huber (2S-AH)
and FZ’s joint estimate (FZ),  at quantile level 5\%.  The mean squared errors (MSEs) of these three methods are 0.1153, 0.1099 and 0.1227 in the normal model,  and 0.9996,  0.6229 and 1.0089 in the $t_{3.5}$ model. We also note that the total number of crossings (over $T$ observations and 1000 replications) of
the FZ method is 256 in the normal model but increases to 1338 in the $t_{3.5}$ model. Without non-crossing constraints, the total numbers of crossings are 22 and 729 for two-step adaptive Huber
versus 92 and 1641 for two-step least squares.  To summarize, the two-step robust method
remains to be statistically and numerically preferable for serially dependent data generated from a quantile autoregression process \citep{KX2006}.
 
Under the same settings, Table~\ref{tab:qar} reports the empirical coverage probability and average width (in parenthesis) of the 95\% normal-based confidence intervals described in Section~\ref{sec:2.2} and Section~\ref{sec:2.3}  for the two slope coefficients in \eqref{sim.coef}.  Because these methods rely crucially on the asymptotic normality of the estimator when observations are independent,   the covariance estimation part suffers from a bias due to serial correlations and leads to undercoverage in confidence intervals.    To fix this,  research into the asymptotic analysis (as $T\to \infty$ with $q$ fixed) of the estimators in \eqref{ar.est}  under the QAR framework is underway.  This could also  bring additional interests to practitioners.

\begin{table}[!htp]
\begin{center}
\caption{Empirical coverage probability and average width (in parenthesis),  based on 1000 replications, of the 95\% normal-based confidence intervals \eqref{eq:twostep:ci} and \eqref{eq:ci} for the two slope coefficients in \eqref{sim.coef} with $\alpha=5\%$.}
\begin{tabular}{c|     c  | c}
  \hline
Normal model &   \multicolumn{1}{c}{$\theta_1(\alpha)=0.5125$} & \multicolumn{1}{c}{$\gamma(\alpha)=0.9625$}   \\ \hline
\texttt{2S-LS}  & 91.5\% (0.1439)  & 93.2\% (1.0996) \\
\texttt{2S-AH}  & 90.8\% (0.1351)  & 92.3\% (1.0356) \\
\hline
\hline
$t_{3.5}$ model &   \multicolumn{1}{c}{$\theta_1(\alpha)=0.5125$} & \multicolumn{1}{c}{$\gamma(\alpha)=0.9625$ }   \\ \hline
\texttt{2S-LS}  & 92.9\% (0.3012)  & 94.8\% (3.0965) \\
\texttt{2S-AH}  & 91.5\% (0.2729)  & 93.4\% (2.5166) \\ 
\hline
\end{tabular}
\label{tab:qar}
\end{center}
\end{table}

\section{Proofs of Main Results}

\subsection{Supporting lemmas}
\label{sec:pre}

We first introduce some basic notations that will be frequently used in the proof. For $i=1,\ldots, n$, define the standardized covariates and quantile regression residuals as  
\#
	W_i = \Sigma^{-1/2} X_i ~~\mbox{ and }~~ \varepsilon_i = Y_i - X_i^\T \beta^* , \label{QR.residuals}
\#
respectively. For $\beta \in \RR^p$, define
\#
\begin{cases}
Z_i(\beta)  = (Y_i - X_i^\T \beta) \mathbbm{1}( Y_i \leq X_i^\T \beta) + \alpha X_i^\T \beta  , \ \ Z_i = Z_i(\beta^*),  \\
  \omega_i(\beta)  = Z_i(\beta) - \alpha X_i^\T \theta^*  , \ \ \omega_i = \omega_i(\beta^*), 
\end{cases}
 \label{def:Zi}
\#
and for some initial estimator $\hat \beta$ of $\beta^*$, write
\$
	   \hat Z_i = Z_i(\hat \beta) , \quad \hat \omega_i = \omega_i(\hat \beta) .
\$
Then, the second-stage robust ES estimator can be equivalently defined as 
\#
	\hat \theta_{ \tau }  \in \argmin_{\theta \in \RR^p}   \bigg\{   \hat \cL_\tau(\theta) :=  \frac{1}{n} \sn \ell_\tau\big( \hat Z_i - \alpha X_i^\T \theta \big)  \bigg\} ,  \label{ES.M-estimator}
\#
which is a Huber's $M$-estimator with generated response variables.

Let $\psi_\tau(u) = \ell_\tau'(u) = \min\{ \max( - \tau,  u),  \tau \}$ denote the score function, which is 1-Lipschitz continuous and differentiable except at $\pm \tau$. To control the estimation error $\| \hat \theta_\tau - \theta^* \|_\Sigma$, the keys are an upper bound for the $\ell_2$-norm of the score 
\$
	\nabla  \hat \cL_\tau(\theta^* ) =  - \frac{\alpha }{n} \sn \psi_\tau\big( \hat Z_i - \alpha X_i^\T \theta^* \big) X_i 
	= - \frac{\alpha }{n} \sn \psi_\tau\big(  Z_i (\hat \beta)  - \alpha X_i^\T \theta^* \big) X_i ,
\$
and the restricted strong convexity property of $\hat \cL_\tau(\cdot)$ in a neighborhood of $\theta^*$.
Conditioned on the event $\{ \hat \beta \in \BB_\Sigma(\beta^*, r_0)\}$, we have
\$
  \| \nabla  \hat \cL_\tau(\theta^* ) \|_{\Sigma^{-1}}  &  \leq  \alpha  \sup_{ \beta \in \BB_\Sigma(\beta^*, r_0) }  \bigg\|   \frac{1 }{n} \sn \psi_\tau\big(  \underbrace{  (Y_i - X_i^\T \beta) \mathbbm{1}( Y_i \leq X_i^\T \beta) + \alpha \langle X_i, \beta - \theta^* \rangle }_{= \, \omega_i(\beta) } \big) W_i \bigg\|_2 \\
  & \leq \alpha   \underbrace{  \sup_{ \beta \in \BB_\Sigma(\beta^*, r_0) }  \bigg\| \frac{1}{n} \sn (1 - \EE) \big\{ \psi_\tau\big( \omega_i(\beta) \big) - \psi_\tau(\omega_i)  \big\} W_i \bigg\|_2 }_{\lesssim_{\PP} \, r_0 \sqrt{\frac{p}{n}} {\rm ~(Lemma~\ref{lem:first-order.error})} } \\
  &~~~~~ + \alpha \underbrace{  \sup_{ \beta \in \BB_\Sigma(\beta^*, r_0) } \big\|  \EE   \psi_\tau\big( \omega_i(\beta) \big) W_i   \big\|_2  }_{\lesssim \,r_0^2 + r_0/\tau {\rm ~(Lemma~\ref{lem:approximate.neyman})} }   + \, \alpha \underbrace{  \bigg\| \frac{1}{n} \sn  (1-\EE)  \psi_\tau(\omega_i) W_i   \bigg\|_2 }_{\lesssim_{\PP} \, \sqrt{\frac{p}{n}} + \frac{\tau  p}{n}  {\rm ~(Lemma~\ref{lem:score.bound})}} .
\$

\begin{lemma}   \label{lem:first-order.error}
Assume Conditions~\ref{cond:density} and \ref{cond:covariate} hold. Then, for any $r_0>0$ and $t\geq 1/2$,  
\$
\sup_{ \beta \in \BB_\Sigma(\beta^*, r_0) }  \bigg\| \frac{1}{n} \sn (1 - \EE) \big\{ \psi_\tau\big( \omega_i(\beta) \big) - \psi_\tau(\omega_i)  \big\} W_i \bigg\|_2 \leq C_1   \upsilon_1^2   \sqrt{\frac{p+ t}{n}} \cdot r_0
\$
holds with probability at least $1-e^{-t}$ as long as $n \geq C_2(p+t )$, where $C_1, C_2>0$ are absolute constants.
\end{lemma}

\begin{lemma} \label{lem:approximate.neyman}
Assume Conditions~\ref{cond:density} and \ref{cond:covariate} hold. For any $\tau>0$ and $r_0 >0$,
\#
	 \sup_{\beta \in \BB_\Sigma(\beta^*, r_0) } \big\|  \EE   \{  \psi_\tau  ( \omega_i(\beta)   ) W_i    \}   \big\|_2    \leq   \frac{ \overbar \sigma^2 }{\tau}  +   \frac{1}{2} \kappa_3 ( \overbar f + 1/\tau)  r_0^2  +      \frac{\overbar \sigma }{ \tau } r_0  , \label{mean.grad.ubd}
\#
where $\omega_i(\beta)$ is given in \eqref{def:Zi}. Moreover,
\#
	 \sup_{\beta \in \BB_\Sigma(\beta^*, r_0) } \big\|  \EE   \{  \psi_\tau ( \omega_i(\beta) ) W_i   \}   \big\|_2    \leq   \frac{ \overbar \sigma^2 }{\tau}  +   \frac{\kappa_3}{2} \overbar f r_0^2 +  (\overbar \sigma^2 + \kappa_3  \overbar \sigma r_0 + \kappa_4 r_0^2/3)    \frac{r_0}{\tau^2}   . \label{mean.grad.ubd2}
\#
\end{lemma}

\begin{lemma}  \label{lem:score.bound}
Assume Conditions~\ref{cond:density} and \ref{cond:covariate}  hold. For any $t > 0$, it holds with probability at least $1-e^{-t}$ that
\$
\bigg\| \frac{1}{n} \sn  (1 -\EE)  \psi_\tau(\omega_i)  W_i   \bigg\|_2  \leq  C_0 \upsilon_1 \bigg(      \overbar \sigma \sqrt{\frac{p + t}{n}}  +  \tau \frac{p + t}{n}  \bigg) , 
\$
where $C_0 > 0$ is an absolute constant.
\end{lemma}

The following lemma provides a form of the restricted strong convexity for the empirical Huber loss $\hat \cL_\tau(\cdot)$ with estimated response variables.

\begin{lemma} \label{lem:RSC}
Assume Conditions~\ref{cond:density} and \ref{cond:covariate} hold.  For any pair of radii  $r \geq r_0 >0$, let the robustification parameter $\tau$ satisfy $\tau^2 \geq  32 \{  \kappa_4 (r^2 + 4 r_0^2) +  \overbar \sigma^2  \}  $. Then, conditioned on the event $\{ \hat \beta \in \BB_\Sigma(\beta^*, r_0) \}$, we have that with probability at least $1-e^{-t}$,
\#
	 \langle  \nabla \hat \cL_\tau(\theta) - \nabla \hat \cL_\tau(\theta^*) , \theta - \theta^* \rangle   \geq \frac{\alpha^2}{4} \| \theta - \theta^* \|_\Sigma^2
\#
holds uniformly over $\theta \in \BB_\Sigma(\theta^*, r/\alpha)$ as long as $n\gtrsim  (\tau / r)^2(p+t)$.
\end{lemma}

\begin{lemma}  \label{lem:first-order.error2}
Assume Conditions~\ref{cond:density} and \ref{cond:covariate} hold. Then, for any $r >0$ and $t\geq 1/2$,  
\$
\sup_{ \theta \in \BB_\Sigma(\theta^*, r/\alpha ) }  \bigg\| \frac{1}{n} \sn  \big\{ \psi_\tau ( Z_i - \alpha X_i^\T \theta   )  & - \psi_\tau(Z_i - \alpha X_i^\T \theta^*)  \big\} W_i  + \alpha \Sigma^{1/2} (\theta - \theta^*) \bigg\|_2  \\
& \leq C_3   \upsilon_1^2  \sqrt{\frac{p+ t}{n}}   \cdot r + (  \overbar \sigma^2 + \kappa_4 r^2 /3 )  \frac{r}{\tau^2}
\$
holds with probability at least $1-e^{-t}$ as long as $n \geq C_4(p+t )$, where $C_3, C_4>0$ are absolute constants.
\end{lemma}

For every $\gamma \in \RR^p$, define the quantile loss difference $\hat D(\gamma) = \hat \cQ(\beta^* + \gamma) - \hat \cQ(\beta^*)$  and its population counterpart $D(\gamma) = \EE \hat D(\gamma)$, where $\hat \cQ(\beta) = (1/n) \sn \rho_\alpha(Y_i - X_i^\T \beta)$.

\begin{lemma}   \label{lem:qr.difference}
Assume Condition~\ref{cond:covariate} holds. For any $r>0$ and $t\geq 0$, the bound
\#
	\sup_{\gamma \in \BB_\Sigma(r) } \{ D(\gamma) - \hat D(\gamma)  \} \leq  C \upsilon_1     \sqrt{\frac{p+t}{n}} \cdot r \label{diff.loss.bound}
\#
holds with probability at least $1- e^{-t}$, where $C>0$ is an absolute constant.
\end{lemma}

To prove Theorem~\ref{thm:huber.ES}, the following convexity lemma (see Lemma~C.1 in \cite{SZF2020}) will be needed.  We reproduce it here for the sake of readability.

\begin{lemma} \label{lem:convexity}
Let $f:\RR^p\to \RR$ be a differentiable convex function, and define the corresponding symmetrized Bregman divergence 
$D_f(\beta_1, \beta_2) = \langle \nabla f(\beta_2) - \nabla f(\beta_1), \beta_2 - \beta_1 \rangle$ for $\beta_1, \beta_2 \in \RR^p$. Then, for any $\beta, \delta \in \RR^p$ and $\lambda\in [0, 1]$,  $D_f( \beta_\lambda , \beta) \leq \lambda \cdot D_f(\beta_1 , \beta )$, where $\beta_\lambda =\beta + \lambda\delta $ and $\beta_1 = \beta + \delta$.
\end{lemma}

\subsection{Proof of Proposition~\ref{prop:lbd}}

The proof is adapted from that of Proposition~6.2 in \cite{C2012} with certain adjustments.
Given $\eta>0$ and $n\geq 1$, let $Y$ follow a discrete distribution with support $\{ - n   \eta , 0, n \eta \}$, satisfying
$$
	\PP(Y = n   \eta ) = \PP(Y = - n  \eta) = \frac{\sigma^2}{2 n^2 \eta^2 }.
$$
It is easy to see that $\EE(Y)=0$ and $\EE(Y^2) = \sigma^2$. Provided $n \eta  > \sigma/  \sqrt{2\alpha} $, we have $Q_\alpha(Y) = 0$. Let $Z_1, \ldots, Z_n$ be independent copies of $Z= Y \mathbbm{1}\{ Y \leq Q_\alpha(Y) \} = Y \mathbbm{1}( Y \leq 0 )$ with mean
$$
	\EE(Z) = -    n \eta  \cdot \frac{\sigma^2}{2 n^2 \eta^2} = -\frac{\sigma^2 }{2   n  \eta } .
$$ 
Then we have 
\$
	& \PP \bigg\{   \frac{1}{n}\sn (Z_i - \EE Z_i) \leq  - \eta + \frac{\sigma^2}{2  n \eta } \bigg\} \\
	& =\PP \bigg(    \frac{1}{n}\sn Z_i   \leq  - \eta   \bigg) \geq \PP \bigg\{    \sn Y_i \mathbbm{1}(Y_i \leq 0)    =  - n \eta \bigg\} \\
	& = n \cdot  \frac{\sigma^2}{2 n^2 \eta^2} \bigg( 1- \frac{ \sigma^2}{2 n^2 \eta^2} \bigg)^{n-1} = \frac{\sigma^2 }{2 n \eta^2 } \bigg( 1- \frac{ \sigma^2}{2 n^2 \eta^2} \bigg)^{n-1}.
\$
Taking $\eta =c_n \sigma/\sqrt{2n\delta}$ with $c_n  = (1-  e \delta /n)^{(n-1)/2 }$ and $0<\delta \leq  e^{-1}$, we obtain that 
\$
	  \frac{1}{n}\sn (Z_i - \EE Z_i) \leq  -   \frac{ c_n \sigma }{ \sqrt{2n\delta}} + \frac{   \sigma   }{  c_n   }  \sqrt{\frac{\delta}{2n}}
\$
holds with probability at least $\delta$. Note that $c_n\in (e^{-1/2} , 1]$ for all $n\geq 1$, the right-hand side is further bounded from above by
$$
	- \sigma \sqrt{\frac{1}{2e n\delta }} +\sigma  \sqrt{\frac{e \delta }{2n}} = -\frac{\sigma (1- e\delta )}{\sqrt{2 e n \delta }} .
$$
This proves the claimed bound.  \qed

\subsection{Proof of Proposition~\ref{prop:bias}}
 
To begin with,  define the ES response variable $Z= \varepsilon \mathbbm{1}( \varepsilon \leq 0) + \alpha X^\T \beta^*$ and residual $\omega = Z - \alpha X^\T \theta^*$, where $ \varepsilon = Y - X^\T \beta^*$ satisfies $\EE_X\{  \varepsilon \mathbbm{1}( \varepsilon \leq 0) \} = \alpha X^\T(\theta^* - \beta^*)$.
Moreover, define the bias vector $h_\tau =   \theta_\tau^* - \theta^*$, and population loss functions 
$\cL_\tau(\theta)  = \EE \ell_\tau( Z - \alpha X^\T \theta)$ and $\cL(\theta)  = \frac{1}{2} \EE ( Z - \alpha X^\T \theta)^2$.  Using the optimality of $\theta^*_\tau$ and the mean value theorem,   we have $\nabla \cL_\tau(\theta^*_\tau)= 0$ and 
\#
    \int_0^1  \langle h_\tau, \nabla^2 \cL_\tau( \theta^* + t h_\tau  ) h_\tau \rangle {\rm d} t  &   = \langle   \nabla \cL_\tau(\theta^*_\tau) -  \nabla \cL_\tau(\theta^*)    ,    \theta^*_\tau - \theta^*  \rangle  \nn \\
    &= - h_\tau^\T \nabla \cL_\tau(\theta^*) =   \alpha  \EE\{ \psi_\tau(\omega)  X^\T h_\tau \} .\label{population.divergence}
\# 
Since $\EE(\omega | X) =0$,  we have $ \EE_X (\omega) - \EE_X \{ \psi_\tau(\omega) \} =  \EE_X  [  \{ \omega - \tau  \sign(\omega) \} \mathbbm{1}( |\omega| > \tau)]$. This together with the fact $\EE_X(\omega^2) = \var_X(  \varepsilon \mathbbm{1}( \varepsilon \leq 0))$ implies
\#
|    \EE\{ \psi_\tau(\omega)  X^\T h_\tau \} | \leq \frac{1}{\tau} \EE ( \omega^2 | X^\T h_\tau| ) \leq \frac{\overbar \sigma^2}{\tau} \EE | X^\T h_\tau |  \leq \frac{\overbar \sigma^2}{\tau} \|  h_\tau \|_\Sigma.  \label{population.divergence.ubd}
\#

Throughout the proof, we assume $\tau  \geq  2 \kappa_4^{1/4}\overbar \sigma \geq 2 \overbar \sigma$. 
For the left-hand side of \eqref{population.divergence},  write $\omega(t)= Z - \alpha X^\T (\theta^* + t h_\tau) = \omega - t \alpha X^\T h_\tau$ and note that
\$
 \langle h_\tau, \nabla^2 \cL_\tau( \theta^* + t h_\tau  ) h_\tau \rangle    = \alpha^2  \EE\{ \mathbbm{1} ( |\omega(t) | \leq \tau ) (X^\T h_\tau)^2 \} = \alpha^2  \|  h_\tau \|_\Sigma^2 - \alpha^2 \EE \{ \mathbbm{1} ( | \omega(t) | > \tau ) (X^\T h_\tau)^2 \} .
\$
By Markov's inequality,
\$
& \EE \{ \mathbbm{1} ( | \omega(t) | > \tau ) (X^\T h_\tau)^2 \}   \leq \frac{1}{\tau^2} \EE (\omega - t \alpha X^\T h_\tau)^2(X^\T h_\tau)^2 \\
& \leq \frac{\overbar \sigma^2}{\tau^2} \| h_\tau \|_\Sigma^2 + \frac{(\alpha t)^2}{\tau^2}  \EE (X^\T h_\tau)^4 \leq \frac{\overbar \sigma^2}{\tau^2} \| h_\tau \|_\Sigma^2 + \frac{ \kappa_4}{\tau^2}  (\alpha t)^2  \| h_\tau \|_\Sigma^4.
\$
Putting these two observations together, we obtain that
\$
   \int_0^1  \langle h_\tau, \nabla^2 \cL_\tau( \theta^* + t h_\tau  ) h_\tau \rangle {\rm d} t   \geq  \alpha^2  \| h_\tau \|_\Sigma^2 \cdot  \bigg(  \frac{3}{4} - \frac{ \kappa_4}{3 \tau^2} \alpha^2 \| h_\tau \|_\Sigma^2 \bigg).
\$
Now set $r_\tau = \| h_\tau \|_\Sigma$.  Then, combining the above inequality with \eqref{population.divergence} and \eqref{population.divergence.ubd} yields 
\#
	\alpha  r_\tau  \bigg(  \frac{3}{4} - \frac{ \kappa_4}{3 \tau^2} \alpha^2 r_\tau^2 \bigg) \leq \frac{\overbar \sigma^2}{\tau}. \label{population.divergence.2}
\#

Assume at the moment that $\alpha r_\tau \leq r^* := \sqrt{3/(4  \kappa_4)} \cdot \tau$.  Hence, it follows immediately from \eqref{population.divergence.2} that $\alpha r_\tau \leq 2 \overbar \sigma^2 / \tau$, as claimed. It remains to show that $\alpha r_\tau > r^*$ cannot be the case.  Otherwise, if $\theta^*_\tau$ satisfies $r_\tau = \| \theta^*_\tau - \theta^*\|_\Sigma > r^*/\alpha$,    then let $\lambda = r^*/(\alpha r_\tau) \in (0, 1)$ and $\wt \theta_\tau = (1-\lambda) \theta^* + \lambda \theta^*_\tau$,  so that $\wt r_\tau:= \| \wt \theta_\tau - \theta^* \|_\Sigma = \lambda r_\tau = r^*/\alpha$. By Lemma~\ref{lem:convexity}, 
\$
 \langle   \nabla \cL_\tau(\wt \theta_\tau) -  \nabla \cL_\tau(\theta^*)    ,    \wt \theta_\tau  - \theta^*  \rangle 
 & \leq \lambda \cdot  \langle   \nabla \cL_\tau(\theta^*_\tau) -  \nabla \cL_\tau(\theta^*)    ,   \theta^*_\tau  - \theta^*  \rangle  \\
 &  =  \langle -\nabla \cL_\tau(\theta^*) , \wt \theta_\tau - \theta^* \rangle \leq \frac{\overbar \sigma^2}{\tau} \alpha  \wt r_\tau,
\$
where the last inequality follows from \eqref{population.divergence.ubd}.  Arguing as above, it can be similarly shown that
\$
 \langle   \nabla \cL_\tau(\wt \theta_\tau) -  \nabla \cL_\tau(\theta^*)    ,    \wt \theta_\tau  - \theta^*  \rangle  \geq ( \alpha  \wt r_\tau)^2  \cdot \bigg( \frac{3}{4} - \frac{\kappa_4}{3\tau^2} ( \alpha  \wt r_\tau)^2 \bigg) =  \frac{1}{2}( \alpha  \wt r_\tau)^2 .
\$
Together, the above upper and lower bounds imply $\alpha \wt r_\tau \leq 2\overbar \sigma^2 / \tau$. This contradicts the assumption $\tau  \geq  2 \kappa_4^{1/4}\overbar \sigma$,  thus completing the proof of the proposition.  \qed

\subsection{Proof of Proposition~\ref{prop:qr}}

Let $\hat \cQ(\beta) = (1/n) \sn \rho_\alpha(Y_i - X_i^\T \beta)$ and $\cQ(\beta) = \EE \hat \cQ(\beta)$ be the sample and population quantile loss functions. For every $\gamma \in \RR^p$, define 
\$
	\hat D(\gamma) = \hat \cQ(\beta^* + \gamma ) - \hat \cQ(\beta^*), \quad  D(\gamma) = \cQ(\beta^* + \gamma) - \cQ(\beta^*) , \quad   R(\gamma) = D(\gamma) - \langle \nabla \cQ(\beta^*) , \gamma \rangle,
\$
and note that $\nabla \cQ(\beta)  = \EE \{ F_{Y_i|X_i}(X_i^\T \beta) - \alpha \}X_i$. Since $\nabla \cQ(\beta^*)=0$, we have
\#
 \hat D(\gamma)  = \langle \nabla \cQ(\beta^*) , \gamma \rangle + R(\gamma)  -  \{  D(\gamma) -  \hat D(\gamma) \}  = R(\gamma) - \{   D(\gamma) - \hat D(\gamma) \}. \label{hatD.decomposition}
\#

To bound $R(\gamma)= \cQ(\beta^* + \gamma) - \cQ(\beta^*)$, note that the population Hessian $\nabla^2 \cQ(\beta)=  \EE\{ f_{\varepsilon_i |X_i} ( \langle X_i, \beta - \beta^* \rangle ) X_i X_i^\T \}$ satisfies for any $\gamma \in \RR^p$ and $t\in [0, 1]$ that
\$
	& \langle \gamma, \nabla^2 \cQ(\beta^* + t \gamma ) \gamma \rangle = \EE  \big\{  f_{\varepsilon_i |X_i} ( t X_i^\T\gamma )  (X_i^\T \gamma)^2 \big\} \\
& =  \EE  \big\{  f_{\varepsilon_i |X_i} (0)  (X_i^\T \gamma)^2 \big\} +  \EE  \big\{ f_{\varepsilon_i |X_i} ( t X_i^\T\gamma )  -  f_{\varepsilon_i |X_i}(0) \big\} (X_i^\T \gamma)^2   \\
& \geq \underbar{$f$} \, \| \gamma \|_\Sigma^2 - l_0 t \cdot  \EE | X_i^\T \gamma |^3 \geq \underbar{$f$}\, \| \gamma \|_\Sigma^2 - l_0 \kappa_3 t  \cdot \| \gamma \|_\Sigma^3, 
\$
where we used the Lipschitz continuity of $f_{\varepsilon | X}(\cdot)$ in the first inequality. This, together with the fundamental theorem of calculus, implies
\$
R(\gamma) & =  \cQ(\beta^* + \gamma) - \cQ(\beta^*)  = \int_0^1 \langle \nabla \cQ(\beta^* + s\gamma ) - \nabla \cQ(\beta^*) , \gamma \rangle  {\rm d} s \\
& =     \int_0^1  \int_0^1  s \langle \gamma,  \nabla^2 \cQ(\beta^* + t s \gamma) \gamma \rangle  {\rm d} s {\rm d} t  \geq \frac{1}{2}\underbar{$f$}\,\| \gamma \|_\Sigma^2 - \frac{1}{6} l_0 \kappa_3 \| \gamma \|_\Sigma^3   .
\$
For some $r_0>0$ to be determined, Lemma~\ref{lem:qr.difference} states that, with probability at least $1-e^{-t}$,
\$
	\sup_{\gamma \in \BB_\Sigma(r_0) } \{ D(\gamma) - \hat D(\gamma) \} \leq C \upsilon_1  r_0 \sqrt{\frac{p+t}{n}} .
\$
Together, the previous two inequalities and \eqref{hatD.decomposition} show that, for any $\gamma \in \partial \BB_\Sigma(r_0)$, i.e. $\| \gamma \|_\Sigma= r_0$,  
\#
	\hat D(\gamma) \geq  \frac{r_0}{2}  \Bigg(  \underbar{$f$}  r_0   -\frac{1}{3} l_0 \kappa_3 r_0^2  - 2 C \upsilon_1  \sqrt{\frac{p+t}{n}} \Bigg) . \label{hatD.lbd}
\#

In view of \eqref{hatD.lbd}, we choose $r_0 = 4 C   \upsilon_1 \underbar{$f$}^{-1}  \sqrt{(p+t)/n}$ and let the sample size $n$ satisfy $ \underbar{$f$}^2 >\frac{8}{3}C l_0 \kappa_3\upsilon_1 \sqrt{(p+t)/n}$. Then, with probability at least $1-e^{-t}$, $\hat D(\gamma) >0$ for all $\gamma \in \partial \BB_\Sigma(r_0)$. On the other hand, let $\hat \gamma = \hat \beta - \beta^*$. Hence, $\hat D(\hat \gamma) \leq 0$ by  the optimality of $\hat \beta$. Finally, using Lemma~9.21 in \cite{W2019} and the convexity of $\hat \cQ(\cdot)$, we have $\hat \gamma \in \BB_\Sigma(r_0)$, proving the claim. \qed

\subsection{Proof of Theorem~\ref{thm:ES}}

To begin with, note that the two-step ES regression estimator $\hat \theta$ defined in \eqref{ES.est} satisfies the first-order condition
\$
	0 = \frac{1}{n} \sn S_i( \hat \beta , \hat \theta) X_i =  \frac{\alpha}{n} \sn X_i X_i^\T(\hat \theta - \theta^*) + \frac{1}{n} \sn S_i( \hat \beta , \theta^* ) X_i ,
\$
which implies
\#
\hat \theta - \theta^* = \bigg(  \frac{\alpha}{n} \sn X_i X_i^\T \bigg)^{-1} \frac{-1}{n}\sn   S_i( \hat \beta  , \theta^* ) X_i =    \bigg(  \frac{\alpha}{n} \sn X_i X_i^\T \bigg)^{-1} \frac{1}{n}\sn   \omega_i( \hat \beta   ) X_i , \label{ES.closed.form}
\#
where $\omega_i(\cdot)$ is defined in \eqref{def:Zi}. Recall that $W_i = \Sigma^{-1/2} X_i$ and $\EE(W_i W_i^\T) = {\rm I}_p$. Then, it follows from \eqref{ES.closed.form} that
\#
	\| \hat \theta - \theta^* \|_\Sigma  = \Bigg\| \bigg(  \frac{\alpha}{n} \sn W_i W_i^\T \bigg)^{-1} \frac{1}{n}\sn   \omega_i( \hat \beta   ) W_i \Bigg\|_2  \leq \frac{\| (\alpha n )^{-1} \sn   \omega_i( \hat \beta   ) W_i \|_2 }{\lambda_{\min}((1/n) \sn W_i W_i^\T)}  . \label{ES.ubd1}
\#
To bound $\lambda_{\min}( (1/n) \sn W_i W_i^\T )$ from below,  using Theorem~1.1 in \cite{O2016} we  obtain that for a sufficiently large sample size $n\geq  C_0 \kappa_4 \{ p + 2\log(2/\xi)\}$,
\#
	\PP\Bigg\{ \lambda_{\min}\bigg( \frac{1}{n} \sn W_i W_i^\T \bigg) \geq \frac{1}{2} \Bigg\} \geq 1- \xi. \label{gram.matrix.lbd}
\#

To upper bound $\|  (1/n) \sn   \omega_i( \hat \beta   ) W_i \|_2$ conditioned on the event $\{ \hat \beta \in \BB_\Sigma(\beta^*, r_0)\}$, consider the decomposition
\$
	  \frac{1}{n}\sn   \omega_i( \hat \beta   ) W_i   = \frac{1}{n}\sn  (1-\EE)   \{ \omega_i( \hat \beta   ) - \omega_i  \} W_i + \EE \{ \omega_i(\beta) W_i \} \big|_{\beta=\hat \beta} + \frac{1}{n} \sn \omega_i W_i ,
\$
where $\omega_i = \omega_i(\beta^*)$ satisfying $\EE(\omega_i | X_i)=0$. It follows that
\#
	\bigg\| \frac{1}{n}\sn   \omega_i( \hat \beta   ) W_i \bigg\|_2 &  \leq  \sup_{\beta \in \BB_\Sigma(\beta^*, r_0)} \Bigg\| \frac{1}{n}\sn  (1-\EE)    \{ \omega_i(   \beta   ) - \omega_i   \} W_i \Bigg\|_2  \nn  \\
	& ~~~~~~ +  \sup_{\beta \in \BB_\Sigma(\beta^*, r_0)} \big\| \EE \{ \omega_i(\beta) W_i   \} \big\|_2  + \Bigg\| \frac{1}{n}\sn  \omega_i  W_i \Bigg\|_2 \nn \\
	&= : \Xi_1 +  \Xi_2 + \Xi_3  . \label{score.ubd.decomposition}
\#

From the proof of Lemma~\ref{lem:first-order.error} we see that the stated bound remains valid if $\tau = \infty$ for which $\psi_\infty(t) = t$.  It follows that with probability at least $1- \xi $,
 \$
 	\Xi_1 \leq C_1 \upsilon_1^2 r_0 \sqrt{\frac{p+ \log (1/\xi)}{n}}   .
 \$
For $\Xi_2  =  \sup_{\beta \in \BB_\Sigma(\beta^*, r_0)}  \| \EE  \{ \omega_i(\beta) W_i  \} \|_2$, following the proof of Lemma~\ref{lem:approximate.neyman} it can be similarly shown that $\Xi_2 \leq    \overbar f\kappa_3  r_0^2 /2$. Turning to $\Pi_3$, note that $\EE\Xi_3^2 = (1/n^2) \sn \EE ( \omega_i^2 \| W_i \|_2^2 ) \leq \overbar \sigma^2 p /n$. By Markov's inequality, we have that for any $\xi \in (0, 1)$,
\$
	\PP \Bigg(\Xi_3 \geq  \overbar \sigma  \sqrt{\frac{  p }{n \xi }}  \Bigg) \leq  \frac{\EE \Xi_3^2}{ \overbar \sigma^2 p / (n \xi )}   \leq  \xi.
\$ 
Substituting these bounds into \eqref{score.ubd.decomposition}, we find that with probability at least $1-2\xi$,
\$
\bigg\| \frac{1}{n}\sn   \omega_i( \hat \beta   ) W_i \bigg\|_2 \leq  \overbar \sigma \sqrt{\frac{  p}{n \xi}} +  \frac{1}{2}  \overbar f \kappa_3 r_0^2 + C_1   \upsilon_1^2   \sqrt{\frac{p + \log(1/\xi) }{n}} \cdot r_0  .
\$
Combining this with \eqref{ES.ubd1} and \eqref{gram.matrix.lbd} proves the claim \eqref{ES.est.bound}. \qed

\subsection{Proof of Theorem~\ref{thm:uniform.clt}}

\noindent
{\sc Step 1.   A high-level Gaussian approximation result for $\max_{1\leq j\leq p} |\hat \theta_j - \theta^*_j|$}.  From \eqref{ES.closed.form} we obtain
\$
	\Sigma^{1/2} \, \alpha  (\hat \theta - \theta^* )   - \frac{1}{n} \sn \omega_i W_i  
= \frac{1}{n} \sn  \{ \omega_i(\hat \beta) - \omega_i  \}  W_i -    ( \Sigma^{-1/2} \hat \Sigma \Sigma^{-1/2} - {\rm I}_p ) \Sigma^{1/2}  \alpha( \hat \theta - \theta^* ) ,
\$
and hence
\$
&  \bigg\| \alpha (\hat \theta - \theta^*) -   \frac{1}{n} \sn \omega_i \Sigma^{-1} X_i \bigg\|_\Sigma \\
& \leq   \bigg\|  \frac{1}{n} \sn  \{ \omega_i(\hat \beta) - \omega_i   \} W_i \bigg\|_2 +    \| \Sigma^{-1/2} \hat \Sigma   \Sigma^{-1/2} - {\rm I}_p  \|_2 \cdot  \alpha  \| \hat \theta - \theta^* \|_\Sigma .
\$
By condition~\eqref{var.lbd},  $\Omega = \EE( \omega^2 X X^\T ) \succeq  \underbar{$\sigma$}^2 \Sigma$.  Thus,  $\| u \|_{\Sigma  \Omega^{-1} \Sigma}  = \sqrt{ u^\T \Sigma  \Omega^{-1} \Sigma u } \leq \underbar{$\sigma$}^{-1}  \| u \|_\Sigma $ for any $u\in \RR^p$.  Moreover,  observe that for any invertible matrix $A \in \RR^{p\times p}$ and $u\in \RR^p$, 
\$
	 \| u \|_A = \sqrt{u^\T A u } = \max_{v \in \mathbb{S}^{p-1} } \frac{|u^\T v| }{\sqrt{v^\T A^{-1} v}} \geq \max_{1\leq j\leq p} \frac{| u_j | }{ { \sqrt{ (A^{-1})_{jj} }}}.
\$ 
Putting together the pieces, we conclude that
\#
& \max_{1\leq j\leq p} \Bigg|  \frac{  \alpha (   \hat \theta_j - \theta_j^* ) - n^{-1} \sn \psi_{ij} }{ \sqrt{(\Sigma^{-1}  \Omega \Sigma^{-1} )_{jj}}  } \Bigg|   \nn \\
& \leq \bigg\|  \frac{1}{  \underbar{$\sigma$} n } \sn  \{ \omega_i(\hat \beta) - \omega_i   \} W_i \bigg\|_2 +  \| \Sigma^{-1/2} \hat \Sigma   \Sigma^{-1/2} - {\rm I}_p  \|_2 \cdot  \frac{\alpha}{ \underbar{$\sigma$}}  \| \hat \theta - \theta^* \|_\Sigma , \label{distribution.approximation.1}
\#
where $\psi_{ij} := \omega_i ( \Sigma^{-1} X_i )_j$ satisfies $\EE(\psi_{ij} )=0$ and $\EE(\psi_{ij}^2) = (\Sigma^{-1} \Omega \Sigma^{-1})_{jj}$.
Given two sequences $\eta_{1n}, \eta_{2n} >0$, define the events
\$
	\cE_{1n} & =  \Bigg\{ \bigg\|  \frac{1}{ n } \sn \{ \omega_i(\hat \beta) - \omega_i   \} W_i \bigg\|_2 \leq \eta_{1n} \Bigg\} ,  \\
	 \cE_{2n} & = \big\{   \| \Sigma^{-1/2} \hat \Sigma   \Sigma^{-1/2} - {\rm I}_p  \|_2 \cdot   \alpha    \| \hat \theta - \theta^* \|_\Sigma \leq \eta_{2n} \big\}.
\$
Moreover,  let $G=(G_1, \ldots, G_p)^\T$ be a centered Gaussian vector with covariance matrix
$\Cov(G) = {\rm Corr}( \omega  \Sigma^{-1} X  )$.  Define the Gaussian approximation error (under the Kolmogorov distance) for the maximum statistic $\max_{1\leq j\leq p} |n^{-1/2}  \sn \psi_{ij} | / \sqrt{(\Sigma^{-1} \Omega \Sigma^{-1} )_{jj}} $ as 
\#
	\Delta_{n, p} : = \sup_{t\geq 0} \Bigg| \PP\bigg(  \max_{1\leq j\leq p} \bigg| \frac{n^{-1/2} \sn \psi_{ij}}{  \sqrt{(\Sigma^{-1} \Omega \Sigma^{-1} )_{jj} } } \bigg|  \leq t  \bigg) -  \PP\bigg(  \max_{1\leq j\leq p} |G_j| \leq t  \bigg)  \Bigg| . \label{distribution.approximation.2}
\#
For the Gaussian maximum $\max_{1\leq j\leq p} |G_j|  = \max_{1\leq j\leq p} \max(G_j, -G_j)$,  it follows from Nazarov's inequality \citep{N2003} that for any $\epsilon >0$, 
\#
	\sup_{t\geq 0,  \,\epsilon >0 }   \frac{1}{\epsilon } \PP\bigg( t-\epsilon \leq   \max_{1\leq j\leq p} |G_j| \leq t+\epsilon \bigg) \leq \sqrt{2 \log(2p)} + 2. \label{distribution.approximation.3}
\#
Together,   inequalities \eqref{distribution.approximation.1}--\eqref{distribution.approximation.3} imply 
\#
 & 	\sup_{t\geq 0}  \Bigg| \PP \bigg(   \max_{1\leq j\leq p} \bigg|  \frac{  \alpha \sqrt{n} (   \hat \theta_j - \theta_j^* )   }{  \sqrt{(\Sigma^{-1}  \Omega \Sigma^{-1} )_{jj}}  } \bigg|  \leq t \bigg)   -  \PP\bigg(  \max_{1\leq j\leq p} |G_j| \leq t  \bigg)  \Bigg|   \nn \\
	& \leq 	\Delta_{n, p} +  \big\{ \sqrt{2 \log(2p)} + 2 \big\} \frac{\sqrt{n}}{\underbar{$\sigma$} } (\eta_{1n} + \eta_{2n} )   +   \PP(\cE_{1n}^{\rm c} ) + \PP(\cE_{2 n}^{\rm c} ) .   \label{GAR.error1}
\#

\medskip
\noindent
{\sc Step 2.  Control the Gaussian approximation error $\Delta_{n, p}$.}
By Condition~\ref{cond:density} and \eqref{var.lbd}, 
$$
 	 \rho_0 = \lambda_{\min}\big(  {\rm Corr}( \omega  \Sigma^{-1} X ) \big)  \geq \frac{ \lambda_{\min}(\Sigma^{-1} \Omega \Sigma^{-1} ) }{\max_{1\leq j\leq p} (\Sigma^{-1} \Omega \Sigma^{-1} )_{jj} }  \geq  ( \underbar{$\sigma$} / \overbar \sigma )^2 \frac{ \lambda_{\min}(\Sigma^{-1} )}{\max_{1\leq j\leq p} (\Sigma^{-1}   )_{jj}}     .
$$
Applying Theorem~1 of \cite{KR2020} yields the following Berry-Esseen-type bound
\#
	\Delta_{n, p} \leq  C   \frac{ \rho_0^{-3/2}   (\log n)^{1/2} \log p + \rho_0^{-1} (\log n)^{3/2}    (\log p)^2}{\sqrt{n}}  \EE\bigg\{  \max_{1\leq j\leq p}  |\psi_{ij}  /\varrho_j |^3   \bigg\}  , \label{moment.ineq1}
\#
where $C>1$ is a universal constant and $\varrho^2_j := (\Sigma^{-1}  \Omega \Sigma^{-1} )_{jj}$. 
Noting that  $\omega_i =  \varepsilon_{i, -} - \EE_{X_i} (  \varepsilon_{i, -})$ with $ \varepsilon_{i, -} = \min\{ \varepsilon_i, 0 \} \leq  0$, we have $|\omega_i|^3 \leq \max\{  | \varepsilon_{i, -} |^3,  | \EE_{X_i} \varepsilon_{i, -} |^3 \} \leq  | \varepsilon_{i, -} |^3 + \EE_{X_i} |\varepsilon_{i, -}|^3$.  From condition~\eqref{var.lbd} it follows that
\#
 & \EE\big\{  \max\nolimits_{1\leq j\leq p}  |\psi_{ij}  /\varrho_j |^3   \big\} \nn  \\
& \leq   \EE \big\{  | \varepsilon_{i,-} |^3 \max\nolimits_{1\leq j\leq p}  |(\Sigma^{-1} X_i)_j / \varrho_j |^3 \big\} 
+    \EE   \big\{  \EE_{X_i} |\varepsilon_{i, -} |^3    \max\nolimits_{1\leq j\leq p}  |(\Sigma^{-1} X_i)_j / \varrho_j |^3 \big\}  \nn \\
& \leq 2  \alpha_3 \EE \big\{  \max\nolimits_{1\leq j\leq p}  |(\Sigma^{-1} X_i)_j / \varrho_j |^3 \big\}  . \label{moment.ineq2}
\#
For  $p$  arbitrary non-negative random variables $U_1, \ldots, U_p$,  it holds for any $q \geq 1$ that
\$
	 \EE \big( \max\nolimits_{1\leq j\leq p}  U_j \big) \leq  \big\{    \EE \big( \max\nolimits_{1\leq j\leq p}  U_j^q \big) \big\}^{1/q} \leq p^{1/q} \max\nolimits_{1\leq j\leq p}  ( \EE U_j^q )^{1/q} ,
\$
provided $\EE( U_j^q)$'s are finite.  Applying this moment inequality with $U_j =  |(\Sigma^{-1} X_i)_j / \varrho_j |^3$ and $q = 4/3$, we obtain
\$
 \EE \big\{  \max\nolimits_{1\leq j\leq p}  |(\Sigma^{-1} X_i)_j / \varrho_j |^3 \big\} \leq p^{3/4}  \max\nolimits_{1\leq j\leq p} \big\{ \EE |(\Sigma^{-1} X_i)_j / \varrho_j |^4  \big\}^{3/4} .
\$
For each $1\leq j\leq p$,  let ${\rm e}_j = (0, \ldots, 1,\ldots, 0)^\T \in \RR^p$ be the canonical basis vector whose $j$-th entry equals 1 and remaining entries are 0, such that $(\Sigma^{-1} X_i)_j = (\Sigma^{-1/2} {\rm e}_j)^\T W_i $.  Then, Condition~\ref{cond:covariate} ensures that $\EE (\Sigma^{-1} X_i)_j^4 =  \EE \{ (\Sigma^{-1/2} {\rm e}_j)^\T W_i \}^4 \leq \kappa_4  \| \Sigma^{-1/2} e_j \|_2^4$. On the other hand,  using \eqref{var.lbd} we have $\varrho_j^2 =  {\rm e}_j^\T \Sigma^{-1}\Omega \Sigma^{-1} {\rm e}_j \geq \underbar{$\sigma$}^2    {\rm e}_j^\T\Sigma^{-1} {\rm e}_j = \underbar{$\sigma$}^2 \| \Sigma^{-1/2} {\rm e}_j \|_2^2$. Substituting these bounds into the above inequality yields
\$
 \EE \big\{  \max\nolimits_{1\leq j\leq p}  |(\Sigma^{-1} X_i)_j / \varrho_j |^3 \big\} \leq p^{3/4}   ( \kappa_4 / \underbar{$\sigma$}^4 )^{3/4}  =  p^{3/4}  \kappa_4^{3/4} / \underbar{$\sigma$}^3  .   
\$
Combining with the earlier bounds \eqref{moment.ineq1} and \eqref{moment.ineq2}, we conclude that
 \#
\Delta_{n, p} \leq  2 C\kappa_4^{3/4 }    \frac{  \alpha_3}{\underbar{$\sigma$}^3}  \{  \rho_0^{-3/2}   (\log n)^{1/2} \log p  + \rho_0^{-1} (\log n)^{3/2}    (\log p)^2 \}  \frac{p^{3/4}}{\sqrt{n}}  . \label{moment.ineq3}
 \#

\medskip
\noindent
{\sc Step 3.  Control the events $\cE_{1n}$ and $\cE_{2n}$.} 
For any $\xi\in (0, 1)$, recall from the proof of Theorem~\ref{thm:ES} that conditioned on $ \cE_{1n} $,
\$
	 \alpha  \|  \hat \theta - \theta^* \|_\Sigma \leq    \frac{1}{ \lambda_{\min}( \Sigma^{-1/2} \hat \Sigma \Sigma^{-1/2} ) }  \Bigg(    \eta_{1n} +     \bigg\| \frac{1}{n} \sn \omega_i W_i \bigg\|_2 \Bigg) .
\$
Under the moment condition \eqref{var.lbd},  applying Theorem~3.1 in \cite{EL2008} to $\| \sn \omega_i W_i \|_2$, we obtain that for any $t>0$,
\$
	\PP\Bigg(  \bigg\|   \sn \omega_i W_i  \bigg\|_2 \geq  2 \EE \bigg\|   \sn \omega_i W_i  \bigg\|_2  + t \Bigg) \leq e^{-t^2/(3 n  \upsilon) } + C_0 n \frac{  \EE  \| \omega_i W_i \|_2^3}{t^3} ,
\$
where $\upsilon = \sup_{u\in \mathbb{S}^{p-1}} \EE     (\omega_i W_i^\T u)^2  \leq \overbar \sigma^2$ and $C_0>0$ is an absolute constant.  Similar to \eqref{moment.ineq2}, we bound the third moment as $
\EE  \| \omega_i W_i \|_2^3 \leq 2 \alpha_3 \EE \| W_i \|_2^3 \leq 2 \kappa_3  \alpha_3 \cdot  p^{3/2} $.
Re-organizing the constants,  it follows that with probability $1-2\xi$,
\$
 \bigg\|  \frac{1}{n} \sn \omega_i W_i  \bigg\|_2 \leq 2 \overbar \sigma \sqrt{\frac{p}{n}} + \max\bigg\{  \overbar \sigma \sqrt{\frac{3\log(1/\xi)}{n}} ,      (2 C_0  \kappa_3 \alpha_3)^{1/3}    \frac{p^{1/2}}{n^{2/3}\xi^{1/3}}  \bigg\}.
\$

To bound $\lambda_{\min}( \Sigma^{-1/2} \hat \Sigma \Sigma^{-1/2} )$ and $\|  \Sigma^{-1/2} \hat \Sigma \Sigma^{-1/2} - {\rm I}_p \|_2$,   it follows from Theorem~1 and Example~1 in \cite{Z2022} that with probability at least $1-  \xi$,  
$$
\| \Sigma^{-1/2} \hat \Sigma \Sigma^{-1/2} - {\rm I}_p \|_2 \lesssim  \upsilon_1^2 \sqrt{\frac{p+ \log(1/\xi)}{n}}
$$
and hence $\lambda_{\min}( \Sigma^{-1/2} \hat \Sigma \Sigma^{-1/2} ) \geq 1/2$ as long as $n\gtrsim p + \log(1/\xi)$.
To control the event $ \cE_{1n} $,  combining Proposition~\ref{prop:qr} with Lemmas~\ref{lem:first-order.error} and \ref{lem:approximate.neyman} we obtain that with probability at least $1- 2\xi$,  $\| \hat \beta -\beta \|_\Sigma \leq r_0 \asymp \underbar{$f$}^{-1} \sqrt{ (p + \log(1/\xi)) / n }$ and
 \$
  \bigg\|  \frac{1}{  n } \sn \{ \omega_i(\hat \beta) - \omega_i   \} W_i \bigg\|_2 &  \lesssim  \upsilon_1^2   \sqrt{\frac{p + \log(1/\xi) }{n}}  r_0  +  \kappa_3 \overbar f  \,r_0^2   
 \$
as long as $n\gtrsim p+ \log(1/\xi)$.

Under Condition~\ref{cond:covariate} we have $\kappa_3 \leq \sqrt{\kappa_4} \lesssim \upsilon_1^2$.
Ignoring constant factors that depend only on $\upsilon_1$,  we choose
\$
	\eta_{1n}  \asymp  \frac{ \overbar f}{\underbar{$f$}^2  }  \frac{p+ \log(1/\xi) }{ n}  ,  \quad 
  \eta_{2n}   \asymp    \sqrt{\frac{p+ \log(1/\xi) }{n}}    \bigg\{ 
\eta_{1n} +  \overbar \sigma   \sqrt{\frac{p + \log(1/\xi) }{n  }} +   \alpha_3^{1/3}  \frac{ p^{1/2} }{  n^{2/3} \xi^{1/3} }   \bigg\}  ,
\$
so that $ \PP(\cE_{1n}^{\rm c} ) + \PP(\cE_{2 n}^{\rm c} ) \leq 5 \xi$. In view of \eqref{GAR.error1},  we take $\xi = (p^{3/2} / n)^{1/2} <1$ to minimize $ \sqrt{n}  \eta_{2n}   +  \xi$ as a function of $\xi$,  implying
\$
  \sqrt{n \log p} \,  (\eta_{1n} + \eta_{2n} )   +   \PP(\cE_{1n}^{\rm c} ) + \PP(\cE_{2 n}^{\rm c} ) 
\lesssim    ( \overbar f  /\underbar{$f$}^2   \vee \alpha_3^{1/3} )  \sqrt{  \log p} \,    \frac{ p + \log n }{ \sqrt{n} }   .
\$
Combining this with \eqref{GAR.error1} and \eqref{moment.ineq3} proves the claim \eqref{uniform.clt}. \qed

\subsection{Proof of Theorem~\ref{thm:CLT}}

Recall from the proof of Theorem~\ref{thm:uniform.clt} that with probability at least $1- 6\xi$,
\$
&  \bigg\| \alpha \Sigma^{1/2} (\hat \theta - \theta^*) -   \frac{1}{n} \sn \omega_i W_i \bigg\|_2     \\
& \leq   \bigg\|  \frac{1}{n} \sn  \{ \omega_i(\hat \beta) - \omega_i   \} W_i \bigg\|_2 +    \| \Sigma^{-1/2} \hat \Sigma   \Sigma^{-1/2} - {\rm I}_p  \|_2 \cdot  \alpha  \| \hat \theta - \theta^* \|_\Sigma    \\ 
& \leq  \eta_{1n} (\xi)  +   \eta_{2n}   (\xi)  
\$
as long as $n\gtrsim p + \log(1/\xi)$,  where 
\#
\begin{cases}
 \eta_{1n} (\xi)   \asymp  \frac{ \overbar f}{\underbar{$f$}^2}     \frac{p+ \log(1/\xi) }{  n}   ,  \\
 \eta_{2n}(\xi)    \asymp   \sqrt{\frac{p+ \log(1/\xi) }{n}}    \big\{ 
\eta_{1n}(\xi) +    \overbar \sigma   \sqrt{\frac{p + \log(1/\xi) }{n  }} + \alpha_3^{1/3}  \frac{  p^{1/2} }{   n^{2/3} \xi^{1/3} }   \big\}  . 
\end{cases}  \label{eta12.order}
\#

For any deterministic  vector $a\in \RR^p$, define the partial sum $S_a = n^{-1/2} \sn \omega_i\langle a, \Sigma^{-1}X_i \rangle$, satisfying $\EE(S_a)=0$ and $\var(S_a) = \varrho_a^2 = a^\T \Sigma^{-1} \Omega \Sigma^{-1}a$. Hence, with probability at least $1-6 \xi$,
\#
	|  \alpha \sqrt{n} \langle a, \hat \theta - \theta^* \rangle - S_a | \leq \| a \|_{\Sigma^{-1}}  \sqrt{n}  \{ \eta_{1n} (\xi) + \eta_{2n}(\xi)  \}  . \label{es.clt.1}
\#
Next, applying the Berry-Esseen inequality (see, e.g. \cite{S2014}) to $S_a$ yields
$$
	\sup_{t\in \RR} | \PP( S_a \leq  t ) - \Phi(t/ \varrho_a) | \leq \frac{\EE |\omega_i \langle a, \Sigma^{-1}X_i \rangle|^3  }{2   \varrho_a^3 \sqrt{n} }.  
$$
From moment condition \eqref{var.lbd} it follows that $ \varrho_a^2 \geq \underbar{$\sigma$}^2 \| a \|^2_{\Sigma^{-1}}$, and by Condition~\ref{cond:covariate},
$$
\EE |\omega_i \langle a, \Sigma^{-1}X_i \rangle|^3 \leq 2 \alpha_3  \EE  |\langle a, \Sigma^{-1}X_i \rangle|^3 \leq 2 \kappa_3  \alpha_3 \cdot \| a \|_{\Sigma^{-1}}^3. 
$$
Putting these bounds together and applying the Berry-Esseen inequality, we obtain
\#
\sup_{t\in \RR} | \PP( S_a \leq  t ) - \Phi(t/ \varrho_a ) | \leq \frac{ \kappa_3  \alpha_3}{\underbar{$\sigma$}^3 \sqrt{n}} . \label{es.clt.2}
\#

Let $G\sim \cN(0,1)$. To complete the proof,  applying \eqref{es.clt.1} and \eqref{es.clt.2} we see that
\$
	& \PP(   \alpha \sqrt{n}  \langle a, \hat \theta  -\theta^* \rangle \leq t )  \\
	& \leq \PP \big\{   S_a \leq t + \| a \|_{\Sigma^{-1}}  \sqrt{n} ( \eta_{1n}  + \eta_{2n} ) \big\} + 6 \xi \\
	& \leq \PP \big\{  \varrho_a  G  \leq t + \| a \|_{\Sigma^{-1}} \sqrt{n} ( \eta_{1n}  + \eta_{2n} )   \big\} + 6\xi + \frac{ \kappa_3  \alpha_3}{\underbar{$\sigma$}^3 \sqrt{n}}  \\
	& \leq \Phi(t/  \varrho_a ) + 6 \xi +  \frac{ \kappa_3  \alpha_3}{\underbar{$\sigma$}^3 \sqrt{n}}  + \frac{1}{\sqrt{2\pi } \underbar{$\sigma$}} \sqrt{n} ( \eta_{1n}  + \eta_{2n} ) 
\$
for any $t\in \RR$.   A lower bound can be obtained by the same argument.
In view of the order of $\eta_{1n}(\xi), \eta_{2n}(\xi)$ stated in \eqref{eta12.order}, we choose $\xi = (p^{3/2} / n)^{1/2}$ so that
$$
 \xi + \frac{\sqrt{n}}{\underbar{$\sigma$}} ( \eta_{1n}  + \eta_{2n} )  \lesssim   \frac{\overbar f}{\underbar{$f$}^2}  \frac{p + \log n}{\underbar{$\sigma$} \sqrt{n}} +  \overbar \sigma  \frac{p + \log n}{\underbar{$\sigma$} \sqrt{n}} +  \alpha_3^{1/3}  \frac{p^{1/4} (p+\log n)^{1/2} }{  \underbar{$\sigma$} \sqrt{n} }
$$
This proves the claimed bound \eqref{univariate.clt}. \qed

\subsection{Proof of Theorem~\ref{thm:huber.ES}}

For simplicity, we write $\hat \theta = \hat \theta_\tau$ throughout the proof.
For some $r\geq r_0>0$ to be determined (in the end of the proof), we construct an intermediate ``estimator" $\wt \theta = (1-\lambda_r ) \theta^* + \lambda_r \hat \theta$, where $\lambda_r = \sup\{ \lambda \in [0, 1] : \theta^* +   \lambda(\hat \theta - \theta^*) \in \BB_\Sigma(r/\alpha)\}$. If $\hat \theta \in \BB_\Sigma(\theta^*, r/\alpha)$, $\lambda_r = 1$ and hence $\wt \theta = \hat \theta$; otherwise if $\hat \theta \notin \BB_\Sigma(\theta^*, r/\alpha)$, $\lambda_r$ is strictly less than 1 and $\wt \theta$ will lie at the boundary of $\BB_\Sigma(\theta^*, r/\alpha)$, i.e. $\alpha \| \wt \theta - \theta^* \|_\Sigma = r$. 

Since the loss function $\theta \mapsto \hat \cL_\tau(\theta)$ is convex and continuously differentiable,  by the convexity lemma---Lemma~\ref{lem:convexity}, and the first-order optimality condition that $\nabla \hat \cL_\tau( \hat \theta) = 0$, we have
\#
	\langle \nabla \hat \cL_\tau(\wt \theta) - \nabla \hat \cL_\tau( \theta^*) , \wt \theta - \theta^* \rangle  & \leq \lambda_r \cdot \langle \nabla \hat \cL_\tau(\hat \theta) - \nabla \hat \cL_\tau( \theta^*) , \hat \theta - \theta^*  \rangle  \nn \\
& = \lambda_r \cdot \langle - \nabla \hat \cL_\tau( \theta^*) , \hat \theta - \theta^*  \rangle \leq  
 \| \nabla \hat \cL_\tau( \theta^*) \|_{\Sigma^{-1} } \| \wt \theta - \theta^* \|_\Sigma .  \label{convexity.property}
\#
 
Let $\psi_\tau(u) = \ell_\tau'(u)$ be the score function.
Conditioned on the event $\{ \hat \beta \in \BB_\Sigma(\beta^*, r_0)\}$, we see that
\$
  \| \nabla  \hat \cL_\tau(\theta^* ) \|_{\Sigma^{-1}}  &  \leq  \alpha  \sup_{ \beta \in \BB_\Sigma(\beta^*, r_0) }  \bigg\|   \frac{1 }{n} \sn \psi_\tau\big(  \underbrace{  (Y_i - X_i^\T \beta) \mathbbm{1}( Y_i \leq X_i^\T \beta) + \alpha \langle X_i, \beta - \theta^* \rangle }_{= \, \omega_i(\beta)  {\rm ~by~\eqref{def:Zi}}} \big) W_i \bigg\|_2 \\
  & \leq \alpha  \sup_{ \beta \in \BB_\Sigma(\beta^*, r_0) }  \bigg\| \frac{1}{n} \sn (1 - \EE) \big\{ \psi_\tau\big( \omega_i(\beta) \big) - \psi_\tau(\omega_i)  \big\} W_i \bigg\|_2  \\
  &~~~~~ + \alpha   \sup_{ \beta \in \BB_\Sigma(\beta^*, r_0) } \big\|  \EE   \psi_\tau\big( \omega_i(\beta) \big) W_i   \big\|_2    +   \alpha   \bigg\| \frac{1}{n} \sn  (1-\EE)  \psi_\tau(\omega_i) W_i   \bigg\|_2 ,
\$
where $\omega_i = \omega_i(\beta^*) = \varepsilon_i \mathbbm{1}(\varepsilon_i\leq 0) +\alpha X_i^\T(\beta^* - \theta^*)$. 
Applying Lemmas~\ref{lem:first-order.error}, \ref{lem:approximate.neyman} and \ref{lem:score.bound}, we obtain that with probability at least $1 - 2 e^{-t}$,
\$
   \alpha^{-1} \| \nabla  \hat \cL_\tau(\theta^* ) \|_{\Sigma^{-1}}  
 & \leq \underbrace{ C_0 \upsilon_1 \bigg(  \overbar \sigma \sqrt{ \frac{p+t }{n}} + \tau \frac{p+t}{n} \bigg) }_{{\rm variance~upper~bound}} +  \underbrace{   \frac{\overbar \sigma^2}{\tau }  }_{{\rm robustification~bias~upper~bound}}  \\
 &~~~~~+  \underbrace{  C_1 \upsilon_1^2   \sqrt{\frac{p+t}{n}} \, r_0  +      \frac{\overbar \sigma}{\tau}  r_0 +  \frac{1}{2} \kappa_3 ( \overbar f  + 1/\tau ) r_0^2  }_{{\rm accumulated~estimation~error~bound}}   ,
\$
where  $C_0, C_1 >0 $ are absolute constants.
For the left-hand side of \eqref{convexity.property}, it follows from Lemma~\ref{lem:RSC} and the fact $\wt \theta \in \BB_{\Sigma}(\theta^*, r/\alpha )$ that, with probability at least $1-e^{-t}$ conditioned on $\{ \hat \beta \in \BB_\Sigma(\beta^*, r_0)\}$,
\$
\langle \nabla \hat \cL_\tau(\wt \theta) - \nabla \hat \cL_\tau( \theta^*) , \wt \theta - \theta^* \rangle \geq \frac{\alpha^2}{4} \| \wt \theta - \theta^* \|_\Sigma^2
\$
as long as $\tau^2 \geq 32 \{  \kappa_4 (r^2 + 4r_0^2) +  \overbar \sigma^2  \}$ and $n\gtrsim (\tau /r)^2 (p+t)$. With $\tau \asymp \overbar \sigma \sqrt{ n/(p+t)}$, putting these upper and lower bounds together and applying \eqref{convexity.property}, we obtain 
\$
 \alpha \| \wt \theta - \theta^* \|_\Sigma \leq   C_3 \upsilon_1 \overbar \sigma \sqrt{\frac{p+t }{n}} + C_4  \upsilon_1^2   \sqrt{\frac{p+t}{n}}    r_0  +2  \kappa_3( \overbar f + 1/\tau)  r_0^2    . 
\$ 

To complete the proof, we choose $r \asymp   \tau /\kappa_4^{1/2}$ so that $r\geq r_0$ under the sample size requirement $ n\gtrsim   p+t$.  
Note further that $\kappa_3 \leq \kappa_4^{1/2} \lesssim \upsilon_1^2$.
Under the stated upper bounds on $r_0$ and with $\tau \asymp \overbar \sigma \sqrt{ n/(p+t)}$,  it holds with probability at least $1- 3 e^{-t}$ that
\#
	\alpha \| \wt \theta - \theta^* \|_\Sigma \leq  C_3 \upsilon_1   \overbar \sigma  \sqrt{\frac{ p+t  }{n}} + C_5 \upsilon_1^2 \bigg(   \sqrt{\frac{p+t}{n}}   r_0  +   \overbar f   r_0^2  \bigg)   .   \label{intermediate.bound}
\#
Provided $n\gtrsim p+t$, we are guaranteed that $\alpha \| \wt \theta - \theta^* \|_\Sigma < r$ (with high probability), that is, $\wt \theta$ falls in the interior of $\BB_\Sigma(\theta^*, r/\alpha)$. By its construction in the first paragraph of the proof, we must have $\hat \theta = \wt \theta$ so that the above bound \eqref{intermediate.bound} also applies to $\hat \theta$, which completes the proof. \qed

\subsection{Proof of Theorem~\ref{thm:ES.bahadur}}

Define the vector-valued random process
\$
	\Pi_n(\beta, \theta) = \frac{1}{n} \sn  \big\{  \psi_\tau\big( \omega_i(\beta, \theta) \big) -  \psi_\tau\big( \omega_i(\beta^*, \theta^*) \big) \big\} W_i  +  \alpha \Sigma^{1/2}(\theta - \theta^*) ,
\$
for $\beta , \theta \in \RR^p$, where $\omega_i(\beta, \theta) = (\varepsilon_i - \Delta_i) \mathbbm{1}(\varepsilon_i \leq \Delta_i) + \alpha X_i^\T (\beta - \theta)$ and $\Delta_i = \Delta_i(\beta) = X_i^\T (\beta - \beta^*)$. Note that the two-step ES estimator $\hat \theta_\tau$ satisfies  the first-order condition $\nabla \hat \cL_\tau(\hat \theta_\tau ) = (-1/n) \sn \psi_\tau ( \omega_i(\hat \beta, \hat \theta_\tau )  ) X_i = 0$. The key is then to bound the supremum $\sup_{ (\beta, \theta) \in \BB_\Sigma (\beta^*, r_0) \times \BB_\Sigma(\theta^*, r/\alpha ) } \| \Pi_n(\beta, \theta) \|_2$.

For any $\epsilon \in (0, r)$, there exists an $\epsilon$-net $\cN(\epsilon, r) = \{ \theta_1, \ldots, \theta_N\}$ of $ \BB_\Sigma(\theta^*, r/\alpha)$ with $N\leq (1+2r/\epsilon)^p$. For any $(\beta, \theta) \in \BB_\Sigma (\beta^*, r_0) \times \BB_\Sigma(\theta^*, r/\alpha)$ given, there exists some $1\leq j\leq N$ such that $\alpha \| \theta - \theta_j \|_\Sigma \leq \epsilon$. Recall that $\psi_\tau(\cdot)$ is 1-Lipschitz continuous, we have
\$
	  \| \Pi_n(\beta, \theta) - \Pi_n(\beta, \theta_j)   \|_2 &  = \bigg\|  \frac{1}{n} \sn  \big\{  \psi_\tau\big( \omega_i(\beta, \theta) \big) -   \psi_\tau\big( \omega_i(\beta, \theta_j) \big)\big\} W_i +  \alpha \Sigma^{1/2}(\theta - \theta_j ) \bigg\|_2 \\
& \leq  \bigg\|  \frac{1}{n} \sn  \big\{  \psi_\tau\big( \omega_i(\beta, \theta) \big) -   \psi_\tau\big( \omega_i(\beta, \theta_j) \big)\big\} W_i   \bigg\|_2 +   \epsilon \\
& \leq \sup_{u\in \mathbb{S}^{p-1}} \frac{1}{n} \sn | W_i^\T u \cdot \alpha X_i^\T (\theta - \theta_j) | +  \epsilon  \\
& \leq  \sup_{u\in \mathbb{S}^{p-1}} \bigg( \frac{1}{n} \sn \langle W_i, u\rangle^2 \bigg)^{1/2} \cdot \bigg( \frac{\alpha^2}{n} \sn \langle X_i, \theta-\theta_j\rangle^2 \bigg)^{1/2} +  \epsilon  \\
& \leq \bigg\| \frac{1}{n} \sn W_i W_i^\T \bigg\|_2  \cdot    \epsilon +  \epsilon ,
\$
which further implies 
\#
	&  \sup_{(\beta, \theta ) \in  \BB_\Sigma (\beta^*, r_0) \times  \BB_\Sigma (\theta^*, r/\alpha ) } \| \Pi_n(\beta, \theta) \|_2 \nn \\
	&  \leq \max_{1\leq j\leq N} \sup_{ \beta \in  \BB_\Sigma (\beta^*, r_0)   }  \| \Pi_n(\beta, \theta_j ) \|_2 + \bigg\| \frac{1}{n} \sn W_i W_i^\T \bigg\|_2  \cdot     \epsilon +  \epsilon . \label{Pin.discretization}
\#

We first bound $\max_{1\leq j\leq N} \sup_{ \beta \in  \BB_\Sigma (\beta^*, r_0)   }  \| \Pi_n(\beta, \theta_j ) \|_2$. By the triangle inequality,
\$
  \| \Pi_n(\beta, \theta_j ) \|_2  & \leq   \underbrace{ \bigg\| \frac{1}{n} \sn  \big\{  \psi_\tau\big( \omega_i(\beta, \theta_j) \big) -  \psi_\tau\big( \omega_i(\beta^*, \theta_j ) \big) \big\} W_i  \bigg\|_2 }_{=:\,  \Lambda_1(\beta, \theta_j )  } \\
&~~~~~~ + \underbrace{  \bigg\| \frac{1}{n} \sn  \big\{  \psi_\tau\big( \omega_i(\beta^*, \theta_j) \big) -  \psi_\tau\big( \omega_i(\beta^*, \theta^*)  \big) \big\} W_i + \alpha \Sigma^{1/2}(\theta_j - \theta^*) \bigg\|_2 }_{=:\, \Lambda_2 (\theta_j)  } ,
\$
and hence
\#
\max_{1\leq j\leq N} \sup_{ \beta \in  \BB_\Sigma (\beta^*, r_0)   }  \| \Pi_n(\beta, \theta_j ) \|_2 \leq \max_{1\leq j\leq N} \sup_{\beta \in \BB_\Sigma(\beta^* , r_0) } \Lambda_1(\beta, \theta_j ) + \max_{1\leq j\leq N} \Lambda_2(\theta_j) .    \nn
\#
Following the proof of Lemma~\ref{lem:approximate.neyman}, and noting that $\EE_{X_i} \{ \omega_i(\beta^*, \theta_j) \} = \alpha X_i^\T (\theta^* - \theta_j)$,  it can be similarly shown that
\$
 	\sup_{\beta \in \BB_\Sigma(\beta^* , r_0)}   \big\|  \EE  \big\{  \psi_\tau\big( \omega_i(\beta, \theta_j) \big) -  \psi_\tau\big( \omega_i(\beta^*, \theta_j ) \big) \big\} W_i  \big\|_2  \leq  \frac{1}{2} \kappa_3 ( \overbar f + 1/\tau)  r_0^2  +  (\overbar \sigma + \kappa_3  \epsilon )     \frac{  r_0  }{ \tau } .
\$
Then, applying Lemmas~\ref{lem:first-order.error} and \ref{lem:approximate.neyman} to $\sup_{\beta \in \BB_\Sigma(\beta^*, r_0) } \Lambda_1(\beta, \theta_j )$ for each $j$, and taking the union bound over $j=1,\ldots, N$, we obtain that with probability at least $1-e^{-t}$,
\#
\max_{1\leq j\leq N} \sup_{\beta \in \BB_\Sigma(\beta^*, r_0) } \Lambda_1(\beta, \theta_j ) \leq C_1  \upsilon_1^2  \sqrt{\frac{p \log(3 r/\epsilon) + t}{n}}  r_0     +   \frac{1}{2} \kappa_3 ( \overbar f + 1/\tau)  r_0^2  +  (\overbar \sigma +  \kappa_3 \epsilon )     \frac{  r_0  }{ \tau }.  \nn
\#
Moreover, it follows from Lemma~\ref{lem:first-order.error2} that with probability at least $1-e^{-t}$,
\$
	 \max_{1\leq j\leq N} \Lambda_2(\theta_j)  \leq C_2  \upsilon_1^2 \sqrt{\frac{p + t}{n}} \cdot   r   +  (   \overbar \sigma^2 + \kappa_4 r^2 /3 ) \frac{r}{\tau^2} 
\$
Together, the previous three bounds imply that with probability at least $1-2 e^{-t}$,
\#
&  \max_{1\leq j\leq N} \sup_{ \beta \in  \BB_\Sigma (\beta^*, r_0)   }  \| \Pi_n(\beta, \theta_j ) \|_2 \nn \\
& \leq C_3  \upsilon_1^2 \bigg\{ r_0 \sqrt{\frac{p \log(3 r/\epsilon) + t}{n}}  + r \sqrt{\frac{p + t}{n}}  \,\bigg\}  \nn \\
& ~~~~~  +   \frac{1}{2} \kappa_3 ( \overbar f + 1/\tau)  r_0^2  +  (\overbar \sigma +  \kappa_3   \epsilon )     \frac{  r_0  }{ \tau } +( \overbar \sigma^2  + \kappa_4 r^2 /3 )  \frac{  r}{\tau^2} . \label{Pin.ubd}
\#

Turning next to $\| (1/n) \sn W_i W_i^\T  \|_2$,  from Example~1 in \cite{Z2022} we see that $\| (1/n) \sn W_i W_i^\T - {\rm I}_p \|_2 \leq  20\upsilon_1^2\sqrt{(4p+t)/n}$ with probability at least $1-e^{-t}$ as long as $n\geq 4p+t$.  Provided that $n\gtrsim p + t$, we have $\| (1/n) \sn W_i W_i^\T  \|_2 \leq  2$. Combining this with \eqref{Pin.discretization} and \eqref{Pin.ubd}, and taking $\epsilon=r/n$, we conclude that with probability at least $1-3e^{-t}$,
\#
&  \sup_{ (\beta, \theta) \in \BB_\Sigma (\beta^*, r_0) \times \BB_\Sigma(\theta^*, r/\alpha)  } \| \Pi_n(\beta, \theta) \|_2 \nn \\ & \lesssim 
  \bigg( \sqrt{\frac{p \log n + t}{n}} + \overbar f  r_0 + \frac{\overbar \sigma}{\tau} \bigg) \cdot r_0   +   \bigg( \sqrt{\frac{p+t}{n}} +  \frac{ \overbar \sigma^2 +  r^2}{\tau^2} + \frac{r_0}{n \tau} \bigg) \cdot r   \label{Pin.sup.ubd}
\#
as long as $n\gtrsim p + t$.

In view of Theorem~\ref{thm:huber.ES}, the two-step ES estimator $\hat \theta_\tau$ with $\tau \asymp \overbar \sigma \sqrt{ n /(p+t)}$ satisfies the bound  $\alpha \| \hat \theta_\tau - \theta^* \|_\Sigma \lesssim \overbar \sigma \sqrt{(p + t)/n}$ with probability at least $1-3e^{-t}$ conditioned on $\{ \hat \beta \in \BB_\Sigma(\beta^*, r_0) \}$.
This together with \eqref{Pin.sup.ubd} proves \eqref{ES.bahadur}. \qed

\subsection{Proof of Theorem~\ref{thm:ES.CLT}}

To begin with,  applying Proposition~\ref{prop:qr} and Theorem~\ref{thm:ES.bahadur} with $t=\log n$ we obtain that with probability at least $1-7 n^{-1}$,  $\| \hat \beta - \beta^* \|_\Sigma \leq r_0 \asymp  \underbar{$f$}^{-1}  \sqrt{(p+\log n)/n}$ and 
\$
 \bigg\| \alpha \Sigma^{1/2} ( \hat \theta_\tau - \theta^*  ) - \frac{1}{n} \sn \psi_\tau(\omega_i) W_i \bigg\|_2 \leq r_1  \asymp  \overbar \sigma \frac{p+ \log n}{ n}   + \frac{\overbar f}{\underbar{$f$}^2} \frac{(p\log n)^{1/2} (p + \log n)^{1/2}}{ n} .
\$
For any deterministic vector $a \in \RR^p$,  define $S_a= n^{-1/2} \sn \psi_\tau(\omega_i) \langle a,  \Sigma^{-1} X_i \rangle$ and $S_a^0 = S_a - \EE (S_a)$. 
Noting that $|\psi_\tau(t) - t | \leq |t|^{1+q} /\tau^q$ for any $q >0$,  and similar to \eqref{moment.ineq2}, we have
\$
 | \EE \psi_\tau(\omega_i) \langle a, \Sigma^{-1} X_i \rangle |   & \leq \frac{1}{\tau^2} \EE\big(  |\omega_i|^3 |\langle \Sigma^{-1/2} a,  W_i \rangle | \big)  \\
& \leq \frac{2\alpha_3 }{\tau^2} \EE  |\langle \Sigma^{-1/2} a,  W_i \rangle |  \leq \frac{  2 \alpha_3 }{\tau^2}  \| a \|_{\Sigma^{-1}}  .
\$
Hence, with probability at least $1-7 n^{-1}$,
\#
	|     \alpha \sqrt{n}  \langle a, \hat \theta_\tau - \theta^* \rangle - S_a^0 |  \leq \| a \|_{\Sigma^{-1}}    \sqrt{n}  \big(r_1 + 2 \alpha_3  \tau^{-2} \big) . \label{clt.approx.1}
\#

Next, define $\xi_i =  \psi_\tau(\omega_i) \langle a,  \Sigma^{-1} X_i \rangle$, so that $S_a^0 = n^{-1/2} \sn (\xi_i - \EE \xi_i) $.   The Berry-Esseen inequality (see, e.g. \cite{S2014}) states that
\$
		\sup_{t\in \RR} \big| \PP\big\{ S_a^0 \leq \var(S_a^0)^{1/2} t \big\} - \Phi(t) \big| \leq     \frac{\EE |\xi_i - \EE \xi_i |^3}{2  \var(\xi_i)^{3/2} \sqrt{n} }.
\$
Note that the mean satisfies $| \EE \xi_i | \leq \tau^{-1} \EE( \omega_i^2 |\langle a,  \Sigma^{-1} X_i \rangle| ) \leq  \| a \|_{\Sigma^{-1}}  \overbar \sigma^2/ \tau$. For the second moments,  recall that  
$$
	\varrho_{a, \tau}^2 = \EE \{ \psi_\tau( \omega_i )  \langle a,  \Sigma^{-1} X_i \rangle \}^2 ~\mbox{ and }~  \varrho_a^2 = \EE ( \omega_i  \langle a,  \Sigma^{-1} X_i \rangle )^2    = a^\T  \Sigma^{-1}  \Omega \Sigma^{-1}   a 
$$
with $\Omega = \EE ( \omega^2 X X^\T )$, satisfying
\#
  0 & \leq   \varrho_a^2 - 	\varrho_{a, \tau}^2    \leq       \EE \omega_i^2 \mathbbm{1}(|\omega_i| >\tau)  \langle a,  \Sigma^{-1} X_i \rangle^2 \nn\\
 &  \leq \frac{2 \alpha_3 }{\tau}   \EE   \langle a,  \Sigma^{-1} X_i \rangle^2 =  \frac{2 \alpha_3}{\tau}  \| a \|_{\Sigma^{-1}}^2 . \nn
\#
Moreover,  $ \varrho_a^2\geq \underbar{$\sigma$}^2 \| a \|_{\Sigma^{-1}}^2 $ according to \eqref{var.lbd}.  
Provided  $\tau \geq 2\max\{ 2 \alpha_3  / \underbar{$\sigma$}^2 , \overbar \sigma^2 /\underbar{$\sigma$} \}$,  it follows that
\#
	\var(\xi_i) = \varrho^2_{a, \tau} - ( \EE \xi_i )^2  \geq \| a \|_{\Sigma^{-1}}^2  \big(  \underbar{$\sigma$}^2  - 2 \alpha_3  /\tau - \overbar \sigma^4 /\tau^2 \big) \geq \| a \|_{\Sigma^{-1}}^2    \frac{\underbar{$\sigma$}^2}{4}. \label{var.approxi}
\#
Similarly,  using the inequality $|\psi_\tau(t)|  \leq |t|$   we obtain
 \$
 \EE |\xi_i|^3 \leq \EE |\omega_i \langle a, \Sigma^{-1} X_i\rangle |^3 \leq  2  \alpha_3 \,  \EE     | \langle a, \Sigma^{-1} X_i \rangle|^3    \leq 2 \kappa_3 \alpha_3 \, \| a \|_{\Sigma^{-1}}^3 .
 \$
 Putting these bounds together leads to
\#
 		\sup_{t\in \RR} \big| \PP\big\{ S_a^0 \leq \var(S_a^0)^{1/2} t \big\} - \Phi(t) \big| \leq C_1   \frac{  \kappa_3 \alpha_3  + (\overbar \sigma^2 /\tau)^3}{\underbar{$\sigma$}^3 \sqrt{n}} . \label{clt.approx.2}
\#
Note further that $| \var(S_a^0)  - \varrho_{a }^2 | \leq  \| a \|_{\Sigma^{-1}}^2  ( 2 \alpha_3 / \tau +  \overbar \sigma^4 / \tau^2)$. This together with \eqref{var.approxi} implies
 \#
 		\sup_{t\in \RR} \big|  \Phi(t/ \var(S_a^0)^{1/2})  - \Phi(t/ \varrho_{a } ) \big| \leq  
\frac{C_2}{ \underbar{$\sigma$}^2} 		\bigg(  \frac{\overbar \sigma^4}{\tau^2} + \frac{\alpha_3}{\tau} \bigg)   . \label{clt.approx.3}
\#
Here both $C_1, C_2>0$ are absolute constants. 
 
 Let $G\sim \cN(0, 1)$.  Combing \eqref{clt.approx.1}, \eqref{clt.approx.2} and \eqref{clt.approx.3} we conclude that, for any $t\in \RR$,
 \$
  & \PP \big(  \alpha \sqrt{n} \langle a, \hat \theta_\tau -\theta^* \rangle \leq t \big)  \\
  & \leq \PP\big\{  S_a^0 \leq t +  \| a \|_{\Sigma^{-1}}    \sqrt{n}  \big(r_1 + 2 \alpha_3  \tau^{-2} \big) \big\} + 7n^{-1} \\
  & \leq \PP\big\{ \var(S_a^0)^{1/2} G \leq t +  \| a \|_{\Sigma^{-1}}   \sqrt{n}  \big(r_1 + 2 \alpha_3  \tau^{-2} \big) \big\} + 7 n^{-1} +  C_1   \frac{ \kappa_3 \alpha_3  + (\overbar \sigma^2 /\tau)^3}{\underbar{$\sigma$}^3 \sqrt{n}} \\
  & \leq \PP\big\{  \varrho_{a } G \leq t +  \| a \|_{\Sigma^{-1}}    \sqrt{n}  \big(r_1 + 2 \alpha_3  \tau^{-2} \big) \big\} + 7 n^{-1} +  C_1   \frac{ \kappa_3 \alpha_3  + ( \overbar \sigma^2 /\tau)^3}{\underbar{$\sigma$}^3 \sqrt{n}}  + \frac{C_2}{ \underbar{$\sigma$}^2} 		\bigg(  \frac{\overbar \sigma^4}{\tau^2} + \frac{\alpha_3}{\tau} \bigg)    \\ 
  & \leq \PP(  \varrho_{a } G \leq t) +   7 n^{-1} +  C_1   \frac{ \kappa_3 \alpha_3 + ( \overbar \sigma^2 /\tau)^3}{\underbar{$\sigma$}^3 \sqrt{n}}  + \frac{C_2}{ \underbar{$\sigma$}^2} 		\bigg(  \frac{\overbar \sigma^4}{\tau^2} + \frac{\alpha_3}{\tau} \bigg)    + \frac{\| a \|_{\Sigma^{-1}} }{\sqrt{2\pi}  \varrho_a}   \sqrt{n}  \big(r_1 + 2\alpha_3 \tau^{-2} \big)  \\
  & \leq  \PP(   \varrho_{a } G \leq t) +   7 n^{-1} +  C_1   \frac{ \kappa_3 \alpha_3 + ( \overbar \sigma^2 /\tau)^3}{\underbar{$\sigma$}^3 \sqrt{n}}  +\frac{C_2}{ \underbar{$\sigma$}^2} 		\bigg(  \frac{\overbar \sigma^4}{\tau^2} + \frac{\alpha_3}{\tau} \bigg)   + \frac{ \sqrt{n} }{\sqrt{2\pi} \underbar{$\sigma$}} \big(r_1 + 2\alpha_3  \tau^{-2} \big).
 \$
 Moreover, similar arguments lead to a lower bound for  $ \PP   (  \alpha  \sqrt{n}  \langle a, \hat \theta_\tau -\theta^* \rangle \leq t )$.    This proves \eqref{clt}  by noting that 
$$
	\tau \asymp \overbar \sigma  \sqrt{\frac{ n}{p + \log n}} ~\mbox{ and }~  r_1  \asymp  \overbar \sigma \frac{p+ \log n}{ n}  + \frac{\overbar f}{\underbar{$f$}^2} \frac{(p\log n)^{1/2} (p + \log n)^{1/2}}{n} 
$$ 

Next we consider the oracle Huberized ES estimator $\hat \theta^{{\rm ora}}_\tau  = \argmin_\theta \sn \ell_\tau(Z_i - \alpha X_i^\T \theta)$.  Note that $\omega_i = Z_i - \alpha X_i^\T \theta^*= \varepsilon_i \mathbbm{1}(\varepsilon_i\leq 0) - \EE_{X_i} \{ \varepsilon_i \mathbbm{1}(\varepsilon_i\leq 0)  \}$ and hence $\EE( \omega_i^2 | X_i) \leq \overbar \sigma^2$.  Applying Theorem~2.1 in \cite{CZ2020} we obtain that for any $t>0$, the oracle estimator $\hat \theta^{{\rm ora}}_\tau$ with $\tau \asymp \overbar \sigma \sqrt{n/(p + t)}$ satisfies
$$
\alpha \| \hat \theta^{{\rm ora}}_\tau - \theta^*  \|_\Sigma \leq \overbar \sigma \sqrt{\frac{p+t}{n}} ~~\mbox{ and }~~
\bigg\| \alpha \Sigma^{1/2} ( \hat \theta^{{\rm ora}}_\tau - \theta^* ) - \frac{1}{n} \sn \psi_\tau(\omega_i ) W_i \bigg\|_2 \lesssim \overbar \sigma \frac{p+t}{n}
$$
with probability at least $1-3e^{-t}$ as long as $n\gtrsim p+t$.  The Berry-Esseen bound \eqref{oracle.be} can then be proved using similar arguments as above.  \qed

\subsection{Proof of Theorem~\ref{thm:var.consistency}}

Recall that $\omega_i(\beta, \theta) = ( Y_i - X_i^\T \beta) \mathbbm{1}(Y_i \leq  X_i^\T \beta) + \alpha X_i^\T(\beta - \theta)$ for $\beta, \theta \in \RR^p$, and $\omega_i  = \omega_i( \beta^*, \theta^*) =  \varepsilon_i  \mathbbm{1}(\varepsilon_i \leq 0) + \alpha X_i^\T(\beta^* - \theta^*)$.   Given $r_0, r_1>0$, our goal is to bound
$$
  \sup_{   \| \beta - \beta^* \|_\Sigma\leq r_0,  \atop  \alpha \| \theta - \theta^* \|_\Sigma \leq r_1  }	 \big\|  \hat V_\gamma (\beta, \theta )  -   V_\gamma  (\beta^*, \theta^* ) \big\|_2~\mbox{ and }~  \| V_\gamma(\beta^*, \theta^*) - V \|_2 ,
$$
where 
$$
\hat V_\gamma (\beta, \theta ) =   \frac{1}{n} \sn \psi_\gamma^2( \omega_i(  \beta,  \theta )   )  W_i W_i^\T, \quad V_\gamma  (\beta, \theta )  = \EE \hat V_\gamma (\beta, \theta )    ~\mbox{ and }~ V = \Sigma^{-1/2} \Omega \Sigma^{-1/2} .
$$ 
Noting that $|\psi_\tau^2(\omega_i) - \omega_i^2 | = |(\omega_i^2 - \gamma^2) \mathbbm{1} (|\omega_i| > \gamma ) | \leq \gamma^{-1} |\omega_i|^3$, similarly to \eqref{moment.ineq2} we obtain
$$
\| V_\gamma(\beta^*, \theta^*) - V \|_2 \leq  \gamma^{-1} \sup_{ u \in \mathbb{S}^{p-1} }  \EE \{ |\omega_i |^3 (W_i^\T u)^2 \} \leq 2 \alpha_3 \gamma^{-1}. 
$$

Without loss of generality, we assume $C_0=1$ for brevity in the rest of the proof, that is, $\max_{1\leq i\leq n} \| W_i \|_2 \leq \sqrt{p}$. 
For some $\epsilon_0 \in (0, r_0)$ and  $\epsilon_1 \in (0, r_1)$ to be determined,  there exist $\epsilon_0$-net $\{ \beta_1, \ldots ,\beta_{N_0} \} \subseteq \BB_\Sigma( \beta^*, r_0)$ and $(\epsilon_1/\alpha)$-net $\{ \theta_1 , \ldots ,\theta_{N_1} \} \subseteq \BB_\Sigma( \theta^*, r_1/\alpha)$ such that $N_0 \leq (1 + 2r_0/\epsilon_0)^p$ and  $N_1 \leq (1 + 2r_1/\epsilon_1)^p$.
For any $(\beta, \theta) \in  \BB_\Sigma( \beta^*, r_0) \times \BB_\Sigma( \theta^*, r_1/\alpha)$, there exist some $1\leq j\leq N_0$ and $1\leq k\leq N_1$ such that $\|  \beta - \beta_j \|_\Sigma \leq \epsilon_0$ and $\alpha \| \theta  - \theta_k \|_\Sigma \leq \epsilon_1$.   Consequently,  
\#
		 & \| \hat V_\gamma (\beta, \theta )   -  V_\gamma  (\beta^*, \theta^* )  \|_2 \nn \\
& \leq \underbrace{  \| \hat V_\gamma (\beta, \theta )   - \hat V_\gamma (\beta_j, \theta_k )    \|_2}_{{\rm discretization~error}} +  \underbrace{ \|  \hat V_\gamma (\beta_j, \theta_k )  -   V_\gamma (\beta_j, \theta_k ) \|_2}_{{\rm stochastic~error}} +  \underbrace{ \|   V_\gamma (\beta_j, \theta_k )   - V_\gamma  (\beta^*, \theta^* )  \|_2 }_{{\rm approximation~error}} .   \nn
\#
For the discretization error term,  we have
\$
&  \| \hat V_\gamma (\beta, \theta )   - \hat V_\gamma (\beta_j, \theta_k )    \|_2   \\
& \leq \sup_{u, v \in \mathbb{S}^{p-1} } \frac{1}{n} \sn  | \psi^2_\gamma ( \omega_i(\beta, \theta) ) - \psi^2_\gamma (\omega_i(\beta_j, \theta_k) )  | \cdot | W_i^\T u \cdot W_i^\T v | 
\$
Following the proof of Lemma~\ref{lem:RSC} and since $\sup_u | \psi_\gamma(u) | \leq \gamma$,  it holds
\#
 & | \psi^2_\gamma  ( \omega_i(\beta, \theta) ) - \psi^2_\gamma (\omega_i(\beta_j, \theta_k) )  |  \nn  \\
 & \leq | \psi^2_\gamma  ( \omega_i(\beta, \theta) ) - \psi^2_\gamma (\omega_i(\beta, \theta_k) )  | 
 +   | \psi^2_\gamma  ( \omega_i(\beta, \theta_k) ) - \psi^2_\gamma (\omega_i(\beta_j, \theta_k) )  |   \nn  \\
 & \leq 2 \gamma  \{   | \alpha X_i^\T (\theta - \theta_k) | +  | X_i^\T(\beta - \beta_j ) |  \} \leq 2 \gamma  (\epsilon_0 + \epsilon_1)\sqrt{p} . \nn
\#
Substituting this into the above inequality yields
\#
 & \| \hat V_\gamma (\beta, \theta )   - \hat V_\gamma (\beta_j, \theta_k )    \|_2   \nn \\ 
 & \leq 2\gamma  \sup_{u, v \in \mathbb{S}^{p-1} } \frac{1}{n} \sn  \{   | \alpha X_i^\T (\theta - \theta_k) | +  | X_i^\T(\beta - \beta_j ) |  \}  \cdot | W_i^\T u \cdot W_i^\T v |  \nn \\
 & \leq 2 \gamma( \epsilon_0  + \epsilon_1 ) \sqrt{p}  \bigg\| \frac{1}{n} \sn W_i W_i^\T \bigg\|_2 . \label{part1}
\#
For the random matrix $(1/n) \sn W_i W_i^\T$, it follows from Theorem~1 in \cite{Z2022} that with probability at least $1-1/n$,
$ \|(1/n) \sn W_i W_i^\T  - {\rm I}_p  \|_2 \lesssim \upsilon_1^2 \sqrt{(p+\log n) / n}$ and thus $\| (1/n) \sn W_i W_i^\T \|_2 \leq 2$ provided that $n\gtrsim p+\log n$.

Next we jump to the approximation error. Write 
$$
	\Delta_{ij} = X_i^\T(\beta_j - \beta^*) ~\mbox{ and }~  \Theta_{ik} =  \alpha X_i^\T(\theta_k - \theta^*) 
$$
satisfying $|\Delta_{ij}| \leq \sqrt{p} r_0$ and $|\Theta_{ik}| \leq \sqrt{p} r_1$. Note that $  |\omega_i(\beta_j, \theta_k)  - \omega_i |  \leq  |\Delta_{ij}  | + |  \Theta_{ik} | $.  By the Lipschitz continuity of $\psi_\gamma$, that is, $|\psi_\gamma(u) - \psi_\gamma(v) | \leq |u-v|$, and the fact that $|\psi_\gamma(u) | \leq |u|$, we have
\#	
	| \EE_{X_i}  \{ \psi^2_\gamma (\omega_i(\beta_j, \theta_k)) \} - \EE_{X_i} \{  \psi^2_\gamma (\omega_i ) \} |  \leq 
	 \EE_{X_i } \{  (2 |\omega_i| + |\Delta_{ij}  | + |  \Theta_{ik} | )  ( |\Delta_{ij}  | + |  \Theta_{ik} | )  \} .
 \nn
\#
In particular, Condition~\ref{cond:density} implies $\EE_{X_i} (|\omega_i |) \leq \overbar \sigma$.
Consequently,
\#
	 & \|    V_\gamma (\beta_j, \theta_k )   - V_\gamma  (\beta^*, \theta^* )  \|_2 \nn\\
	 & \leq \sup_{u\in \mathbb{S}^{p-1} }  \EE |  \EE_{X_i} \psi^2_\gamma (\omega_i(\beta_j, \theta_k)) -\EE_{X_i}  \psi_\gamma^2 (\omega_i) | (W_i^\T u )^2  \nn \\
	 & \leq  \sup_{u\in \mathbb{S}^{p-1} }  \EE \{  (2 |\omega_i| + |\Delta_{ij}  | + |  \Theta_{ik} | )  ( |\Delta_{ij}  | + |  \Theta_{ik} | ) (W_i^\T u )^2 \} \nn \\
	 & \leq 2  \kappa_3 \overbar \sigma ( r_0 + r_1 ) + \kappa_4( r_0 + r_1 )^2 .  \label{part3}
\#

It remains to control the stochastic error term $\|  \hat V_\gamma (\beta_j, \theta_k )  -   V_\gamma (\beta_j, \theta_k ) \|_2$. Via a standard covering argument, there exists a $(1/4)$-net $\cN$ of the unit sphere with $| \cN  | \leq 9^p$ such that
$$
\|  \hat V_\gamma (\beta_j, \theta_k )  -   V_\gamma (\beta_j, \theta_k ) \|_2 \leq 2 \max_{u \in \cN } \bigg| \frac{1}{n} \sn (1-\EE) \psi^2_\gamma (\omega_i(\beta_j, \theta_k)) ( W_i^\T u)^2 \bigg| .
$$
Given $u\in \cN$ and for $k=2, 3, \ldots$, note that
\$
	\EE | \psi^2_\gamma (\omega_i(\beta_j, \theta_k)) ( W_i^\T u)^2 |^k \leq \gamma^{2(k-2)}\EE  \{ \psi^4_\gamma (\omega_i(\beta_j, \theta_k)) ( W_i^\T u)^{2k} \} 
\$
and
\$
& \EE_{X_i}  \{ \psi^4_\gamma (\omega_i(\beta_j, \theta_k)) \} \leq \gamma \EE_{X_i}  \{  |\omega_i(\beta_j, \theta_k) |^3 \}  \\
&\leq    6 \gamma  \{ \EE_{X_i} (  |\omega_i|^3) + ( |\Delta_{ij}  | + |  \Theta_{ik} |  )^3 \} \lesssim  \gamma \{ \alpha_3 + ( \sqrt{p} r_0 + \sqrt{p} r_1)^3 \} .
\$
Moreover, the sub-Gaussianity of $W_i$ implies
\$
	\EE ( W_i^\T u / \upsilon_1 )^{2k} & = 2k \int_0^\infty \PP ( | W_i^\T u / \upsilon_1 | \geq  t) {\rm d} t \leq 4k \int_0^\infty t^{2k-1} e^{-t^2/2} {\rm d} t \leq 2^{k+1} k!.
\$
Applying Bernstein's inequality, we find that for any $z\geq 0$,
$$
	\bigg| \frac{1}{n} \sn (1-\EE) \psi^2_\gamma (\omega_i(\beta_j, \theta_k)) ( W_i^\T u)^2 \bigg| \lesssim \sqrt{ \{  \alpha_3 +  ( \sqrt{p} r_0 +   \sqrt{p} r_1)^3 \} \frac{\gamma z}{n}} +  \frac{ \gamma^2 z}{n}
$$
with probability at least $1-2 e^{-z}$. Taking the union bound over $u\in \cN$ and $(j, k) \in \{ 1, \ldots, N_0\} \times \{ 1, \ldots, N_1\}$, and setting $z = \log(9^p)+y$, we conclude that with probability at least $1-2 N_0 N_1 e^{-y}$,
\#
\max_{1\leq j\leq N_0 \atop  1\leq k\leq N_1 }\|  \hat V_\gamma (\beta_j, \theta_k )  -   V_\gamma (\beta_j, \theta_k ) \|_2  \lesssim \sqrt{ \{\alpha_3 +  ( \sqrt{p} r_0 +  \sqrt{p} r_1)^3 \}\gamma  \frac{p+y}{n}} + \gamma^2\frac{p+y}{n} . \label{part2}
\#

Finally, we choose $\epsilon_0 =  r_0/n$ and $\epsilon_1 = r_1/n$ such that $N_0 N_1 \leq (2n+1)^{2p}$. Combining \eqref{part1}--\eqref{part2} and taking $y = (2p+1) \log(2n+1)$ prove the claimed bound. \qed

\section{Proof of Technical Lemmas}

\subsection{Proof of Lemma~\ref{lem:first-order.error}}

With a change of variable $\delta = \Sigma^{1/2} (\beta - \beta^*)$  for $\beta \in \BB_\Sigma(\beta^*, r_0)$, write $r_i(\delta) = \omega_i(\beta) = \phi(\varepsilon_i - W_i^\T \delta) + \alpha W_i^\T \delta +  \alpha X_i^\T (\beta^* - \theta^*)$ where $\phi(t) := t\mathbbm{1}(t\leq 0)$. 
Moreover, define the $\RR^p$-valued random process
$\mathcal{R}(\delta)  = (1/n) \sn (1-\EE)  \{ \psi_\tau(r_i(\delta)) - \psi_\tau(r_i(0)) \} W_i$,
satisfying $\mathcal{R}(0) = 0$ and $\EE \cR(\delta) = 0$.  The goal is to bound the supremum $\sup_{\delta \in \BB(r_0)} \| \cR(\delta) \|_2$. 

Since both $\psi_\tau(\cdot)$ and $\phi(\cdot)$ are Lipschitz continuous that have derivatives $\psi_\tau'(t) = \mathbbm{1} (|t| \leq \tau)$ and $\phi'(t) = \mathbbm{1}(t\leq 0)$ almost everywhere, respectively, the stochastic process $\mathcal{R}(\delta) $ is absolutely continuous.  To apply Theorem~A.3 in \cite{S2013},  in the following we show that its gradient  $\nabla \cR(\delta) = (1/n) \sn \{ w_i(\delta) W_i W_i^\T - \EE w_i(\delta) W_i W_i^\T \}$ has bounded exponential moments, where
\$
  w_i(\delta) := \psi_\tau' \big( r_i(\delta) \big) \big\{ \alpha - \mathbbm{1}(\varepsilon_i \leq W_i^\T \delta ) \big\}   
\$
satisfies $|w_i(\delta)| \leq 1 - \alpha$. For any $u, v \in \mathbb{S}^{p-1}$ and $|\lambda| \leq  \sqrt{n} / 4 $,  using the inequality $| e^u - 1 - u | \leq u^2 e^{|u|} /2$ we obtain
\#
	& \EE \exp \big\{ \lambda \sqrt{n} u^\T  \nabla \cR(\delta) v / \upsilon_1^2 \big\} \nn  \\
	& = \prod_{i=1}^n  \Bigg[ 1 + \frac{\lambda^2 }{2 \upsilon_1^4 n } e^{ | \lambda| / ( \upsilon_1^2\sqrt{n} )  }  \EE \big\{   w_i(\delta) W_i^\T u W_i^\T v - \EE w_i(\delta) W_i^\T u W_i^\T v \big\}^2 e^{   |\lambda|  |W_i^\T u W_i^\T v| / (  \upsilon_1^2 \sqrt{n} ) }   \Bigg]   \nn  \\
	& \leq \prod_{i=1}^n \Bigg[ 1 + \frac{ \lambda^2  e^{1/4}}{\upsilon_1^4 n } \EE  \{ w_i(\delta) W_i^\T u W_i^\T v \}^2 e^{  |W_i^\T u W_i^\T v| / (2\upsilon_1)^2 }  \nn \\
	&~~~~~~~~~~~~~~~~
	+ \frac{\lambda^2 e^{1/4} }{\upsilon_1^4 n }  \{ \EE w_i(\delta) W_i^\T u W_i^\T v \}^2 \EE e^{  |W_i^\T u W_i^\T v| / (2\upsilon_1)^2 }  \Bigg] \nn  \\
	& \leq \prod_{i=1}^n \Bigg\{ 1 +  \frac{\lambda^2 e^{1/4}}{\upsilon_1^4 n }  \EE  ( W_i^\T u W_i^\T v )^2 e^{  |W_i^\T u W_i^\T v| / (2\upsilon_1)^2 } +   \frac{\lambda^2 e^{1/4} }{\upsilon_1^4 n }   \EE e^{  |W_i^\T u W_i^\T v| / (2\upsilon_1)^2 }   \Bigg\} .  \label{mgf.ubd1}
\#
For each $u\in \mathbb{S}^{p-1}$, define the non-negative random variable $\chi_u = ( W_i^\T u)^2 / (2 \upsilon_1)^2$. From Condition~\ref{cond:covariate} we see that $ \PP(  \chi_u \geq t ) \leq 2 e^{-2t}$ for any $t \geq 0 $. A standard calculation shows that
\$
	\EE ( e^{\chi_u} ) = 1 + \int_0^\infty e^t \PP(\chi_u \geq  t ) {\rm d} t \leq 3 ~\mbox{ and }~
	\EE( \chi_u^2 e^{\chi_u}  )  = \int_0^\infty  (t^2 + 2t) e^t \PP( \chi_u \geq t ) {\rm d} t \leq 8 . 
\$
Taking the supremum over $u\in \mathbb{S}^{p-1}$ yields
\$
	\sup_{u \in \mathbb{S}^{p-1}} \EE e^{(W_i^\T u)^2/(2\upsilon_1 )^2 } \leq 3~\mbox{ and }~
	\sup_{u \in \mathbb{S}^{p-1}} \EE (W_i^\T u)^4 e^{(W_i^\T u)^2/(2 \upsilon_1 )^2 }  \leq 128 \upsilon_1^4 .
\$
Substituting these exponential moment bounds into \eqref{mgf.ubd1}, and by H\"older's inequality, we obtain that for any $\delta \in \RR^p$ and $\lambda \in \RR$ satisfying $|\lambda| \leq  \sqrt{n} / 4$,
\$
 \sup_{ u , v \in \mathbb{S}^{p-1} } \EE \exp \big\{ \lambda \sqrt{n}  u^\T  \nabla \cR(\delta) v/ \upsilon_1^2 \big\}  \leq  \prod_{i=1}^n \Bigg( 1 +  128 e^{1/4}   \frac{\lambda^2 }{ n }    +  3e^{1/4}   \frac{\lambda^2 }{ n }    \Bigg) \leq e^{ C_0^2 \lambda^2 /2  } ,
\$
where $C_0 =  e^{1/8} \sqrt{262}$. This verifies condition (A.4) in \cite{S2013} with ${\rm g} = \frac{\sqrt{n}}{4\sqrt{2} }$. Therefore, applying Theorem~A.3 therein to the process $\{ \sqrt{n} \cR(\delta)  / \upsilon_1^2 , \delta \in \BB(r_0)\}$ we have that, with probability at least $1-e^{-t}$ ($t \geq  1/2$),
\$
	 \sup_{\delta \in \BB(r_0)  } \| \cR(\delta) \|_2 \leq 6 C_0 \upsilon_1^2  \sqrt{\frac{4p + 2t }{n}} \cdot r_0 
\$
as long as $n\geq  64 (2p+   t)$. This establishes the claim. \qed

\subsection{Proof of Lemma~\ref{lem:approximate.neyman}}

Note that
$\| \EE \{  \psi_\tau  ( \omega_i(\beta) ) W_i  \}  \|_2 =  \sup_{u\in \mathbb{S}^{p-1} } | \EE   \{ \psi_\tau  ( \omega_i(\beta)   ) W_i^\T u \} |$ and $\psi_\tau(t) = t \mathbbm{1}(|t|\leq \tau) + \tau \sign(t) \mathbbm{1}(|t| >\tau)$. 
Recall that the conditional CDF $F=F_{\varepsilon_i | X_i}$ of $\varepsilon_i$ given $X_i$ is continuously differentiable with $f=F'$. 
Let $\EE_{X_i}$ be the conditional expectation given $X_i$. For $\beta \in \RR^p$ and $u\in \mathbb{S}^{p-1}$, define $\Delta_i = \Delta_i(\beta) = X_i^\T (\beta - \beta^*)$  and  
\$
	& E_i(\beta) = \EE_{X_i} \psi_\tau \big( \omega_i (\beta)  \big)   \\
	&   =   \int_{-\infty}^{\Delta_i} \psi_\tau\big( t - \Delta_i + \alpha X_i^\T (\beta - \theta^*)  \big) f(t) {\rm d} t+ \int_{\Delta_i}^\infty \psi_\tau\big( \alpha X_i^\T (\beta - \theta^* ) \big) f(t) {\rm d}t      .  
\$
Since $\psi_\tau(\cdot)$ is absolutely continuous and has a derivative $\psi'_\tau(t) = \mathbbm{1}(|t| \leq \tau)$ almost everywhere, by the fundamental theorem of Lebesgue integral calculus we have
\$
	E_i(\beta) - E_i(\beta^*) = \int_0^1 \langle  \nabla E_i \big( \beta^* + t(\beta-\beta^*) \big) , \beta - \beta^* \rangle {\rm d} t ,
\$
where 
\$
	& \nabla E_i(\beta) \\
	&  =  \int_{-\infty}^{\Delta_i} \psi'_\tau\big( t - \Delta_i + \alpha X_i^\T (\beta - \theta^*) \big) f(t) {\rm d} t \cdot (\alpha -1) X_i +  f(\Delta_i)  \psi_\tau \big( \alpha X_i^\T ( \beta - \theta^* ) \big) \cdot (\alpha-1) X_i  \\ 
& ~~~~~~ + \psi_\tau'\big(\alpha X_i^\T (\beta - \theta^* ) \big) \{ 1- F(\Delta_i) \} \cdot \alpha X_i - f(\Delta_i) \psi_\tau\big( \alpha X_i^\T (\beta - \theta^* ) \big) \cdot (\alpha - 1 ) X_i \\
& = (\alpha-1) F(\Delta_i) X_i + \alpha \{ 1- F(\Delta_i) \} X_i    + \EE_{X_i} \mathbbm{1}\{ | \omega_i(\beta) | > \tau \}  \{  \mathbbm{1}(\varepsilon_i \leq \Delta_i) - \alpha \} X_i  \\
& = \{ \alpha - F(\Delta_i) \} X_i + \EE_{X_i} \mathbbm{1}\{ | \omega_i(\beta) | > \tau \}  \{  \mathbbm{1}(\varepsilon_i \leq \Delta_i) - \alpha \} X_i . 
\$  
For $t\in [0, 1]$, write $\beta_t = \beta^* + t(\beta-\beta^*)$ so that  $X_i^\T(\beta_t - \beta^*) = t \Delta_i$ and
\$
\langle  \nabla E_i \big( \beta^* + t(\beta-\beta^*) \big) , \beta - \beta^* \rangle = \{ \alpha - F(t\Delta_i) \} \Delta_i + \EE_{X_i} \mathbbm{1}\{ | \omega_i(\beta_t) | > \tau \}  \{  \mathbbm{1}(\varepsilon_i \leq t \Delta_i) - \alpha \} \Delta_i .
\$
By Condition~\ref{cond:density}, $| \alpha - F(t\Delta_i) | \leq \overbar f \cdot t | \Delta_i |$ almost surely. Moreover,  
\#
\EE_{X_i} \mathbbm{1}\{ | \omega_i(\beta_t) | > \tau \}  | \mathbbm{1}(\varepsilon_i \leq t \Delta_i) - \alpha | \leq \frac{1-\alpha}{\tau } \EE_{X_i} | \omega_i(\beta_t)|. \label{ei.first.moment}
\#
Observe that
\$
| \omega_i(\beta_t)  |  & \leq | (\varepsilon_i - t\Delta_i) \mathbbm{1}(\varepsilon_i \leq  t \Delta_i) -\varepsilon_i \mathbbm{1}(\varepsilon_i \leq 0)   +   \alpha t \Delta_i |  + | \omega_i (\beta^*) |  \\
& \leq| \omega_i |  +  \begin{cases}
   |\varepsilon_i \mathbbm{1}(0<\varepsilon_i \leq t \Delta_i) + t \Delta_i \{\alpha - \mathbbm{1}(\varepsilon_i \leq t \Delta_i ) \} |   & {\rm~if }~  \Delta_i \geq  0 \\
   |t  \Delta_i \{ \alpha - \mathbbm{1}(\varepsilon_i \leq t \Delta_i) \}- \varepsilon_i \mathbbm{1} (t\Delta_i < \varepsilon_i \leq 0 ) |   & {\rm~if }~ \Delta_i < 0 
\end{cases} \\
& \leq | \omega_i | +  t  |  \Delta_i |  ,
\$
thus implying
\$
  \EE_{X_i} | \omega_i(\beta_t)|  =  	\EE_{X_i}  |  \omega_i |  + t | \Delta_i |  \leq \big( \EE_{X_i} \omega_i^2 \big)^{1/2} + t | \Delta_i | \leq \overbar \sigma + t | \Delta_i | .
\$
Substituting this into \eqref{ei.first.moment} yields $\EE_{X_i} \mathbbm{1}\{ | \omega_i(\beta_t) | > \tau \}  | \mathbbm{1}(\varepsilon_i \leq t \Delta_i) - \alpha | \leq   \tau^{-1} \big( \overbar \sigma  + t | \Delta_i | \big)$. Putting together the pieces, we conclude that for any $\beta \in \BB_\Sigma(\beta^*, r_0)$,
\#
 & 	\big|  \EE  \big\{ E_i(\beta) - E_i(\beta^*) \big\} W_i^\T u \big|  \nn \\
&\leq     \int_0^1 \EE \big\{  \overbar f \cdot t   \Delta_i^2 +   \tau^{-1} \big(  \overbar \sigma  + t | \Delta_i | \big) \cdot  | \Delta_i | \big\} \cdot | W_i^\T u | \, {\rm d} t  \nn \\ 
& \leq \frac{1}{2} \kappa_3 (\overbar f + 1/\tau)  r_0^2 +  \overbar \sigma  \frac{r_0}{ \tau  }   . \label{Su.diff.bound}
\#
Note also that $\EE_{X_i} \mathbbm{1}\{ | \omega_i(\beta_t) | > \tau \}    \leq    \tau^{-2}  ( \overbar \sigma^2 + 2 \overbar \sigma  t |\Delta_i| +  t^2   \Delta^2_i   )$,  which in turn implies
\#
	& 	\big|  \EE  \big\{ E_i(\beta) - E_i(\beta^*) \big\} W_i^\T u \big| \nn \\
		& \leq \frac{\kappa_3}{2} \overbar f r_0^2 +  (\overbar \sigma^2 + \kappa_3  \overbar \sigma r_0 + \kappa_4 r_0^2/3)    \frac{r_0}{\tau^2}  .\label{Su.diff.bound2}
\#

Finally, for $\| \EE \psi_\tau(\omega_i) W_i \|_2 =\sup_{u\in \mathbb{S}^{p-1} } |\EE \{ \psi_\tau(\omega_i) W_i^\T u\}|$, note that $\EE_{X_i} (\omega_i) = 0$, $\EE_{X_i} (\omega_i^2) \leq \overbar \sigma^2$ and $|\psi_\tau(t) - t| = (|t| - \tau) \mathbbm{1}(|t| > \tau)\leq \tau^{-1} t^2$. Therefore, $| \EE \{ \psi_\tau(\omega_i) W_i^\T u \}   | \leq \tau^{-1} \EE \big( \omega_i^2 |W_i^\T u| \big) \leq \tau^{-1} \overbar \sigma^2$. Combining this with \eqref{Su.diff.bound} and \eqref{Su.diff.bound2}   proves the claims \eqref{mean.grad.ubd}. and \eqref{mean.grad.ubd2}, respectively. \qed

\subsection{Proof of Lemma~\ref{lem:score.bound}}

We apply a standard covering argument and Bernstein's inequality to bound the $\ell_2$-norm of the centered random vector $ (1/n) \sn  (1-\EE ) \psi_\tau(\omega_i ) W_i$. 
For any $\epsilon\in (0, 1)$, there exists an $\epsilon$-net $\cN_\epsilon$ of $\mathbb{S}^{p-1}$ with $|\cN_\epsilon | \leq (1 + 2/\epsilon)^p$ such that 
\#
 \bigg\| \frac{1}{n} \sn  (1-\EE ) \psi_\tau( \omega_i) W_i \bigg\|_2 \leq  \frac{1}{1 - \epsilon} \max_{u\in \cN_\epsilon } \frac{1}{n} \sn (1-\EE ) \psi_\tau( \omega_i ) W_i^\T u .  \label{score.discretization}
\#
Recall that $| \psi_\tau( \omega_i ) | \leq \tau$ and $\EE_{X_i} \psi^2_\tau( \omega_i ) \leq \EE_{X_i} (\omega_i^2) \leq \overbar \sigma^2$.
To bound the higher-order moments, Condition~\ref{cond:covariate} ensures that for each $k\geq 3$,
$\EE | W_i^\T u |^k \leq 2 \upsilon_1^k  k \int_0^\infty t^{k-1} e^{-t^2/2} {\rm d} t =  \upsilon_2^k k \Gamma(k/2)$, where $\Gamma(\cdot)$ is the Gamma function and $\upsilon_2 = \sqrt{2} \upsilon_1$. If $k=2l$ for some $l\geq 2$, $\EE | W_i^\T u |^k  \leq 2 \upsilon_2^k (k/2)! \leq 2 \upsilon_2^k k!/2^k$; and if $k=2l+1$ for some $l\geq 1$, 
$$
\EE | W_i^\T u |^k  \leq \upsilon_2^{k} k \Gamma(l + 1/2) =   \sqrt{\pi} \upsilon_2^k \frac{ k (2l)!}{4^l l!} = 2 \sqrt{\pi} \upsilon_2^k \frac{k!}{2^k l!}.
$$
Putting together the pieces, we obtain that $\EE  | \psi_\tau( \omega_i )  W_i^\T u  |^2 \leq \overbar \sigma^2$ and 
\$
	\EE | \psi_\tau( \omega_i ) W_i^\T u |^k  & \leq \tau^{k-2} \EE \{ |W_i^\T u|^k \EE_{X_i} ( \omega_i^2 ) \} \leq \tau^{k-2}  \overbar \sigma^2 \EE  ( |W_i^\T u|^k ) \\
	&  \leq \frac{k!}{2}  \cdot  \sqrt{\pi}  \upsilon_2^2 \overbar \sigma^2  \cdot  (\upsilon_2 \tau /2)^{k-2} , \ \ k\geq 3.
\$
By applying Bernstein's inequality and the union bound, we find that with probability at least $1 -  e^{-t}$,
\$
 \max_{u\in \cN_\epsilon} \frac{1}{n} \sn (1-\EE ) \psi_\tau(\omega_i ) W_i^\T u \leq  2 \upsilon_2  \overbar \sigma \sqrt{\frac{ p\log(1+2/\epsilon) + t }{n}}    + \upsilon_2 \tau \frac{ p\log(1+2/\epsilon) + t }{2 n } .
\$ 
Combining this with \eqref{score.discretization}, and taking $\epsilon= 1/2$, we establish the claimed tail bound. \qed

\subsection{Proof of Lemma~\ref{lem:RSC}}

For $\beta, \theta \in \RR^p$, define the joint loss function $\hat \cL_\tau(\beta, \theta) = (1/n) \sn \ell_\tau\big( Z_i ( \beta) - \alpha X_i^\T \theta \big)$, and $\Delta_i = \Delta_i(\beta) = X_i^\T (  \beta - \beta^* )$.
Note that
\$
| Z_i(  \beta) - \alpha X_i^\T \theta^* |  & \leq | (\varepsilon_i - \Delta_i) \mathbbm{1}(\varepsilon_i \leq  \Delta_i) -\varepsilon_i \mathbbm{1}(\varepsilon_i \leq 0)   +   \alpha \Delta_i |  + | \omega_i |  \\
& \leq| \omega_i |  +  \begin{cases}
   |\varepsilon_i \mathbbm{1}(0<\varepsilon_i \leq \Delta_i) +\Delta_i \{\alpha - \mathbbm{1}(\varepsilon_i \leq \Delta_i ) \} |   & {\rm~if }~  \Delta_i \geq  0 \\
   | \Delta_i \{ \alpha - \mathbbm{1}(\varepsilon_i \leq \Delta_i) \}- \varepsilon_i \mathbbm{1} (\Delta_i < \varepsilon_i \leq 0 ) |   & {\rm~if }~ \Delta_i < 0 
\end{cases} \\
& \leq | \omega_i | + | \Delta_i | .
\$
For each $i=1,\ldots, n$,  define the event 
\#
	\cE_i(\beta, \theta) &  = \big\{  |\omega_i | \leq  \tau  /4 \big\}  \cap \big\{ | X_i^\T (\beta - \beta^*) |  \leq \tau /4 \big\}  \cap  \big\{ | X_i^\T (\theta - \theta^*) |/ \| \theta - \theta^* \|_\Sigma \leq \tau /(2 r ) \big\} , \nn
\#
such that conditioning on $\cE_i(\beta, \theta)$ with $\beta \in \BB_\Sigma(\beta^* , r_0)$ and $\theta \in \BB_\Sigma(\theta^* , r / \alpha )$,
\$ 
 | Z_i ( \beta) - \alpha X_i^\T \theta | \leq | Z_i(  \beta )  - \alpha X_i^\T \theta^* | +  \alpha  | X_i^\T (\theta - \theta^* ) | \leq  \frac{ \tau}{4}    + \frac{ \tau}{4} + \frac{ \tau }{2}  =  \tau. 
\$
Consequently,
\#
	& \langle \partial_\theta \hat \cL_\tau (\beta,   \theta) -  \partial_\theta  \hat \cL_\tau ( \beta,  \theta^*) , \theta - \theta^* \rangle  \nn \\
	& = \frac{\alpha}{n} \sn \big\{ \psi_\tau \big( Z_i (  \beta) -\alpha  X_i^\T \theta^* \big) -  \psi_\tau \big( Z_i (  \beta) - \alpha X_i^\T \theta  \big) \big\} X_i^\T (\theta - \theta^* ) \nn \\
	& \geq \frac{\alpha}{n} \sn \big\{ \psi_\tau \big( Z_i (  \beta) - \alpha X_i^\T \theta^* \big) - \psi_\tau\big( Z_i (\beta) - \alpha X_i^\T \theta  \big) \big\} X_i^\T(\theta -\theta^*) \mathbbm{1}_{\cE_i(\beta,\theta)  } \nn \\
	& \geq \frac{\alpha^2 }{n} \sn \langle   X_i, \theta -  \theta^* \rangle^2   \mathbbm{1}_{\cE_i(\beta, \theta)  } . \label{RSC.lbd1}
\#

For any $R>0$, define the functions
\$
\varphi_R(t) = t^2 \mathbbm{1}(|t| \leq R/2) + ( t - \sign(t) R)^2 \mathbbm{1}(R/2 < |t| \leq R  ) \\
\mbox{ and }~ \phi_R(t) = \mathbbm{1}(|t|\leq R/2) + \{ 2 - (2t/R) \sign(t) \} \mathbbm{1}( R/2 <  |t| \leq R ) ,
\$
which are smoothed proxies of $t\mapsto t^2 \mathbbm{1} (|t| \leq R)$ and $t\mapsto  \mathbbm{1} (|t| \leq R)$, respectively.
Moreover, $\varphi_R(\cdot)$ is $R$-Lipschitz continuous  and satisfies (i) $t^2 \mathbbm{1}(|t| \leq R/2) \leq \varphi_R(t) \leq  t^2 \mathbbm{1}(|t| \leq R)$ and (ii) $\varphi_{cR}(c t)  = c^2 \varphi_R(t) $ for any $c>0$; $\phi_R(\cdot)$ is $(2/R)$-Lipschitz continuous and satisfies $\mathbbm{1}(|t| \leq R/2) \leq \phi_R(t) \leq   \mathbbm{1}(|t| \leq R)$.

For $\beta \in \BB_\Sigma(\beta^* , r_0)$ and $\theta \in \BB_\Sigma(\theta^*, r / \alpha)$, consider the following reparametrizations
$$
	\gamma  = \Sigma^{1/2} (\beta - \beta^*) \in \BB(r_0) ~\mbox{ and }~ \delta = \Sigma^{1/2}(\theta-\theta^*) / \|\theta - \theta^* \|_\Sigma   \in \mathbb{S}^{p-1}  
$$ 
throughout the rest of the proof.  Then 
\# 
 \langle \partial_\theta \hat \cL_\tau (\beta,   \theta) -  \partial_\theta  \hat \cL_\tau ( \beta,  \theta^*) , \theta - \theta^* \rangle  &  \geq \frac{\alpha^2 }{n} \sn \chi_i \cdot  \varphi_{ \tau  \| \theta - \theta^* \|_\Sigma / (2r) } (\langle  X_i, \theta-\theta^* \rangle )  \phi_{\tau /4 }(W_i^\T \gamma )  \nn \\
& =  \alpha^2  \| \theta - \theta^* \|_\Sigma^2 \cdot \underbrace{ \frac{1}{n} \sn \chi_i \cdot   \varphi_{ \tau   / (2r) } (   W_i^\T \delta  )  \phi_{\tau /4 }(W_i^\T \gamma)  }_{=: \, G_n(\beta, \theta) },  \label{RSC.lbd2}
\#
where $\chi_i = \mathbbm{1} \{   |\omega_i | \leq  \tau  /4    \}$. In the following, we bound $G_n(\beta, \theta) - \EE G_n(\beta, \theta)$ and $\EE G_n(\beta, \theta)$, respectively.

Noting that $\varphi_R(t) \geq t^2 \mathbbm{1}(|t| \leq R/2)$ and $\phi_R(t) \geq \mathbbm{1}(|t|\leq R/2)$, we have
\#
  \EE G_n(\gamma, \delta) 
	 & \geq \EE  (W_i^\T \delta)^2 \mathbbm{1}  \{ | W_i^\T \delta | \leq \tau /(4r)     \}  \mathbbm{1}  \{ | W_i^\T \gamma | \leq \tau  / 8   \}  \mathbbm{1}  \{   | \omega_i | \leq  \tau  / 4 \}     \nn \\
	& = \EE (   W_i^\T \delta)^2  - \EE  (  W_i^\T \delta)^2 \mathbbm{1} \{  |   W_i^\T  \delta | > \tau /(4r)   \} -  \EE  (  W_i^\T \delta)^2 \mathbbm{1} \{  |   W_i^\T  \gamma | > \tau / 8    \}      \nn \\
	&~~~~~~ -  \EE (   W_i^\T \delta)^2 \mathbbm{1} \{ | \omega_i | >   \tau  /4   \}   \nn \\
	& \geq  1   -  \Big( \frac{4 r}{\tau  } \Big)^2  \EE (W_i^\T \delta)^4 -  \Big( \frac{8}{\tau } \Big)^2 \EE (   W_i^\T \delta)^2  (   W_i^\T \gamma )^2    -   \Big( \frac{4}{\tau } \Big)^2 \EE (\omega_i W_i^\T \delta)^2     \nn \\
	& \geq 1 - \kappa_4  ( 4r / \tau  )^2 - \kappa_4 (8  r_0 / \tau  )^2 -  (4 \overbar \sigma /\tau )^2 \nn \\
	& \geq \frac{1}{2},  \label{Gn.mean}
\#
where the last inequality holds if $\tau^2  \geq  32 \{  \kappa_4 (r^2 + 4 r_0^2) +  \overbar \sigma^2  \}$.

Next, consider the supremum $
	\Lambda_n = \sup_{ ( \beta, \theta ) \in \BB_\Sigma(\beta^*, r_0) \times \BB_\Sigma(\theta^*, r/\alpha)   }   \{ - G_n(\beta, \theta) + \EE G_n(\beta, \theta )   \}$. For each pair $(\beta, \theta)$, write $g_{\beta, \theta} ( X_i, \varepsilon_i) =  \chi_i  \varphi_{\tau /(2 r)} (W_i^\T \delta ) \phi_{\tau /4} (W_i^\T \gamma) $, so that 
$$
	\Lambda_n = \sup_{ ( \beta, \theta ) \in \BB_\Sigma(\beta^*, r_0) \times \BB_\Sigma(\theta^*, r/\alpha) } \frac{1}{n}\sn  \{  \EE g_{\beta, \theta} ( X_i, \varepsilon_i) - g_{\beta, \theta} ( X_i, \varepsilon_i) \}. 
$$	
Since $0 \leq \varphi_R(t) \leq \min \{ (R/2)^2, t^2 \}$ and $0\leq \phi_R(t) \leq 1$ for any $t\in \RR$, 
\$
   0   \leq   g_{\beta, \theta} ( X_i, \varepsilon_i) \leq (\tau  / 4r )^2 ~\mbox{ and }~  \EE g^2_{\beta, \theta} ( X_i, \varepsilon_i) \leq  \kappa_4 . 
\$
Applying Theorem~7.3 in \cite{B2003}, we have that for any $t\geq 0$,
\#
	\Lambda_n  \leq \EE \Lambda_n   + ( \EE \Lambda_n  )^{1/2} \frac{\tau  }{2 r} \sqrt{\frac{t}{n}} + (2 \kappa_4 )^{1/2} \sqrt{\frac{t}{n}} + \Big(\frac{\tau }{4r} \Big)^2 \frac{t}{3n} \label{Lambdan.concentration}
\#
with probability at least $1-e^{-t}$. To bound $\EE \Lambda_n$, using symmetrization techniques  and by the
connection between Gaussian and Rademacher complexities (see, e.g. Lemma~4.5 in \cite{LT1991}), we see that
\$
	\EE \Lambda_n \leq \sqrt{2\pi} \cdot \EE \bigg\{   \sup_{ ( \beta, \theta ) \in \BB_\Sigma(\beta^*, r_0) \times \BB_\Sigma(\theta^*, r/\alpha)   } \mathbb{G}_{ \beta, \theta} \bigg\}  ,
\$
where $\mathbb{G}_{\beta, \theta } = (1/n) \sn g_i    \chi_i  \varphi_{\tau /(2 r)} (W_i^\T \delta ) \phi_{\tau /4} (W_i^\T \gamma)$ and $g_i$'s are independent standard normal random variables. Let $\EE^*$ be the conditional expectation given $\{ (X_i, \varepsilon_i ) \}_{i=1}^n$. Note that $\{ \mathbb{G}_{\beta, \theta } \}$ is a (conditional) Gaussian process.
For $(\beta, \theta), (\beta', \theta') \in \BB_\Sigma(\beta^*, r_0) \times \BB_\Sigma(\theta^*, r/\alpha)$, define $(\gamma', \delta')$ accordingly, and consider the decomposition
\$
 \mathbb{G}_{\beta, \theta }  - \mathbb{G}_{\beta', \theta' }  &= \mathbb{G}_{\beta, \theta } - \mathbb{G}_{\beta  , \theta' } + \mathbb{G}_{ \beta , \theta' } - \mathbb{G}_{\beta', \theta' }  \\
& = \frac{1}{n} \sn g_i  \chi_i \phi_{\tau /4} (W_i^\T \gamma) \big\{   \varphi_{\tau /(2r)} ( W_i^\T \delta) -  \varphi_{\tau /(2r)} ( W_i^\T \delta')  \big\}      \\
&~~~~~ + \frac{1}{n} \sn g_i \chi_i \varphi_{\tau /(2r)} ( W_i^\T \delta’) \big\{ \phi_{\tau /4 } (W_i^\T \gamma) - \phi_{\tau /4} (W_i^\T \gamma') \big\}   .
\$
Note that $\EE^*(\mathbb{G}_{\beta, \theta }  - \mathbb{G}_{\beta', \theta' })^2 \leq 2 \EE^*(\mathbb{G}_{\beta, \theta } - \mathbb{G}_{\beta  , \theta' } )^2 + 2 \EE^* (\mathbb{G}_{ \beta , \theta' } - \mathbb{G}_{\beta', \theta' } )^2$.
By the Lipschitz properties of $\varphi_R$ and $\phi_R$, 
\$
	\EE^* (\mathbb{G}_{\beta, \theta } - \mathbb{G}_{\beta  , \theta' })^2  \leq \frac{1}{n^2} \sn   \big\{   \varphi_{\tau /(2r)} ( W_i^\T \delta) -  \varphi_{\tau /(2r)} ( W_i^\T \delta')  \big\}^2  \leq \Big(\frac{\tau }{2r} \Big)^2 \frac{1}{n^2} \sn \{ W^\T_i ( \delta - \delta' )   \}^2  
\$
and 
\$
	\EE^* (\mathbb{G}_{\beta  , \theta' } - \mathbb{G}_{\beta'  , \theta' })^2 \leq \frac{1}{n^2} \sn \Big( \frac{\tau }{4r}\Big)^4 \Big( \frac{8}{\tau } \Big)^2 \{ W_i^\T  (\gamma - \gamma')  \}^2 = \Big( \frac{\tau  }{2 r^2 }\Big)^2 \frac{1}{n^2} \sn  \{ W_i^\T  (\gamma - \gamma')  \}^2  .
\$
Define another (conditional) Gaussian process $\{ \mathbb{Z}_{\beta,\theta}\}$ as 
\$
\mathbb{Z}_{\beta,\theta}  =  \frac{\sqrt{2} \tau }{2 r} \cdot \frac{1}{n} \sn  g_i'  W_i^\T \delta  +  \frac{\sqrt{2} \tau  }{2 r^2 } \cdot \frac{1}{n} \sn g_i'' W_i^\T \gamma   ,
\$
where $g_1', g_1'', \ldots, g_n', g_n''$ are independent standard normal random variables that are independent of all the other variables. From the above calculations we have $\EE^* (\mathbb{G}_{\beta, \theta }  - \mathbb{G}_{\beta', \theta' })^2 \leq \EE^* (\mathbb{Z}_{\beta, \theta }  - \mathbb{Z}_{\beta', \theta' })^2$. Then, applying Sudakov-Fernique’s Gaussian
comparison inequality (see, e.g. Theorem~7.2.11 in \cite{V2018}), we obtain
\$
 \EE \bigg\{   \sup_{ ( \beta, \theta ) \in \BB_\Sigma(\beta^*, r_0) \times \BB_\Sigma(\theta^*, r/\alpha )   } \mathbb{G}_{  \beta, \theta } \bigg\}  \leq  \EE^* \bigg\{   \sup_{ ( \beta, \theta ) \in \BB_\Sigma(\beta^*, r_0) \times \BB_\Sigma(\theta^*, r/\alpha )   } \mathbb{Z}_{  \beta, \theta } \bigg\} ,
\$
which remains valid if $\EE^*$ is replaced by $\EE$. For the latter, it is easy to see that
\#
 & \EE \bigg\{   \sup_{ ( \beta, \theta ) \in \BB_\Sigma(\beta^*, r_0) \times \BB_\Sigma(\theta^*, r / \alpha )   } \mathbb{Z}_{  \beta, \theta } \bigg\}  \nn  \\
 & \leq  \frac{\sqrt{2} \tau }{2 r}  \EE \bigg( \sup_{  \delta  \in  \mathbb{S}^{p-1}   }  \frac{1}{n} \sn  g_i'  W_i^\T \delta  \bigg) +  \frac{\sqrt{2} \tau  }{2r^2 }  \EE \bigg\{  \sup_{  \gamma  \in \BB(r_0)   }  \frac{1}{n} \sn g_i'' W_i^\T \gamma \bigg\}    \nn  \\
 & \leq \frac{\sqrt{2}}{2} \frac{\tau }{r} \EE \bigg\| \frac{1}{n} \sn g_i'  W_i \bigg\|_2  +  \frac{\sqrt{2}}{2}  \frac{\tau  r_0}{r^2} \EE \bigg\| \frac{1}{n} \sn g_i'' W_i \bigg\|_2  \nn  \\
 & \leq    \frac{\tau }{r}  \bigg(  \frac{1}{2} +  \frac{r_0 }{2 r} \bigg)    \sqrt{\frac{2 p}{n}} \leq   \frac{\tau }{r}  \sqrt{\frac{2p}{n}} ,\label{Lambdan.mean}
\#
where the last step uses the condition that $r_0 \leq r$.

Together, \eqref{Lambdan.concentration} and \eqref{Lambdan.mean} imply that with probability at least $1-e^{-t}$,
\$
	\Lambda_n \leq \frac{5}{4} \EE \Lambda_n + (2 \kappa_4)^{1/2} \sqrt{\frac{t}{n}} + \Big( \frac{\tau }{r} \Big)^2 \frac{t}{3n } \leq \frac{1}{4}
\$
as long as $n\gtrsim   (\tau /r)^2 (p+t)$. Combined with \eqref{Gn.mean}, this further implies that with high probability,
\$
G_n(\beta, \theta) = \EE G_n(\beta, \theta) - \big\{ \EE G_n(\beta, \theta) - G_n(\beta, \theta) \big\} \geq \frac{1}{2} - \frac{1}{4} = \frac{1}{4}
\$
holds uniformly over $\beta \in \BB_\Sigma(\beta^*, r_0)$ and $\theta \in \BB_\Sigma(\theta^*, r/\alpha )$. Substituting this into \eqref{RSC.lbd2}   establishes the claim. \qed

\subsection{Proof of Lemma~\ref{lem:first-order.error2}}

The proof is based on similar arguments to those employed in the proof of Lemma~\ref{lem:first-order.error}. With slight abuse of notation, define the vector random process
\$
	\cR(\theta) = \frac{1}{n}  \sn \big\{ \psi_\tau(Z_i - \alpha X_i^\T \theta) - \psi_\tau(Z_i - \alpha X_i^\T \theta^* ) \big\} W_i +  \alpha \Sigma^{1/2} (\theta -\theta^* ) , \ \ \theta \in \RR^p ,
\$
where $Z_i = \varepsilon_i \mathbbm{1} (\varepsilon_i \leq 0 ) + \alpha X_i^\T \beta^*$. 
By the mean value theorem for vector-valued functions,
\$
	\EE \cR(\theta) & = \EE \big\{ \psi_\tau(Z_i - \alpha X_i^\T \theta) W_i \big\} - \EE \big\{  \psi_\tau(Z_i - \alpha X_i^\T \theta^*) W_i  \big\} + \alpha \Sigma^{1/2}(\theta-\theta^*)  \\
&  =  \alpha \Sigma^{1/2} (\theta-\theta^*) - \alpha  \EE \int_0^1 \psi_\tau'(Z_i - \alpha X_i^\T \theta_t )  {\rm d} t \cdot  W_i X_i^\T (\theta - \theta^*) \\
& = \alpha  \Bigg\{ {\rm I}_p -  \EE \int_0^1 \psi_\tau'(Z_i - \alpha X_i^\T \theta_t) {\rm d} t \cdot W_i W_i^\T \Bigg\}  \Sigma^{1/2}(\theta - \theta^* ) \\
& = \int_0^1 \EE\big\{ \mathbbm{1}\big( | Z_i - \alpha X_i^\T \theta_t | > \tau \big) W_i W_i^\T  \big\}  \, {\rm d} t  \cdot  \alpha \Sigma^{1/2}(\theta - \theta^* )
\$
where $\theta_t = ( 1-t) \theta^* + t\theta$. By Markov's inequality, 
\$
	& \PP_{X_i}\big(| Z_i - \alpha X_i^\T \theta_t | > \tau\big)   \leq \tau^{-2} \EE_{X_i} | Z_i - \alpha X_i^\T \theta_t |^2 \\
	& = \tau^{-2} \EE_{X_i} | \omega_i - \alpha t X_i^\T(\theta -\theta^*) |^2 = \tau^{-2} \EE_{X_i}(\omega_i^2) + ( t /\tau)^2 |\alpha X_i^\T (\theta-\theta^* ) |^2 .
\$
Putting the above two observations together and applying Condition~\ref{cond:density}, we obtain that for any $\theta \in \BB_\Sigma(\theta^*, r/\alpha)$,
\#
	& \| \EE \cR(\theta) \|_2 = \sup_{u \in \mathbb{S}^{p-1} }   \EE \{  u^\T \cR(\theta)   \}  \nn \\
	& \leq \frac{\overbar \sigma^2}{\tau^2} \alpha  \| \theta - \theta^* \|_\Sigma + \frac{\alpha^3}{3\tau^2}  \sup_{u \in \mathbb{S}^{p-1} }   \EE \big\{ | X_i^\T(\theta - \theta^* )|^3 |W_i^\T u |  \big\} \leq   (  \overbar \sigma^2 + \kappa_4 r^2 /3 )  \frac{r}{\tau^2} .   \label{mean.Rn.ubd}
\#

Turning to $\cR(\theta)  - \EE \cR(\theta)$, consider the change of variable $\delta = \alpha\Sigma^{1/2}(\theta-\theta^*)$, and define the centered process $\cR_0(\delta)  = \cR(\theta)  - \EE \cR(\theta) =  (1/n) \sn (1- \EE) \{ \psi_\tau( \omega_i -   W_i^\T \delta) - \psi_\tau(\omega_i ) \} W_i$. Since  $\cR_0(\delta) $ is absolutely continuous, following the proof of Lemma~\ref{lem:first-order.error} we can show that its gradient $\nabla  \cR_0(\delta)  =   (-1/n) \sn  (1-\EE)  \psi_\tau'( \omega_i -  W_i^\T \delta ) W_i W_i^\T$ has bounded exponential moments. That is,  for any $\delta\in \RR^p$ and $|\lambda | \leq \sqrt{n}/4$,
\$
	\sup_{u, v\in \mathbb{S}^{p-1} } \EE \exp\big\{ \lambda \sqrt{n}  u^\T \nabla  \cR_0(\delta)  v / \upsilon_1^2 \big\} \leq  e^{ C_0^2  \lambda^2 / 2} ,
\$
where $C_0>0$ is an absolute constant.  Applying Theorem~A.3 of \cite{S2013} to the process $\{ \cR_0(\delta) / \upsilon_1^2  , \delta \in \BB(r)\}$, we obtain that for any $t \geq 1/2$,
\$
  \PP\Bigg\{ 	\sup_{\theta \in \BB_\Sigma(\theta^*, r/\alpha ) } \| \cR(\theta)  - \EE \cR(\theta)  \|_2 \geq  C_1 \upsilon_1^2 \sqrt{\frac{p+t}{n}} \cdot r  \Bigg\} \leq e^{-t}.
\$ 
Combining this with \eqref{mean.Rn.ubd} proves the claim. \qed

\subsection{Proof of Lemma~\ref{lem:qr.difference}}

For each sample $Z_i = (X_i, Y_i)$ and $\gamma \in \RR^p$, define the loss difference $r(\gamma; Z_i) = \rho_\alpha(Y_i - X_i^\T (\beta^* + \gamma)) - \rho_\alpha(Y_i - X_i^\T \beta^*) = \rho_\alpha(\varepsilon_i - X_i^\T  \gamma)  - \rho_\alpha(\varepsilon_i)$, so that $\hat D(\gamma) = (1/n) \sn r(\gamma; Z_i)$. By the Lipschitz continuity of $\rho_\alpha(\cdot)$, it is easy to see that $r(\gamma; Z_i)$ is $\bar \alpha$-Lipschitz continuous in $X_i^\T \gamma$, where $\bar \alpha = \max(\alpha, 1-\alpha)$.

Given $r>0$, define the random variable $\Delta(r) = \sqrt{n} \sup_{\gamma \in \BB_\Sigma(r)} \{ D(\gamma) -\hat D(\gamma) \}/( 4 \upsilon_2 \bar \alpha r)$, where $\upsilon_2 = \sqrt{2}\upsilon_1$. For any $s > 0$, using Chernoff's inequality gives
\#
	\PP\big\{ \Delta (r) \geq s \big\} \leq \exp \Bigg[ - \sup_{\lambda \geq 0}  \big\{ \lambda s - \log \EE e^{\lambda \Delta (r) } \big\} \Bigg] . \label{chernoff}
\#
The key is to bound the exponential moment $\EE e^{\lambda \Delta(r) }$. 
Applying first Rademacher symmetrization, and then the Ledoux-Talagrand contraction inequality (see, e.g. (4.20) in \cite{LT1991}), we obtain 
\$
	\EE e^{\lambda \Delta(r) } & \leq \EE \exp \Bigg\{ 2\lambda \sup_{\gamma\in \BB_\Sigma(r) } \frac{1}{4\upsilon_2 \bar \alpha r \sqrt{n} } \sn e_i \cdot r(\gamma; Z_i) \Bigg\} \\
	& \leq \EE \exp \Bigg\{ \frac{\lambda}{2 \upsilon_2  r}  \sup_{\gamma\in \BB_\Sigma(r) } \frac{1}{  \sqrt{n} } \sn e_i  X_i^\T \gamma    \Bigg\} \leq \EE \exp\Bigg( \frac{\lambda}{2\upsilon_2 }\bigg\| \frac{1}{   \sqrt{n} } \sn e_i W_i  \bigg\|_2 \Bigg) ,
\$
where $e_1, \ldots, e_n$ are independent Rademacher random variables.  Moreover, there exists a $(1/2)$-net $\cN$ of $\mathbb{S}^{p-1}$ with $|\cN|\leq 5^p$ such that $\|  \sn e_i W_i \|_2 \leq 2 \max_{u \in \cN}   \sn e_i W_i^\T u$. Recall from the proof of Lemma~\ref{lem:score.bound} that
\$
	\EE ( e_i W_i^\T u )^k  = 
	\begin{cases}
	 0 & \mbox{ if $k$ is odd}  \\
	 \EE(  W_i^\T u)^k  & \mbox{ if $k$ is even}
	\end{cases} 
	\leq \begin{cases}
	 0 & \mbox{ if $k$ is odd}  \\
	 \upsilon_2^k \cdot k \Gamma(k/2)  & \mbox{ if $k$ is even}
	\end{cases} .
\$
Hence, for any $\lambda \geq 0$,
\$
	 & \EE e^{\lambda e_i W_i^\T u/ \upsilon_2 } = 1 + \frac{\lambda^2}{2} \EE (W_i^\T u/\upsilon_2)^2 + \sum_{k=3}^\infty \frac{\lambda^k}{k!} \EE( e_i W_i^\T u /\upsilon_2)^k \\
	& \leq 1 + \frac{\lambda^2}{2} + \sum_{l=2}^\infty \frac{\lambda^{2l}}{(2l)!} 2 l \cdot (l-1)! \leq 1 + \frac{\lambda^2}{2}+ \sum_{l=2}^\infty \frac{(\lambda^2/\sqrt{2})^l}{l!} \leq e^{ \lambda^2/\sqrt{2} } . 
\$
This further implies
\$
\EE e^{\lambda \Delta(r) }&  \leq  \EE \exp\Bigg( \frac{\lambda}{2 \upsilon_2 }\bigg\| \frac{1}{\sqrt{n}} \sn e_i W_i \bigg\|_2 \Bigg)    \leq \EE \exp\Bigg(  \max_{u\in \cN} \frac{\lambda}{ \sqrt{n} } \sn e_i W_i^\T u /\upsilon_2  \Bigg) \\
& \leq \sum_{u \in \cN} \prod_{i=1}^n \EE e^{\lambda e_i W_i^\T u / ( \upsilon_2 \sqrt{n}) }\leq \sum_{u \in \cN} \prod_{i=1}^n   e^{\lambda^2/(\sqrt{2} n) } \leq  5^p \cdot e^{\lambda^2/\sqrt{2} } .
\$
Substituting this into \eqref{chernoff}, we obtain that 
\$
	\PP\big\{ \Delta(r) \geq s \big\} \leq  \exp \Bigg\{ - \sup_{\lambda \geq 0} \bigg( \lambda s -\frac{\lambda^2}{\sqrt{2}} \bigg) +  p \log (5) \Bigg\}  = \exp\Bigg\{ p \log (5) - \frac{s^2}{2 \sqrt{2}}  \Bigg\}.
\$
Finally, taking $s^2 = 2\sqrt{2} \{ \log(5) p + t\}$ proves \eqref{diff.loss.bound}.  \qed



\begin{thebibliography}{9}



\bibitem[{Abadie, Angrist and Imbens(2002)}]{AAI2002}
	{\sc Abadie, A., Angrist, J.} and {\sc Imbens, G.} (2002). 
	Instrumental variables estimates of the effect of subsidized training on the quantiles of trainee earnings.
	{\it Econometrica}  {\bf 70} 91--117.

\bibitem[{Acerbi and Sz\'ekely(2014)}]{AS2014}
	{\sc Acerbi, C.} and {\sc Sz\'ekely, B.} (2014). 
	Backtesting expected shortfall.
	{\it Risk}  {\bf 27} 76--81.
	
\bibitem[{Acerbi and Tasche(2002)}]{AT2002}
	{\sc Acerbi, C.} and {\sc Tasche, D.} (2002). 
	On the coherence of expected shortfall.
	{\it J. Bank. Finance}  {\bf 26} 1487--1503.

\bibitem[{Andrews(1991)}]{A1991}
	{\sc Andrews, D.\,W.\,K.} (1991). 
	Asymptotic normality of series estimators for nonparametric and semiparametric regression models.
	{\it Econometrica}  {\bf 59} 307--345.

\bibitem[{Barendse(2020)}]{B2020}
	{\sc Barendse, S.} (2020).
	Efficiently weighted estimation of tail and interquantile expectations.
	{\it Preprint}. \href{https://papers.ssrn.com/sol3/papers.cfm?abstract_id=2937665}{https://dx.doi.org/10.2139/ssrn.2937665.}
	
\bibitem[Barzilai and Borwein(1988)]{BB1988}	
	{\sc Barzilai, J.} and {\sc Borwein, J.\,M.} (1988).
 	Two-point step size gradient methods.
 	{\it IMA J. Numer. Anal.} {\bf 8} 141--148.
 	
\bibitem[{Basel Committee(2016)}]{BC2016}
	{\sc Basel Committee.} (2016).
	Minimum capital requirements for market risk. 
	Technical report,  Bank for International Settlements.
	\href{https://www.bis.org/bcbs/publ/d352.pdf}{https://www.bis.org/bcbs/publ/d352.pdf.}
		
\bibitem[{Basel Committee(2019)}]{BC2019}
	{\sc Basel Committee.} (2019).
	Minimum capital requirements for market risk. 
	Technical report,  Bank for International Settlements.
	\href{https://www.bis.org/bcbs/publ/d457.pdf}{https://www.bis.org/bcbs/publ/d457.pdf.}

\bibitem[{Bassett, Koenker and Kordas(2004)}]{BKK2004}
	{\sc Bassett,  G., Koenker, R.} and {\sc Kordas,  G.} (2004).
	Pessimistic portfolio allocation and Choquet expected utility.
	{\it J. Financ. Econom.} {\bf 2} 477--492.

\bibitem[{Bayer and Dimitriadis(2019)}]{BD2019}
	{\sc Bayer, S.} and {\sc Dimitriadis, T.} (2019).
	{\it esreg: Joint Quantile and Expected Shortfall Regression.}
	R package version 0.5.0, available at \href{https://cran.r-project.org/package=esreg}{https://cran.r-project.org/package=esreg}.

\bibitem[Belloni {\it et al.}(2019)]{BCCF2019}
	{\sc Belloni,  A.,  Chernozhukov,  V., Chetverikov, D.} and {\sc Fern\'andez-Val, I.} (2019).
	Conditional quantile processes based on series or many regressors.
	{\it J. Econom.}  {\bf 213} 4--29.

\bibitem[{Ben-Tal and Teboulle(1986)}]{BT1986}
	{\sc Ben-Tal, A.} and {\sc Teboulle, M.} (1986). 
	Expected utility, penalty functions, and duality in stochastic nonlinear programming.
	{\it Management Science}  {\bf 32} 1371--1520.

\bibitem[{Bloom {\it et al.}(1997)}]{BOBCDLB1997}
	{\sc Bloom, H., Orr, L., Bell, S., Cave, G., Doolittle, F., Lin, W.} and {\sc Bos, J.} (1997).
	The benefits and costs of JTPA Title II-A programs: key findings from the national job training partnership act study.
	\textit{J. Human Resources} \textbf{32} 549--576.	


\bibitem[Cai and Wang(2008)]{CW2008}
	{\sc Cai,  Z.} and {\sc Wang, X.} (2008).
	Nonparametric estimation of conditional VaR and expected shortfall.
	{\it J. Econom.}  {\bf 147} 120--130.

\bibitem[Camaschella(2015)]{camaschella2015iron}
	{\sc Camaschella, C.}  (2015).
	Iron-deficiency anemia.
	{\it N. England J. Med.}  {\bf 372} 1832--1843.

\bibitem[{Chen(2007)}]{C2007}
	{\sc Chen, X.} (2007).
	Large sample sieve estimation of semi-nonparametric models.
	{\it Handbook of Econometrics} {\bf 6} 5549--5632.
	
\bibitem[{Chernoff(1952)}]{C1952}
	{\sc Chernoff, H.} (1952).
	A measure of asymptotic efficiency for tests of a hypothesis based on the sum of observations.
	{\it Ann. Math. Statist.} {\bf 23} 493--507.


\bibitem[{Chernozhukov {\it et al.}(2018)}]{CCDDHNR2018}
	{\sc Chernozhukov, V., Chetverikov, D., Demirer, M., Duflo, E., Hansen, C., Newey, W.} and {\sc Robins,  J.} (2018).
	Double/debiased machine learning for treatment and structural parameters.
	{\it Econometrics Journal} {\bf 21} C1--C68. 

\bibitem[Chernozhukov and Hansen(2008)]{CH2008}
	{\sc Chernozhukov,  V.} and {\sc Hansen, C.} (2008).
	Instrumental variable quantile regression: A robust inference approach.  
	{\it J. Econom.}  {\bf 142} 379--398.
			
\bibitem[{Cont(2001)}]{C2001}
	{\sc Cont, R.} (2001).
	Empirical properties of asset returns: Stylized facts and statistical issues.
	{\it Quant. Finance} {\bf 1} 223--236.	

\bibitem[{de Haan and Ferreira(2006)}]{HF2006}
	{\sc de Haan, L.} and {\sc Ferreira, A.} (2006).
	{\it  Extreme Value Theory: An Introduction.}
	 Springer-Verlag, New York.
	
\bibitem[{Dimitriadis and Bayer(2019)}]{DB2019}
	{\sc Dimitriadis, T.} and {\sc Bayer, S.} (2019).
	A joint quantile and expected shortfall regression framework.
	{\it Electron. J. Statist.} {\bf 13} 1823--1871.
	
\bibitem[{Du and Escanciano(2017)}]{DE2017}
	{\sc Du, Z.} and {\sc Escanciano, J.\,C.} (2017). 
	Backtesting expected shortfall: Accounting for tail risk.
	{\it Management Science}  {\bf 63} 901--1269.
 
\bibitem[Eddelbuettel and Sanderson(2014)]{ES2014}
	{\sc Eddelbuettel, D.} and {\sc Sanderson, C.} (2014).
	RcppArmadillo: Accelerating R with high-performance C++ linear algebra.
	{\it Comput. Statist. Data Anal.} {\bf 71} 1054--1063.
	

\bibitem[{Eubank and Spiegelman(1990)}]{ES1990}
	{\sc Eubank, R.\,L.} and {\sc Spiegelman, C.\,H.} (1990). 
	Testing the goodness of fit of a linear model via nonparametric regression techniques.
	{\it J. Amer. Statist. Assoc.}  {\bf 85} 387--392.
	
	
\bibitem[Farrell,  Liang and Misra(2021)]{FLM2021}
	{\sc Farrell, M.\,H., Liang, T.} and {\sc Misra,  S.} (2021).
	Deep neural networks for estimation and inference.
	{\it Econometrica}  {\bf 89} 181--213.
	
\bibitem[{Fernandes, Guerre and Horta(2021)}]{FGH2021}
	{\sc Fernandes, M., Guerre, E.} and {\sc Horta, E.} (2021).
	Smoothing quantile regressions.
 	{\it J. Bus. Econ. Statist.} {\bf 39} 338--357.
 	
\bibitem[{Fissler and Ziegel(2016)}]{FZ2016}
	{\sc Fissler,  T.} and {\sc Ziegel,  J.\,F.} (2016).
	Higher order elicitability and Osband's principle.
	{\it Ann.  Statist.} {\bf 44} 1680--1707.
	
\bibitem[{Gneiting(2011)}]{G2011}
	{\sc Gneiting, T.} (2011). 
	Making and evaluating point forecasts.
	{\it J. Amer. Statist. Assoc.}  {\bf 106} 746--762.

\bibitem[{Hahn, Kuelbs and Weiner(1990)}]{HKW1990}
	{\sc Hahn, M.\,G., Kuelbs, J.} and {\sc Weiner, D.\,C.} (1990).
	The asymptotic joint distribution of self-normalized censored sums and sums of squares. 
	{\it Ann. Probab.} {\bf 18} 1284--1341.
	
\bibitem[{He, Hsu and Hu(2010)}]{HHH2010}
	{\sc He, X.}, {\sc Hsu, Y.-H.} and {\sc Hu, M.} (2010).
	Detection of treatment effects by covariate-adjusted expected shortfall.
	{\it Ann. Appl. Stat.} {\bf 4} 2114--2125.
	
\bibitem[He {\it et al.}(2021)]{HPTZ2021}
	{\sc He, X., Pan, X., Tan, K.\,M.} and {\sc Zhou, W.-X.} (2021).
	Smoothed quantile regression with large-scale inference.
	{\it J. Econom.} \href{https://www.sciencedirect.com/science/article/pii/S0304407621001950}{{https://doi.org/10.1016/j.jeconom.2021.07.010}}.	
	
\bibitem[He {\it et al.}(2022)]{conquer2022}
	{\sc He, X., Pan, X., Tan, K.\,M.} and {\sc Zhou, W.-X.} (2022).
	{\it conquer: Convolution-Type Smoothed Quantile Regression.}
	R package version 1.2.2, available at \href{https://cran.r-project.org/package=conquer}{https://cran.r-project.org/package=conquer}.
		
\bibitem[{He and Shao(2000)}]{HS2000}
	{\sc He, X.} and {\sc Shao, Q.-M.} (2000).
	On parameters of increasing dimensions.
	{\it J. Multivariate Anal.} {\bf 73} 120--135.		
 
\bibitem[{Hsu, Kakade and Zhang(2012)}]{HKZ2012}
{\sc Hsu, D., Kakade, S.\,M.} and {\sc Zhang, T.} (2012).
A tail inequality for quadratic forms of subgaussian random vectors.
\textit{Electron. Commun. Probab.} {\bf 52} 1--6.
	
\bibitem[{Huber(1973)}]{H1973}
	{\sc Huber, P.\,J.} (1973).
	{Robust estimation: Asymptotics, conjectures and Monte Carlo.}
	\textit{Ann. Statist.} \textbf{1} 799--821. 

\bibitem[{Kato(2012)}]{K2012}
	{\sc Kato, K.} (2012).
	Weighted Nadaraya-Watson estimation of conditional expected shortfall.
	{\it J. Financ. Econom.} {\bf 10} 265--291.

\bibitem[{Kratovil {\it el al.}(2007)}]{Ketal2007}
	{\sc Kratovil, T., DeBerardinis, J., Gallagher, N., Luban, N., Soldin, S.} and {\sc Wong, E.} (2007).
	Age specific reference intervals for soluble transferrin receptor (sTfR).
	{\it Clin. Chimica Acta} {\bf 1} 222--224.

	
\bibitem[Koenker(2022)]{K2022}
	{\sc Koenker, R.} (2022).
	Package ``\texttt{quantreg}", version $5.88$.
	Reference manual: \href{https://cran.r-project.org/web/packages/quantreg/quantreg.pdf}{https://cran.r-project.org/web/packages/quantreg/quantreg.pdf}.

\bibitem[Koenker and Bassett(1978)]{KB1978}	
	\textsc{Koenker, R.} and \textsc{Bassett, G.} (1978).
 	Regression quantiles.
 	\textit{Econometrica} \textbf{46} 33--50.	


\bibitem[{Linton and Xiao(2013)}]{LX2013}
	{\sc Linton, O.} and {\sc Xiao, Z.} (2013).
	Estimation and inference about the expected shortfall for time series with infinite variance.
	{\it Econom. Theory} {\bf 29} 771--807.

\bibitem[{Mandelbrot(1963)}]{M1963}
	{\sc Mandelbrot, B.} (1963).
	The variation of certain speculative prices.
 	{\it J. Bus.} {\bf 36} 394--419.

\bibitem[{Martins-Filho, Yao and Torero(2018)}]{MYT2018}
	{\sc Martins-Filho, C., Yao, F.} and {\sc Torero,  M.} (2018).
	Nonparametric estimation of conditional value-at-risk and expected shortfall based on extreme value theory.
	{\it Econom. Theory} {\bf 34} 23--67.

\bibitem[{Mast {\it et al.}(1998)}]{Mast1998}
	{\sc Mast, A., Blinder, M., Gronowski, A., Chumley, C.} and {\sc Scott, M.} (1998).
	Clinical utility of the soluble transferrin receptor and comparison with serum ferritin in several populations.
	{\it Clin. Chem.} {\bf 44} 45--51.




	
\bibitem[{McNeil, Frey and Embrechts(2015)}]{MFE2015}
	{\sc McNeil,  A.\,J., Frey, R.} and {\sc Embrechts, P.} (2015).
	{\it  Quantitative Risk Management: Concepts,  Techniques and Tools. 2nd Ed. }
	Princeton University Press,  Princeton.	

\bibitem[{Nemirovski and Yudin(1983)}]{NY1983}
	{\sc Nemirovski, A.} and {\sc Yudin, D.} (1983).
	{\it  Problem Complexity and Method Efficiency in Optimization.}
	Wiley.

\bibitem[Newey(1997)]{N1997}
	{\sc Newey, W.} (1997).
	Convergence rates and asymptotic normality for series estimators.
	{\it J. Econom.}  {\bf 79} 147--168.

\bibitem[{Neyman(1979)}]{N1979}
	{\sc Neyman, J.} (1979).
	$C(\alpha)$ tests and their use. 
	{\it Sankhya} {\bf 41} 1--21.

\bibitem[{Pan and Zhou(2022)}]{PZ2022}
	{\sc Pan, X.} and {\sc Zhou, W.-X.} (2022).
	{\it adaHuber: Adaptive Huber Estimation and Regression.}
	R package version 1.1, available at \href{https://cran.r-project.org/package=adaHuber}{https://cran.r-project.org/package=adaHuber}.
	
\bibitem[Patton,  Ziegel and Chen(2019)]{PZC2019}
	{\sc Patton, A.\,J., Ziegel, J.\,F.} and {\sc Chen,  R.} (2019).
	Dynamic semiparametric models for expected shortfall (and Value-at-Risk).
	{\it J. Econom.}  {\bf 211} 388--413.

\bibitem[Portnoy(1986)]{P1986}
	{\sc Portnoy, S.} (1986).
	On the central limit theorem in $\RR^p$ when $p\to \infty$.
	{\it Probab. Theory Relat. Fields} {\bf 73} 571--583.

\bibitem[{Rockafellar and Royset(2014)}]{RR2013}
	{\sc Rockafellar, R.\,T.} and {\sc Royset, J.\,O.} (2014). 
	Superquantiles and their applications to risk, random variables, and regression.
	In {\it INFORMS TutORials in Operations Research}, 151--167.
	
\bibitem[{Rockafellar, Royset and Miranda(2014)}]{RRM2014}
	{\sc Rockafellar, R.\,T., Royset, J.\,O.} and {\sc Miranda, S.\,I.} (2014). 
	Superquantile regression with applications to buffered reliability, uncertainty quantification, and conditional value-at-risk.
	{\it Eur. J. Oper. Res.}  {\bf 234} 140--154.

\bibitem[{Rockafellar and Uryasev(2000)}]{RU2000}
	{\sc Rockafellar, R.\,T.} and {\sc Uryasev, S.} (2000). 
	Optimization of conditional value-at-risk.
	{\it J. Risk}  {\bf 2} 21--42.

\bibitem[{Rockafellar and Uryasev(2002)}]{RU2002}
	{\sc Rockafellar, R.\,T.} and {\sc Uryasev, S.} (2002). 
	Conditional value-at-risk for general loss distributions.
	{\it J. Bank. Finance}  {\bf 26} 1443--1471.

\bibitem[Rockey and Cello(1993)]{rockey1993evaluation}
	{\sc Rockey, D.} and {\sc Cello, J.}  (1993).
	Evaluation of the gastrointestinal tract in patients with iron-deficiency anemia.
	{\it N. England J. Med.}  {\bf 329} 1691--1695.


\bibitem[{Scaillet(2005)}]{S2005}
	{\sc Scaillet, O.} (2005). 
	Nonparametric estimation of conditional expected shortfall.
	{\it Revue Assurances et Gestion des Risques/Insurance and Risk Management Journal}  {\bf 74} 639--660.
	
\bibitem[{Schmidt-Hieber(2020)}]{SH2020}
	{\sc Schmidt-Hieber, J.} (2020).
	Nonparametric regression using deep neural networks with ReLU activation function.
	{\it Ann. Statist.} {\bf 48} 1875--1897.
	
\bibitem[{Shapiro, Dentcheva and Ruszczynski(2014)}]{SDR2014}
	{\sc Shapiro, A., Dentcheva, D.} and {\sc Ruszczynski, A.} (2014).
	{\it Lectures on Stochastic Programming: Modeling and Theory. Second Edition.}
	 SIAM, 2014.
	 
\bibitem[{Shen {\it et al.}(2021)}]{Shen2021}
	{\sc Shen, G., Jiao, Y., Lin, Y., Horowitz, J.\,L.} and {\sc Huang, J.} (2021).
	Deep quantile regression: Mitigating the curse of dimensionality through composition.
	\textit{arXiv preprint arXiv:2107.04907.}


 
 \bibitem[{Wang {\it et al.}(2012)}]{WLH2012}
	{\sc Wang, H.\,J., Li, D.} and {\sc He, X.} (2012). 
	Estimation of high conditional quantiles for heavy-tailed distributions.
	{\it J. Amer. Statist. Assoc.}  {\bf 107} 1453--1464.

\bibitem[{Wang {\it et al.}(2021)}]{WZZZ2021}
	{\sc Wang, L., Zheng, C., Zhou, W.} and {\sc Zhou, W.-X.} (2021).
	A new principle for tuning-free Huber regression.
	\textit{Statistica Sinica} \textbf{31} 2153--2177.	
	
\bibitem[{Welsh(1989)}]{W1989}
	{\sc Welsh, A.\,H.} (1989).
	On $M$-processes and $M$-estimation.
	{\it Ann. Statist.} {\bf 15} 337--361.
	
\bibitem[{Zhou {\it et al.}(2018)}]{ZBFL2018}
	{\sc Zhou, W.-X., Bose, K., Fan, J.} and {\sc Liu, H.} (2018).
	A new perspective on robust $M$-estimation: Finite sample theory and applications to dependence-adjusted multiple testing.
	\textit{Ann. Statist.} \textbf{46} 1904--1931.
	
\end{thebibliography}

\begin{thebibliography}{9}

\bibitem[{Bousquet(2003)}]{B2003}
	{\sc Bousquet, O.} (2003).
	Concentration inequalities for sub-additive functions using the entropy method.
	{\it In Stochastic Inequalities and Applications. Progress in Probability} {\bf 56} 213--247. Birkh\"auser, Basel.

		
\bibitem[{Boyd and Vandenberghe(2004)}]{BV2004}
	{\sc Boyd,  S.} and {\sc Vandenberghe, L.} (2004).
	{\it Convex Optimization}.
	Cambridge University Press, Cambridge.
	
\bibitem[Catoni(2012)]{C2012}
	{\sc Catoni, O.} (2012).
	Challenging the empirical mean and empirical variance: A deviation study.
	{\it Ann. Inst. Henri Poincar\'e Probab. Stat.} {\bf 48} 1148--1185.

	
\bibitem[{Chen and Zhou(2020)}]{CZ2020}
	{\sc Chen, X.} and {\sc Zhou, W.-X.} (2020).
	Robust inference via multiplier bootstrap.
	\textit{Ann. Statist.} \textbf{48} 1665--1691. 

\bibitem[{Chernozhukov(2005)}]{C2005}
	{\sc Chernozhukov, C.} (2005).
	Extremal quantile regression.
	{\it Ann. Statist.} {\bf 33} 806--839.
	
	
	
\bibitem[{Dimitriadis and Bayer(2019)}]{DB2019}
	{\sc Dimitriadis, T.} and {\sc Bayer, S.} (2019).
	A joint quantile and expected shortfall regression framework.
	{\it Electron. J. Statist.} {\bf 13} 1823--1871.
	
	
\bibitem[Einmahl and Li(2008)]{EL2008}	
	\textsc{Einmahl, U.} and \textsc{Li, D.} (2008).
 	Characterization of LIL behavior in Banach space.
 	\textit{Trans. Am. Math. Soc.} \textbf{360} 6677--6693.

\bibitem[{Fu, Narasimhan and Boyd(2020)}]{FNB2020}
	{\sc Fu, A., Narasimhan, B.} and {\sc Boyd, S.} (2020).
	CVXR: An R package for disciplined convex optimization.
	{\it J. Stat. Softw.} {\bf 94} 1--34. 

\bibitem[{Goldfarb and Idnani(1983)}]{GI1983}
	{\sc Goldfarb, D.} and {\sc Idnani, A.} (1983).
	A numerically stable dual method for solving strictly convex quadratic programs. 
	{\it Math. Program.} {\bf 27} 1--33.
	
\bibitem[{Koenker and Xiao(2006)}]{KX2006}
	{\sc Koenker,  R.} and {\sc Xiao, Z.} (2006). 
	Quantile autoregression.
	{\it J. Amer. Statist. Assoc.}  {\bf 101} 980--990.

\bibitem[Kuchibhotla and Rinaldo(2020)]{KR2020}
	{\sc Kuchibhotla,  A.\,K.} and {\sc Rinaldo, A.} (2020).
	High-dimensional CLT for sums of non-degenerate random vectors: $n^{-1/2}$-rate.
	{\it arXiv preprint arXiv:2009.13673}.
	
\bibitem[{Ledoux and Talagrand(1991)}]{LT1991}
	{\sc Ledoux, M.} and {\sc Talagrand, M.} (1991).
	{\it Probability in Banach Spaces: Isoperimetry and Processes}.
	Springer-Verlag, Berlin.

\bibitem[{McNeil, Frey and Embrechts(2015)}]{MFE2015}
	{\sc McNeil,  A.\,J., Frey, R.} and {\sc Embrechts, P.} (2015).
	{\it  Quantitative Risk Management: Concepts,  Techniques and Tools. 2nd Ed. }
	Princeton University Press,  Princeton.	
	
\bibitem[{Nazarov(2003)}]{N2003}
	{\sc Nazarov,  F.} (2003).
	On the maximal perimeter of a convex set in $\RR^n$ with respect to a Gaussian measure.
	In {\it Geometric Aspects of Functional Analysis. Lecture Notes in Math.} {\bf 1807} 169--187.  Springer, Berlin.	

\bibitem[{Oliveira(2016)}]{O2016}
	{\sc Oliveira, R.\,I.} (2016).
	The lower tail of random quadratic forms with applications to ordinary least squares.
	{\it Probab. Theory Relat. Fields} {\bf 166} 1175--1194.

\bibitem[Patton,  Ziegel and Chen(2019)]{PZC2019}
	{\sc Patton, A.\,J., Ziegel, J.\,F.} and {\sc Chen,  R.} (2019).
	Dynamic semiparametric models for expected shortfall (and Value-at-Risk).
	{\it J. Econom.}  {\bf 211} 388--413.


\bibitem[{Shevtsova(2014)}]{S2014}
	{\sc Shevtsova,  I.\,G.} (2014).
	On the absolute constants in the Berry-Esseen-type inequalities.
	{\it Doklady Mathematics} {\bf 89} 378--381.
	
\bibitem[Spokoiny(2013)]{S2013}
	{\sc Spokoiny, V.} (2013).
	Bernstein--von Mises theorem for growing parameter dimension.
	{\it arXiv preprint  arXiv:1302.3430}.
	
\bibitem[{Sun, Zhou and Fan(2020)}]{SZF2020}
	{\sc Sun, Q., Zhou, W.-X.} and {\sc Fan, J.} (2020). 
	Adaptive Huber regression.
	{\it J. Amer. Statist. Assoc.}  {\bf 115} 254--265.

\bibitem[Velthoen {\it et al.}(2021)]{VDCE2021}	
	{\sc Velthoen, J., Dombry, C., Cai, J.-J.} and {\sc Engelke, S.} (2021).
 	Gradient boosting for extreme quantile regression.
 	\textit{arXiv preprint arXiv:2103.00808.} 
 
\bibitem[{Vershynin(2018)}]{V2018}
	{\sc Vershynin, R.} (2018).
	{\it High-Dimensional Probability}.
	Cambridge University Press, Cambridge.	
	
\bibitem[{Wainwright(2019)}]{W2019}
	{\sc Wainwright, M.\,J.} (2019).
	{\it High-Dimensional Statistics: A Non-Asymptotic Viewpoint}.
	Cambridge University Press, Cambridge.

\bibitem[Zhivotovskiy(2022)]{Z2022}
{\sc Zhivotovskiy, N.} (2022).
Dimension-free bounds for sums of independent matrices and simple tensors via the variational principle.
{\it arXiv preprint arXiv:2108.08198}.

\end{thebibliography}
\end{document}